\documentclass[acmlarge,dvipsnames]{acmart}

\usepackage{booktabs} % For formal tables
\usepackage{gensymb}
\usepackage{multirow}
\usepackage{xspace}
\usepackage{algorithm}
\usepackage[noend]{algpseudocode}
\usepackage{algorithmicx}
\floatname{algorithm}{Template}
\usepackage{siunitx}
\usepackage{fixltx2e}
\usepackage{subfig}
\usepackage{pifont}
\usepackage{hhline}
\usepackage{listings}
\usepackage{enumitem}
\usepackage{xstring}
\usepackage[utf8]{inputenc}
\usepackage[us,12hr]{datetime}

\newif\ifcameraready
\camerareadytrue
\newcommand{\versionnum}[0]{2.4}

\renewcommand\footnotetextcopyrightpermission[1]{}
\ifcameraready
\else
  \fancypagestyle{firststyle}
  {
    \fancyhead[C]{\textcolor{MidnightBlue}{\emph{Version \versionnum~---~\today, \ampmtime}}}
  }
  \fi
\setcopyright{none}

\newcommand{\fullname}[0]{In-Memory PoInter Chasing Accelerator\xspace}
\newcommand{\abbrv}[0]{IMPICA\xspace}

\newcommand{\ignore}[1]{}
\newcommand{\changes}[0]{}
\newcommand{\change}[0]{}
\newcommand{\sg}[0]{}
\newcommand{\changesii}[0]{}
\newcommand{\rachata}[0]{}

\newcommand{\kh}[0]{}
\newcommand{\kii}[0]{}
\newcommand{\kiii}[0]{}
\newcommand{\ch}[0]{}
\newcommand{\chii}[0]{}
\newcommand{\chiii}[0]{}
\newcommand{\chiv}[0]{}
\newcommand{\chv}[0]{}
\newcommand{\chvi}[0]{}
\newcommand{\chvii}[0]{}
\ifcameraready
  \newcommand{\todo}[1][]{}
  \newcommand{\chviii}[0]{}
  \newcommand{\chix}[0]{}
\else
  \newcommand{\todo}[1][]{\textbf{\scriptsize \fcolorbox{black}{red}{\color{white}{TODO}}} \underline{$\overline{\hbox{\emph{#1}}}$}}
  \newcommand{\chviii}[1]{\textcolor{MidnightBlue}{#1}}
  \newcommand{\chix}[1]{\textcolor{BrickRed}{#1}}
\fi

\newcommand{\paratitle}[1]{\vspace{3pt}\textbf{#1.}}

\newcommand{\incircle}[1]{%
    \IfEqCase{#1}{%
        {1}{\ding{182}}%
        {2}{\ding{183}}%
        {3}{\ding{184}}%
        {4}{\ding{185}}%
        {5}{\ding{186}}%
        {6}{\ding{187}}%
        {7}{\ding{188}}%
        {8}{\ding{189}}%
    }[\PackageError{incircle}{Undefined option to incircle: #1}{}]%
}%
\newcommand{\squishlist}{
   \begin{itemize}
   \itemsep 0pt
}
\newcommand{\squishend}{
    \end{itemize}
}

\newcommand{\footnoteref}[1]{\textsuperscript{\ref{#1}}}

\begin{document}
% Title portion
\title[Enabling the Adoption of Processing-in-Memory]{Enabling the Adoption of Processing-in-Memory:\\
Challenges, Mechanisms, Future Research Directions}

\author{Saugata Ghose}
\author{Kevin Hsieh}
\author{Amirali Boroumand}
\author{Rachata Ausavarungnirun}
\affiliation{%
  \institution{Carnegie Mellon University}
}

\author{Onur Mutlu}
\affiliation{%
  \institution{ETH Z{\"u}rich and Carnegie Mellon University}
}

\renewcommand{\shortauthors}{S. Ghose et al.}


% TODO



% !TEX root=../chapter.tex

\begin{abstract}

Performance improvements from DRAM \rachata{technology} scaling have been lagging behind the
improvements from logic \ch{technology} scaling for \rachata{many years. As} application demand for main memory continues to grow, \rachata{DRAM-based main memory} is increasingly becoming a \chvi{larger}
system bottleneck \chv{in terms of both performance and energy consumption}.  A major reason for poor \rachata{memory} performance \chv{and energy efficiency} is \rachata{memory's} inability
to perform computation.  Instead, data stored within DRAM \chv{memory \emph{must}} be moved into
the CPU before any computation can take place.  This data movement is costly,
as it requires a high latency and consumes significant energy to transfer the
data across the pin-limited memory channel.  Moreover, the data moved to the
CPU is often not reused, and thus does not benefit from being cached within the
CPU, which makes it difficult to amortize the overhead of data movement.

Modern 3D-stacked DRAM architectures provide an opportunity to avoid unnecessary
data movement \rachata{between memory and the CPU}. These multi-layer architectures include a \emph{logic layer},
where compute logic can be integrated \chv{underneath} 
\chv{multiple layers of} DRAM cell arrays \chv{(i.e., the \emph{memory layers}) within the same chip}.
Architects can take advantage of \chv{the} logic layer to perform 
\emph{processing-in-memory} (PIM), or \emph{near-data processing}, where some
of the computation is moved from the CPU to the logic \chv{layer underneath the memory layer}. 
In a PIM architecture, the logic \chv{layer} within DRAM has access to the high internal
bandwidth available within 3D-stacked DRAM (which is much greater than the
bandwidth available in the narrow memory channel between DRAM and the CPU).
\chv{Thus,} PIM architectures can effectively free up valuable bandwidth on the \chv{bandwidth-limited} memory
channel while \chv{at the same time} reducing system energy consumption.

A number of \rachata{important} issues arise when we add compute logic
to DRAM.  In particular, logic within DRAM does not have low-latency access to common CPU
structures that are essential for modern application execution, such as the 
\rachata{virtual memory mechanisms, e.g., the}
translation lookaside buffer (TLB) \rachata{or the page table walker,} and the \ch{cache coherence mechanisms, e.g., the coherence} directory.  
\chvi{To ease the widespread adoption of PIM, we ideally \chviii{would like} to maintain traditional
virtual memory abstractions and the shared memory programming model.  
This requires}
\emph{efficient mechanisms} that can
provide logic in DRAM with access to \rachata{virtual memory and cache coherence} without having to
communicate frequently with the CPU, as \chvi{off-chip} communication between the CPU and DRAM
consumes much of the limited bandwidth that PIM aims to avoid using.
\rachata{To this end, we} propose \rachata{and evaluate} two general-purpose \chvi{solutions} that can be used by PIM architectures
to minimize unnecessary \chvi{off-chip communication}.
The first, IMPICA, is an efficient in-memory accelerator for pointer chasing, 
which can handle address translation entirely within DRAM.  The second, LazyPIM,
provides \chvi{coherence support \emph{without}} the need to continually
communicate with the \chvi{CPU}.  We show that both of these
mechanisms provide a significant benefit for a number of \chv{important} memory-intensive
applications, \rachata{thereby both} improving performance and reducing energy consumption.

\end{abstract}

\maketitle

\ifcameraready
\else
  \thispagestyle{firststyle}
\fi

% ===================
% The Main Body
% ===================

% !TEX root=../chapter.tex

\ch{DRAM, the predominant technology used to build main memory,} is a major component of modern computer systems. \rachata{As} the \chvi{data} working set
sizes of modern applications grow~\cite{dean.cacm13, kanev.isca15, ferdman.asplos12, wang.hpca14},
the need for \chv{higher} \ch{memory} capacity and \chv{higher memory} performance continues to grow as well.
However, even though CMOS \rachata{technology} scaling has yet to come to an end, DRAM \ch{technology} scaling
has been unable to keep up with the increasing memory demand from 
applications\chv{~\cite{dean.cacm13, kanev.isca15, mckee.cf04, mutlu.imw13,
mutlu.superfri15, wilkes.can01,kim-isca2014,salp,kang.memoryforum14,yoongu-thesis,raidr,mutlu2017rowhammer,ahn.tesseract.isca15,ahn.pei.isca15,hsieh.isca16,donghyuk-ddma,lee-isca2009,rbla,yoon-taco2014,lim-isca09, wulf1995hitting, chang.sigmetrics16, lee.hpca13, lee.hpca15, chang.sigmetrics17, lee.sigmetrics17,
luo.dsn14, luo.arxiv17,hassan2017softmc,chargecache}}.
For example, if we study the latency and throughput of Double Data Rate (DDR)
DRAM over the last \chv{15--20}~years, we see that neither have been able to keep up
with the growth in \chvi{application} working set size or CPU computational power \chv{~\cite{chang.sigmetrics16,kevinchang-thesis, lim-isca09,lee.thesis16,lee.hpca13}}.

A major bottleneck to improving \chv{the overall system} performance is the high cost of
\emph{data movement}.  Currently, in order to perform an operation on data
that is stored within \ch{memory}, the CPU must issue a request to the memory
controller, which in turn sends a series of commands across an off-chip bus
to the DRAM module.  The data is then read from the DRAM module, at which
point it is returned to the memory controller and typically stored within
the CPU cache.  Only after the data is placed in the CPU cache can the CPU
operate \chvi{(i.e., perform computation) on} the data.  The long latency to retrieve data from DRAM is
exacerbated by two factors.  First, it is difficult to \chv{send} a large
number of requests to \ch{memory} \chv{in parallel}, in part because of the narrow bandwidth of the
off-chip bus between the memory controller and \ch{main memory}.  Second, despite
the time spent on bringing the data into the cache, \ch{which is substantial~\cite{hashemi.isca16,cont-runahead},}
much of \rachata{the data brought into the caches is \emph{not}} reused by the CPU \chv{~\cite{qureshi.isca07,qureshi-hpca07}},
rendering the caching \rachata{either very inefficient or sometimes even unnecessary}.  Ultimately, there is significant time
and energy wasted on moving data between the CPU and \ch{memory}, \rachata{many times} with little
benefit in return, \rachata{especially in workloads where caching is not very effective\chv{~\cite{ahn.tesseract.isca15,ahn.pei.isca15}}.}

Recent advances in \ch{memory} design have unlocked the potential to avoid the
unnecessary data movement.  In an attempt to improve the scalability of
capacity and bandwidth, \chvi{memory} \ch{manufacturers} have turned to 3D-stacked memories,
where multiple layers of \ch{memory} arrays are stacked on top of each other~\cite{loh2008stacked, lee.taco16}.
These layers are connected together using \emph{through-silicon vias} (TSVs),
which provide much greater internal \ch{memory} bandwidth than the narrow off-chip
bus to the CPU.
Some prominent examples of these 3D-stacked \chv{memory} 
architectures\ch{~\cite{jeddeloh2012hybrid,jedec.hbm.spec,lee.taco16, hmc.spec.1.1, hmc.spec.2.0, ramulator}}
include a \emph{logic layer}, which provides an opportunity to embed general-purpose
computational logic \emph{directly within main memory} to take advantage of the
high internal bandwidth available.

The idea of performing \emph{processing-in-memory} (PIM), or \emph{near-data
processing} (NDP), has been proposed for \chvi{at least} several decades\ch{~\cite{stone1970logic,shaw1981non, elliott1992computational,
    kogge1994execube, gokhale1995processing, patterson1997case,
    oskin1998active, kang2012flexram, Draper:2002:ADP:514191.514197,
    Mai:2000:SMM:339647.339673, elliott.dt99}}.
However, these past efforts were \chv{\emph{not}} adopted at large scale due to the
difficulty of integrating processing elements with DRAM.
As a result of the \ch{potential} enabled by the inclusion of a logic layer in 
modern \ch{memory} architectures, \chv{various} recent works explore a range of PIM architectures for 
\chv{multiple different} purposes (e.g., \chviii{\cite{zhu2013accelerating, pugsley2014ndc, zhang.hpdc14,
    farmahini-farahani.hpca15, ahn.tesseract.isca15, ahn.pei.isca15,
    loh2013processing, hsieh.isca16, pattnaik.pact16,
    DBLP:conf/isca/AkinFH15, impica,
    DBLP:conf/sigmod/BabarinsaI15, DBLP:conf/IEEEpact/LeeSK15,
    DBLP:conf/hpca/GaoK16, chi.isca16, gu.isca16, kim.isca16,
    asghari-moghaddam.micro16, boroumand2016pim,
    hashemi.isca16, cont-runahead, GS-DRAM, liu-spaa17,
    gao.pact15, guo2014wondp, sura.cf15,
    morad.taco15, hassan.memsys15, li.dac16, kang.icassp14, aga.hpca17,
    shafiee.isca16, seshadri2013rowclone, Seshadri:2015:ANDOR, 
    chang.hpca16, seshadri.arxiv16, seshadri.micro17, nai2017graphpim,kim.arxiv17,kim.bmc18, li.micro17, kim.sc17, boroumand.asplos18}}).

While PIM avoids the need to move data from \ch{memory} to the CPU for a number
of data-intensive functions, it introduces new challenges for \rachata{system} architects \ch{and programmers}.
In particular, PIM processing logic does \chv{\emph{not}} have quick access to \rachata{important} mechanisms
that exist within the CPU, such as \ch{address translation} and cache
coherence, \chv{which greatly aid the programmer}.  Preserving the functionality \ch{and efficiency} of such mechanisms is essential for
PIM, as these mechanisms can 
\chvi{(1)~preserve the traditional programming models that application developers
rely on to productively write programs, and
(2)~provide significant performance benefits.}. As we \chv{show in this work (see Section~\ref{sec:lazypim})}, 
simply forcing PIM processing logic to send queries
to the CPU \chv{to accomplish} \rachata{address translation and cache coherence} is \chv{very} inefficient, \chv{since} the overhead of a query can \rachata{almost completely}
\chv{\emph{eliminate}} the benefits of moving computation to \ch{memory}.  Therefore, it is
essential that we provide \emph{PIM-specific} mechanisms that 
\emph{efficiently} support the functionality of \rachata{traditional address translation and cache coherence} mechanisms 
\chv{and thus provide support for the use of existing programming models to program PIM architectures}.  
Our goal is to design general-purpose \rachata{address translation and cache coherence}
mechanisms that can be exploited by any \chiii{PIM processing logic} to provide low-overhead
support for common functions, such as pointer chasing \ch{in virtual memory} and cache coherence.

To this end, we propose two mechanisms to support PIM.  The first mechanism,
IMPICA, is an in-memory accelerator for pointer chasing, which exploits the
high bandwidth available within 3D-stacked \ch{memory}.  IMPICA can traverse a chain
of virtual memory pointers within DRAM, \ch{\emph{without}} having to look up
virtual-to-physical address translations in the CPU translation lookaside
buffer (TLB) \rachata{or without using the page walkers within the CPU}.  The second mechanism, LazyPIM, maintains cache coherence
between \chiii{PIM processing logic} and CPU cores \ch{\emph{without}} sending coherence requests for every
memory access. 
\ch{\chii{Instead,} LazyPIM efficiently \chv{provides} coherence by having \chiii{PIM processing logic}
speculatively acquire coherence permissions, and then later sends compressed
\emph{batched} coherence lookups to the CPU to determine \chv{whether or not}}
its speculative permission acquisition \chv{violated} the
memory ordering defined by the programming model.

In Section~\ref{sec:pimarch}, we cover common design principles of modern 
PIM architectures.
In Section~\ref{sec:keyissues}, we discuss key issues that impact the flexibility and adoption of
PIM architectures.
In Section~\ref{sec:impica}, we discuss IMPICA, an accelerator that we propose to efficiently support
pointer-chasing operations within PIM architectures.
In Section~\ref{sec:lazypim}, we discuss LazyPIM, a mechanism that we propose to efficiently support
cache coherence between PIM processing logic and the CPU cores.
In Section~\ref{sec:related}, we discuss related work in the area, \rachata{and in Section~\ref{sec:future} we briefly discuss some future research challenges}, \chv{with a focus on system-level challenges for the 
adoption of PIM architectures}.

% !TEX root=../chapter.tex

\section{Designing Processing-in-Memory Architectures}
\label{sec:pimarch}

Processing-in-memory (PIM) architectures place some form of processing logic
\chvi{(}typically accelerators, simple cores, \chvi{or reconfigurable logic)} inside \chv{the DRAM subsystem}.  
This \chiii{\emph{PIM processing logic}, which we also refer to as 
\emph{PIM cores} or \emph{PIM engines}, interchangeably,} can execute portions
of applications or entire application kernels, depending on the design of the
architecture.  In this section, we first discuss how the \chiii{PIM processing logic is}
integrated within DRAM modules (Section~\ref{sec:pimarch:logic}),
and then we discuss how applications make use of \chiii{this PIM processing logic}
(Section~\ref{sec:pimarch:offloading}).

\subsection{Placing \chiii{Processing} Logic Within \chv{the DRAM Subsystem}}
\label{sec:pimarch:logic}

Modern PIM architectures rely on implementing processing logic
in the DRAM chip itself (e.g., \chviii{\cite{zhu2013accelerating, pugsley2014ndc, zhang.hpdc14,
    farmahini-farahani.hpca15, ahn.tesseract.isca15, ahn.pei.isca15,
    loh2013processing, hsieh.isca16, pattnaik.pact16,
    DBLP:conf/isca/AkinFH15, impica,
    DBLP:conf/sigmod/BabarinsaI15, DBLP:conf/IEEEpact/LeeSK15,
    DBLP:conf/hpca/GaoK16, chi.isca16, gu.isca16, kim.isca16,
    boroumand2016pim,
    GS-DRAM, liu-spaa17,
    gao.pact15, guo2014wondp, sura.cf15,
    morad.taco15, hassan.memsys15,
    seshadri2013rowclone, Seshadri:2015:ANDOR, 
    chang.hpca16, seshadri.arxiv16, seshadri.micro17,seshadri.bookchapter17,seshadri.bookchapter17.arxiv, donghyuk-ddma,
kim.bmc18,kim.arxiv17, li.micro17, kim.sc17, boroumand.asplos18}})
or on the DRAM module \chv{or the DRAM controller\chv{~\cite{asghari-moghaddam.micro16,GS-DRAM, hashemi.isca16, cont-runahead}}}.
DRAM consists of multiple arrays of capacitive \emph{cells}, where each cell
holds one bit of data.  By placing processing logic in close proximity of the
cell arrays, PIM architectures are \chv{\emph{not}} restricted to the \chvi{limited} bandwidth
offered by the \chvi{narrow} off-chip bus between the DRAM module and the CPU.
Instead, PIM processing logic benefits from the \chvi{much} wider buses that are available
within the chip and/or module in modern DRAM architectures.

\begin{figure}[h]
  \centering
  \includegraphics[width=0.5\textwidth]{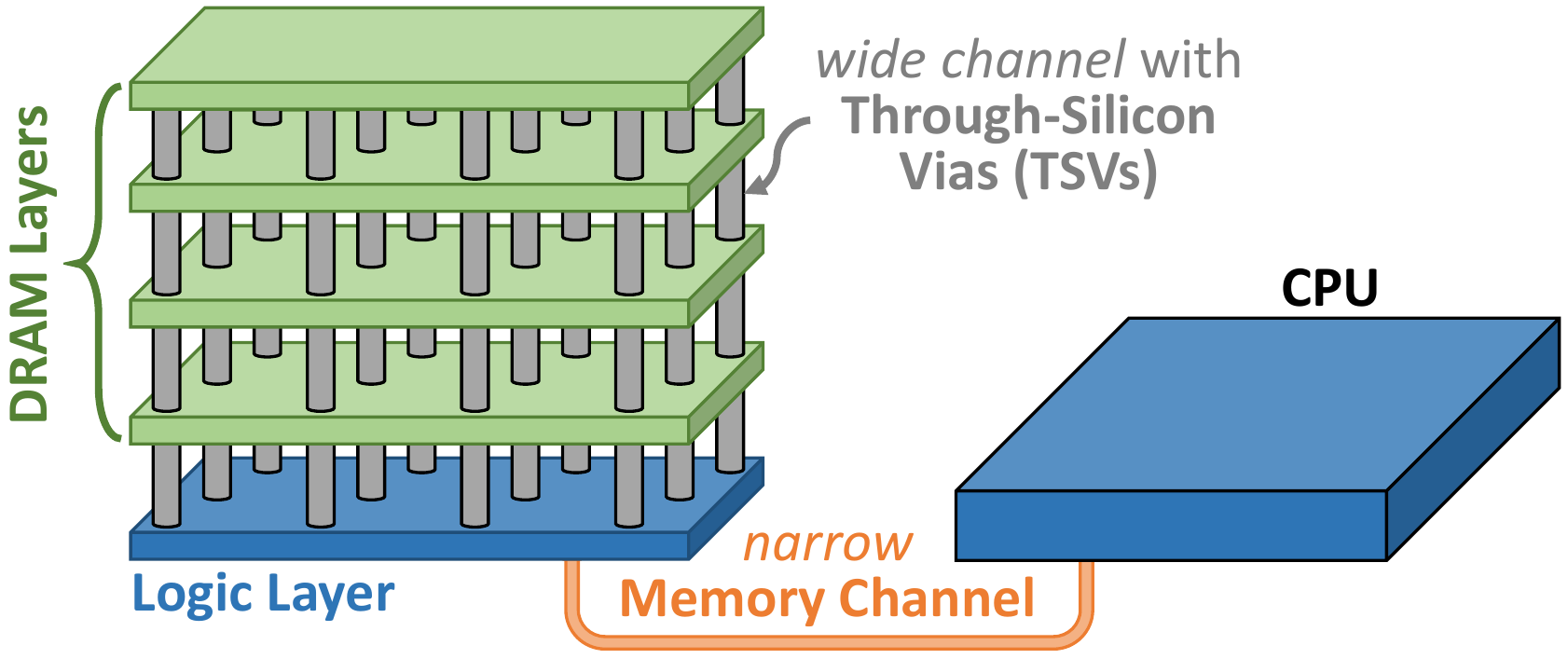}
  \caption{\ch{High-level overview of a 3D-stacked \chii{DRAM based} architecture.}}
  \label{fig:3d-dram}
\end{figure}

Figure~\ref{fig:3d-dram}
shows an overview of a 3D-stacked \chii{DRAM based architecture.  Examples of 3D-stacked DRAM include} High-Bandwidth
Memory (HBM)\ch{~\cite{jedec.hbm.spec,lee.taco16}} \chii{and} the Hybrid Memory Cube (HMC)\chiii{~\cite{hmc.spec.1.1, hmc.spec.2.0, ahn.tesseract.isca15}}.
As the figure shows, a 3D-stacked DRAM consists of multiple layers.
3D-stacked DRAM has a much greater internal data bandwidth than conventional
memory, due to its use of \emph{through-silicon vias} (TSVs), which are 
vertical links that connect the multiple layers of a DRAM stack 
together~\cite{loh2008stacked, lee.taco16}.
In addition to containing multiple layers of DRAM,
a number of 3D-stacked DRAM architectures, such as \chv{HBM~\cite{jedec.hbm.spec,lee.taco16}} and 
HMC~\cite{hmc.spec.1.1, hmc.spec.2.0}, include a \emph{logic layer}, 
typically the bottommost layer, where architects can implement functionality that
interacts with both the processor and the DRAM cells\chv{~\cite{hmc.spec.1.1, hmc.spec.2.0,ahn.tesseract.isca15,ahn.pei.isca15}}.
Currently, 3D-stacked DRAM makes limited use of the logic layer
(e.g., HMC implements command scheduling logic within the logic 
layer~\cite{hmc.spec.1.1, hmc.spec.2.0}).

Recent PIM proposals (e.g., \chviii{\cite{zhu2013accelerating, pugsley2014ndc, zhang.hpdc14,
    farmahini-farahani.hpca15, ahn.tesseract.isca15, ahn.pei.isca15,
    loh2013processing, hsieh.isca16, pattnaik.pact16,
    DBLP:conf/isca/AkinFH15, impica,
    DBLP:conf/sigmod/BabarinsaI15, DBLP:conf/IEEEpact/LeeSK15,
    DBLP:conf/hpca/GaoK16, chi.isca16, gu.isca16, kim.isca16,
    boroumand2016pim, liu-spaa17,
    gao.pact15, guo2014wondp, sura.cf15,
    morad.taco15, hassan.memsys15, nai2017graphpim, seshadri.micro17, 
    kim.bmc18, kim.arxiv17, kim.sc17, boroumand.asplos18}})
add processing logic to the logic layer to exploit the high bandwidth available \rachata{between the logic layer and the DRAM cell arrays}.
The proposed \chiii{PIM processing logic} design varies based on the specific architecture, and can
range from fixed-function accelerators
to simple \emph{in-order} cores,
\chvi{and to reconfigurable logic}.
The complexity of the \chiii{processing logic} that can be added to the logic layer is currently
limited by the manufacturing process technology \rachata{and thermal design points,
which may prevent} highly-sophisticated processors (e.g., out-of-order processor
cores with large caches and sophisticated \chv{instruction-level parallelism} techniques) from being 
implemented within the logic layer at this time\ch{~\cite{pattnaik.pact16, eckert.wondp14, zhang.hpdc14}}.

\subsection{Using PIM \chiii{Processing} Logic \rachata{Functionality} in Applications}
\label{sec:pimarch:offloading}

In order for applications to make use of \chiii{PIM processing logic that resides} within DRAM,
each PIM architecture exposes an interface to the CPU.  While there currently is no
standardization of this interface, most contemporary works on PIM architectures
follow similar models for CPU--PIM interactions.  \chiii{PIM processing logic} is typically 
treated as a coprocessor, and executes only \ch{when some} code (which we refer 
to as a \emph{PIM kernel}) is launched by the CPU on the \chiii{PIM processing logic}.  PIM kernels
vary widely in current proposals \rachata{in terms of granularity}.
Some works (e.g., \cite{ahn.tesseract.isca15, DBLP:conf/hpca/GaoK16})
treat an \chv{\emph{entire application}} thread as a PIM kernel, in order to minimize the 
amount of synchronization and data sharing that takes place \chvi{between the
main CPU and main compute-capable memory}.
Many works (e.g., \chviii{\cite{hsieh.isca16, gao.pact15, farmahini-farahani.hpca15, 
asghari-moghaddam.micro16, zhang.hpdc14, seshadri2013rowclone, Seshadri:2015:ANDOR, 
liu-spaa17, guo2014wondp, pattnaik.pact16, DBLP:conf/isca/AkinFH15, seshadri.arxiv16, seshadri.micro17,
kim.bmc18, kim.arxiv17, kim.sc17, li.micro17, boroumand.asplos18}})
treat only portions of an application thread (e.g., \chvi{\emph{functions}}) as a PIM kernel,
and launch the kernel when a CPU core reaches the PIM kernel call.
Yet other works (e.g., \chii{\cite{ahn.pei.isca15,nai2017graphpim, DBLP:conf/IEEEpact/LeeSK15}}) \rachata{use a much finer granularity
for offloading code to PIM: they offload only a \chvi{\emph{single instruction}} as the PIM kernel,
which is completed atomically.}

An open question for all of these architectures is how a PIM kernel is
identified and demarcated, and who is responsible for identification and
demarcation.  Current works on PIM expect the compiler or programmer to mark
sections of the code and/or data that are to be dispatched to the \chiii{PIM processing logic}.
When a program reaches a point at which a PIM kernel should be executed, the
CPU uses the off-chip memory channel to dispatch the kernel to a free PIM core.
The PIM core then executes the kernel, and upon completing the kernel, notifies
the CPU using the memory channel. \rachata{Several works~\cite{hsieh.isca16,boroumand.arxiv17,ahn.pei.isca15,ahn.tesseract.isca15} provide
a detailed explanation of this process.}

Due to the simple nature of \chiii{PIM processing logic} (i.e., \rachata{that}
\chiii{PIM processing logic} is expected to be 
\chvi{fixed-function accelerators, small in-order general-purpose cores, or
simple reconfigurable logic}), current
PIM architectures do \ch{\emph{not replace}} the CPU cores with PIM cores\rachata{; they} 
instead \emph{augment} the existing CPU cores.  OS threads continue to run on
the CPU cores, and key structures to support application execution, such as
large caches, translation lookaside buffers (TLBs), \rachata{page walkers}, and cache coherence
\chvi{hardware}, are expected to remain within the CPU.  While these decisions
minimize the changes that need to be made to write programs for PIM 
architectures, the decisions introduce new issues that PIM architectures must
address \rachata{to maintain ease of adoption}.  We discuss \rachata{two such critical} issues in the next section.

% !TEX root=../chapter.tex

\section{Key Issues in Enabling Processing-in-Memory}
\label{sec:keyissues}

Pushing some or all of the computation for a program from the CPU to the DRAM introduces
new challenges for \rachata{system architects (as well as programmers)} to overcome.  In particular, \chiii{PIM processing logic does} \chv{\emph{not}}
have direct access to structures within the CPU that are essential to memory
operations, such as \rachata{address translation and cache coherence} hardware.  A naive
solution is to simply have \chiii{PIM processing logic} access these structures remotely over the
off-chip memory channel.  Unfortunately, \rachata{such likely-frequent} remote accesses can introduce a high performance and
energy overhead, and often undermine many, if not all, of the benefits that
PIM architectures provide.  A second naive solution is to limit the \rachata{functionality of the} \chiii{PIM processing logic} such
that \chiii{it} \ch{\emph{cannot}} perform address translation or cache coherence, and to expose
these limitations to programmers.  However, this alters the programming model
of the system, and can lead to great difficulty for the \rachata{widespread adoption of PIM as an execution model}.
In this section, we focus on the \rachata{address translation and cache coherence} challenges, and
discuss why naive solutions are not practical.  We discuss \rachata{new PIM-specific} solutions
that can overcome these challenges in Sections~\ref{sec:impica} and
\ref{sec:lazypim}.

\subsection{Address Translation}
\label{sec:keyissues:translation}

A large amount of code relies on pointers, which are stored as
\emph{virtual} memory addresses.  
\ch{When the application follows a pointer, a core must perform \emph{address translation},
which converts the pointer's stored virtual address into a \emph{physical} address within main memory.}
If \chiii{PIM processing logic} relies on
  existing CPU-side address translation mechanisms, any performance
  gains from performing pointer chasing in memory could easily be
  nullified, as \kh{the \chiii{processing logic} needs to send a
    long-latency translation request to the CPU via the off-chip channel for each memory access}.  
    \changes{The translation can sometimes \rachata{require} a page table walk, where
    the CPU must issue \ch{\emph{multiple}} memory requests to read the page table, which
    further increases traffic on the memory channel.}

  A naive solution is to
  simply duplicate the TLB and page walker within memory \rachata{(i.e., within the \chiii{PIM processing logic})}.
  Unfortunately, this is
  \kh{prohibitively difficult} \rachata{or expensive} for three reasons: (1)~coherence would
  have to be maintained between the CPU and memory-side TLBs,
  introducing extra \kh{complexity and} off-chip requests; (2)~the \rachata{duplicated hardware} is very
  costly in terms of \rachata{storage overhead and complexity}; and (3)~a memory module can be used in
  conjunction with many different processor architectures, which use
  different page table implementations and formats, \changes{and ensuring
  compatibility between the in-memory TLB/page walker and all of these \rachata{different}
  designs is difficult.}

We explore a tractable solution for PIM address translation as part of our
in-memory pointer chasing accelerator, which we discuss in 
Section~\ref{sec:impica}.

\subsection{Cache Coherence}
\label{sec:keyissues:coherence}

\rachata{\chiii{PIM processing logic} can modify the data \chiii{it processes}, and this data may also be needed by CPU cores.}
In a traditional multithreaded execution model \rachata{that uses shared memory between threads}, writes to memory must be coordinated
between multiple cores, to ensure that \chv{threads do not operate on stale data values.}
Due to the per-core caches used in CPUs, this requires that when one
core writes data to a memory address, cached copies of the data held within the caches of other
cores must be updated or invalidated, which is known as \emph{cache coherence}.
Cache coherence involves a protocol that is designed to handle write permissions
for each core, invalidations and updates, and arbitration when multiple cores
request \chv{exclusive} access to the same memory address.  Within a chip multiprocessor (CMP),
the per-core caches \chv{can} perform coherence actions over a \chvi{shared interconnect}. \chv{Both snoopy~\cite{papamarcos.isca84,goodman.isca83} and directory-based~\cite{censier.tc78} coherence mechanisms
are employed in existing multiprocessor systems.}

Cache coherence is a major system challenge for \rachata{enabling} PIM architectures \rachata{as general-purpose execution engines}. 
If \chiii{PIM processing logic is} coherent with the processor,
the PIM programming model is relatively simple, as it
remains similar to conventional \ch{shared memory} multithreaded programming\ch{, which makes PIM architectures easier to adopt in general-purpose systems}.
\rachata{Thus, allowing} \chiii{PIM processing logic} to maintain such a simple \rachata{and traditional shared memory} programming model can
facilitate the \ch{widespread} adoption of PIM. However, it is impractical for PIM to perform
traditional \rachata{fine-grained cache} coherence, as this forces a large number of coherence messages to
traverse a narrow off-chip \chvi{interconnect}, \changes{potentially} undoing the benefits of
high-bandwidth \chvi{and low-latency} PIM \chv{execution, as} we show in Section~\ref{sec:lazypim}.  Prior works have proposed intermediate solutions
that \chv{\emph{sidestep}} coherence \rachata{by either requiring the programmer to ensure data coherence or making PIM data \chv{non-cacheable} in the CPU} (e.g., \chiii{\cite{ahn.tesseract.isca15, farmahini-farahani.hpca15, zhang.hpdc14,
gao.pact15, Seshadri:2015:ANDOR, seshadri2013rowclone, guo2014wondp,
ahn.pei.isca15, morad.taco15, DBLP:conf/hpca/GaoK16, impica, nai2017graphpim,
hsieh.isca16, seshadri.arxiv16, seshadri.micro17, chang.hpca16, pattnaik.pact16,
DBLP:conf/isca/AkinFH15}}). \chv{Unfortunately,} these solutions either place some restrictions
on the programming model or limit the performance \ch{and energy} gains achievable by a PIM
architecture.

\rachata{In this chapter, we} \chii{describe} a new coherence protocol, which allows \chiii{PIM processing logic} to efficiently
perform coherence \chv{\emph{without}} incurring high overhead or changing the programming
model, which we discuss in Section~\ref{sec:lazypim}.

% IMPICA
% !TEX root=../../chapter.tex

\section{IMPICA: An In-Memory Pointer-Chasing Accelerator}
\label{sec:impica}

Linked data structures, such as trees, hash tables, and linked lists
are commonly used in many important
applications\chvi{~\cite{elmasri2007fundamentals,
  giampaolo1998practical,fitzpatrick2004distributed,mao2012cache,wilson1992uniprocessor,
  hashemi.isca16,MutluKP05,CookseyJG02,DBLP:journals/tc/MutluKP06, alkan.naturegenetics09, xin.bmcgenomics13}}. For
example, many databases use B/B$^{+}$-trees to efficiently index large
data sets~\cite{elmasri2007fundamentals, giampaolo1998practical},
key-value stores use linked lists to handle collisions in hash
tables~\cite{fitzpatrick2004distributed, mao2012cache}, and \change{graph processing
  workloads\kh{~\cite{ligra,ahn.tesseract.isca15, ahn.pei.isca15}} use pointers to represent graph edges}.
These structures link nodes using pointers, where each node points to
at least one other node by storing its address. Traversing the link
requires serially accessing consecutive nodes by retrieving the
address(es) of the next node(s) from the pointer(s) stored in the current
node. This fundamental operation is called \textit{pointer chasing} in
linked data structures.

Pointer chasing is currently performed by the CPU cores, as part of an
application thread. While this approach eases the integration of
pointer chasing into larger programs, pointer chasing can be
inefficient within the CPU, as it introduces several sources of
performance degradation: (1)~dependencies exist between memory
requests to the linked nodes, resulting in serialized memory accesses
and limiting the available instruction-level and memory-level
parallelism\kii{~\cite{roth1998, DBLP:journals/micro/MutluSWP03,
    MutluKP05, DBLP:conf/isca/MutluKP05, hashemi.isca16}}; (2)~irregular allocation
or rearrangement of the connected nodes leads to access pattern
irregularity\ch{~\cite{JosephG97, YangL00, KarlssonDS00, ebrahimi2009,
CookseyJG02, MutluKP05,DBLP:journals/tc/MutluKP06}}, causing frequent cache and TLB misses; and (3)~link
traversals in data structures that diverge at each node (e.g., hash
tables, B-trees) frequently go down different paths during different
iterations, resulting in little reuse, further limiting cache
effectiveness~\cite{LukM96}.  Due to these inefficiencies, a
significant \emph{memory bottleneck} arises when executing pointer
chasing operations in the CPU, \kh{which stalls on a large number of
memory requests that suffer from the long round-trip latency between
the CPU and the memory.}

Many prior works
(e.g.,\ch{~\cite{LipastiSKR95, LukM96, JosephG97, roth1998, RothS99,
  YangL00, KarlssonDS00, ZillesS01, Luk01, CollinsWTHLLS01,
  SolihinTL02, Wu02, CollinsSCT02, CookseyJG02, DBLP:conf/hpca/HuMK03,
  HughesA05, MutluKP05, ebrahimi2009,
    DBLP:conf/micro/YuHSD15,DBLP:journals/tc/MutluKP06}}) proposed
mechanisms to predict and prefetch the next node(s) of a linked data
structure early enough to hide the memory latency. Unfortunately,
prefetchers for linked data structures suffer from several
shortcomings: (1)~\kh{they} usually do \ch{\emph{not}} provide \kh{significant}
benefit for data structures that diverge at each
node\ch{~\cite{KarlssonDS00, MutluKP05,DBLP:journals/tc/MutluKP06}}, due to low prefetcher accuracy and low miss
coverage; (2)~aggressive prefetchers can consume too much of the
limited off-chip memory bandwidth and, as a result, slow down the
system\ch{~\cite{JosephG97,ebrahimi2009,SrinathMKP07,pa-micro08,ebrahimi-isca2011,lee-tc2011,cont-runahead,seshadri-taco2015, ebrahimi.micro09}}; and (3)~a prefetcher that works
well for some pointer-based data structure(s) and access patterns
(e.g., a Markov prefetcher designed for mostly-static linked
lists~\cite{JosephG97}) usually does not work efficiently for \kh{different}
data structures \kh{and/or} access patterns.  Thus, it is \kh{important}
to explore new solution directions to alleviate \rachata{the significant} performance and
efficiency loss due to pointer chasing.

\textbf{Our goal} in this \ch{section} is to accelerate pointer chasing by
\emph{directly minimizing the memory bottleneck} caused by pointer
chasing operations.
To this end, we propose to perform pointer chasing \textit{inside main
  memory} by leveraging processing-in-memory (PIM) mechanisms, {\em
  avoiding the need to move data to the CPU}. In-memory pointer
chasing greatly reduces (1)~the latency of the operation, as an
address does not need to be brought all the way into the CPU before it
can be dereferenced; and (2)~the reliance on caching and prefetching
in the CPU, which are \kh{largely} ineffective for pointer chasing \chv{over large data structures}.
In this \rachata{section}, we \emph{\ch{describe} an
  in-memory accelerator for chasing pointers} in any linked data
structure, called the \textit{\fullname} (\abbrv)\ch{~\cite{impica}}.  \kh{\abbrv
  leverages the low memory access latency at the logic layer
  of 3D-stacked memory to speed up pointer chasing operations}.

We identify \emph{two fundamental challenges that we believe exist for a wide range
of in-memory accelerators}, and evaluate them as part of a case study in
designing a pointer chasing accelerator in memory. \changes{These} fundamental
challenges are (1) how
to achieve high parallelism \kii{in the accelerator (in the presence of serial accesses in
pointer chasing)}, and (2) how to effectively perform virtual-to-physical
address translation on the memory side without \kii{performing costly} accesses to
the CPU's memory management unit. We call these, respectively, the {\em
  parallelism challenge} and the {\em address translation challenge}.

\paratitle{The Parallelism Challenge} \sg{Parallelism is challenging
  to exploit in an in-memory accelerator even with the reduced latency
  and higher bandwidth available within 3D-stacked memory, as the
  performance of pointer chasing is limited by \emph{dependent sequential
  accesses}. The serialization problem can be exacerbated when the
  accelerator traverses multiple streams of links:
  while traditional out-of-order or multicore CPUs can service memory
  requests from multiple streams in parallel due to their ability to
  exploit high levels of instruction- and memory-level
  parallelism\chiii{~\cite{tomasulo.ibmjrd67, glew1998mlp,
    DBLP:conf/hpca/MutluSWP03, DBLP:conf/isca/MutluM08, DBLP:journals/micro/MutluSWP03,
    DBLP:conf/isca/MutluKP05, DBLP:journals/micro/MutluKP06, hashemi.isca16}}, simple accelerators
      are unable to exploit such parallelism unless they are carefully
  designed \chv{(e.g., \cite{ahn.tesseract.isca15, pugsley2014ndc, farmahini-farahani.hpca15, zhu2013accelerating})}.
  
  We observe that accelerator-based pointer chasing is
  primarily bottlenecked by memory access \kh{latency}, and that the
  \kh{address generation} computation for link traversal takes only a small fraction of the
  total traversal time, leaving the accelerator idle for \kii{a} majority of
  the traversal time. In \abbrv, we exploit this idle time by
  \emph{decoupling} link address generation from \kh{the issuing and
    servicing of a memory request}, which allows the accelerator to generate addresses for
  one link traversal stream while waiting on the \kh{request} associated with a
  different link traversal stream to return from memory.
  We call this
  design \emph{address-access decoupling}.}
\change{Note that this form of decoupling bears resemblance to 
  \ch{decoupled access/execute architectures}\ch{~\cite{smith1982decoupled, smith.tc86, smith.computer89}}, and
  we in fact take inspiration from past
  works\ch{~\cite{smith1982decoupled, DBLP:conf/isca/KurianHC92, DBLP:conf/isca/CragoP11, smith.tc86, smith.computer89}}, except our design
  is \chv{\emph{specialized}} for building a pointer chasing accelerator in 
  3D-stacked memory, and this paper solves specific challenges} within the 
  context of pointer chasing acceleration.

\paratitle{The Address Translation Challenge} \sg{An in-memory pointer
  chasing accelerator must be able to perform address translation, as
  each pointer in a linked data structure node stores the
  \emph{virtual} address of the next node, even though main memory is
  \emph{physically} addressed.  To determine the next address in the
  pointer chasing sequence, the accelerator must resolve the
  virtual-to-physical address mapping.
As we discuss in Section~\ref{sec:keyissues:translation}, 
relying on existing CPU-side address translation mechanisms or
duplicating the TLB and page walker within DRAM are impractical solutions.

  We observe that traditional address
  translation techniques do \ch{\emph{not}} need to be employed for pointer
  chasing, as link traversals are (1)~limited to linked data structures,
  and (2)~touch only \kii{certain data structures in} memory. We exploit this in \abbrv by
  \changesii{allocating data structures accessed by \abbrv into contiguous
  \emph{regions} within the virtual memory space, and}
  designing a new \ch{address} translation mechanism, the \emph{region-based page
    table}, \kh{which} is optimized for in-memory acceleration.  Our approach provides
  translation within memory at low latency and \kh{low} cost, while minimizing
  the cost of maintaining TLB coherence.  }

\paratitle{Evaluation}
By solving both key challenges, \abbrv provides significant \kii{performance and energy
benefits for pointer chasing operations and applications that use such
operations. First we examine three microbenchmarks, each of which
performs pointer chasing on a \chv{widely-used} data structure (linked list,
hash table, B-tree), and find }that \abbrv improves their performance by \kh{92\%,
29\%, and 18\%, respectively,} on a quad-core system over 
a state-of-the-art baseline. 
Second, we evaluate \abbrv on a \kii{real database workload,
  DBx1000~\cite{yustaring},} on a quad-core system, and \kii{show that} \kh{\abbrv increases \emph{overall} database
transaction throughput by 16\% and reduces transaction latency by 13\%}.
Third, \abbrv reduces \emph{overall} system energy, by \kh{by 41\%, 23\%, and 10\%} for
the three microbenchmarks and by 6\% for DBx1000.
These benefits come at a very small hardware cost:
\change{our evaluations show that \abbrv comprises
  only \kh{7.6\% of} the area of a small embedded core (the ARM Cortex-A57\kii{~\cite{CortexA57}}).}

\sg{\rachata{Our IMPICA proposal, originally published in the ICCD 2016 conference~\cite{impica},} \chii{makes} the following major contributions}: \squishlist 
\item This is the first work to propose an in-memory accelerator for
  \kii{pointer chasing.}
  % in 3D-stacked memory. 
  Our proposal,
  \abbrv, accelerates linked data structure traversal by chasing 
  pointers \kh{inside} the logic layer of 3D-stacked memory, thereby
  eliminating \kii{inefficient,} high-latency \kh{serialized} data transfers between the
  CPU and main memory.

\item We identify two \change{fundamental} challenges in designing an
  efficient in-memory pointer chasing accelerator
  (Section~\ref{sec:challenges}). \change{These challenges can greatly
    hamper performance if the accelerator is not designed
    \emph{carefully} to overcome them.  First, multiple streams of
    link traversal can \kh{unnecessarily} get serialized at the
    accelerator, \rachata{thereby} degrading performance (the \emph{parallelism
      challenge}).} \kh{Second, an in-memory accelerator needs to
    perform virtual-to-physical address translation for each pointer,
    but this critical functionality does \ch{\emph{not}} exist on the memory side (the
    \emph{address translation challenge}).}

\item \abbrv solves the \emph{parallelism challenge} by decoupling link
  address generation from memory accesses, and utilizes the idle time
  during memory accesses to service \emph{multiple} pointer chasing streams
  simultaneously. We call this approach \textit{address-access
    decoupling} (Section~\ref{sec:core_architecture}).

\item \change{\abbrv solves the \emph{address translation challenge} by
  \changesii{allocating data structures \kii{it accesses} into contiguous virtual memory
  regions,
  and using an optimized \kh{and low-cost}
   \textit{region-based
    page table} \kii{structure for} address translation} (Section~\ref{sec:virtual_memory})}.

\item \change{We evaluate \abbrv extensively using both microbenchmarks
  and a real database workload. Our results (Section~\ref{sec:results})
show that \abbrv improves both system performance and energy
efficiency \changesii{for all of these \kh{workloads}}, while
requiring \kii{only very} modest hardware overhead in the
logic layer of 3D-stacked DRAM.}

\squishend
% !TEX root=../../chapter.tex

\subsection{Motivation}

\changesii{
To motivate the need for a pointer chasing accelerator, we first 
examine the usage of pointer chasing in contemporary workloads. We then
discuss opportunities for acceleration within 3D-stacked memory.}

\subsubsection{Pointer Chasing in Modern Workloads} \label{sec:pointer_chasing}
Pointers are ubiquitous in fundamental data structures such as linked
lists, trees, and hash tables, \changesii{where the nodes of the data structure are linked
together by storing the addresses (i.e., pointers) of neighboring nodes.
Pointers make it easy to dynamically add/delete nodes in these
data structures, but link traversal is often serialized, as the
address of the next node can be known \kii{only after} the current node is
fetched. The serialized link traversal is commonly
referred to as \textit{pointer chasing}.}

\sg{Due to the flexibility of insertion/deletion, pointer-based data structures 
and link traversal algorithms are essential building blocks in programming, and
\kh{they} enable a very wide range of workloads.}
For instance, \change{at least \chii{seven}
  different types of modern data-intensive applications} rely \emph{heavily} on linked data
structures: (1)~\textbf{databases and file systems} use
B/B$^{+}$-trees for indexing tables or
metadata\chv{~\cite{elmasri2007fundamentals, giampaolo1998practical, ghemawat.sosp03, rodeh.tos13}};
\changesii{%
(2)~\textbf{in-memory caching} applications based on key-value stores, 
such as Memcached~\cite{fitzpatrick2004distributed} and Masstree~\cite{mao2012cache}, 
use linked lists to resolve hash table
collisions and trie-like B$^{+}$-trees as \kh{their} main data structures;
(3)~\textbf{graph processing workloads} use pointers to represent the edges 
that connect the vertex data structures together\kii{~\cite{ligra, ahn.tesseract.isca15}};}
(4)~\textbf{garbage collectors} in high level languages typically maintain
reference relations using trees\chv{~\cite{wilson1992uniprocessor,joao.isca09,jones.gcbook1996}};
(5)~\textbf{3D video games} use binary space partitioning trees to
determine the objects that need to be rendered\ch{~\cite{naylor1990, joao.isca09}}; 
\chii{(6)~\textbf{dynamic routing tables} \ch{used by} networks} employ balanced search
trees for high-performance IP address lookups~\cite{waldvogel1997}\chii{; and 
(7)~\textbf{hash table based DNA read mappers} that} store and find potential locations of a read in a reference genome 
index\chv{~\cite{alkan.naturegenetics09, xin.bmcgenomics13, xin.shd.bioinformatics15, alser.bioinformatics17, lee.methods14, kim.arxiv17,kim.bmc18}}.

While linked data structures are widely used in many modern
applications, chasing pointers is very inefficient in general-purpose
processors.  There are three major reasons behind the
inefficiency. \changesii{
First, the inherent serialization that occurs when accessing consecutive nodes
limits the available instruction-level and memory-level parallelism\kii{~\cite{LipastiSKR95,
LukM96, JosephG97, roth1998, RothS99, MutluKP05,
DBLP:journals/micro/MutluSWP03, DBLP:conf/hpca/MutluSWP03,
DBLP:conf/isca/MutluKP05, DBLP:journals/tc/MutluKP06,
DBLP:journals/micro/MutluKP06, DBLP:conf/isca/MutluM08}}. As a result,
out-of-order execution provides only limited \kiii{performance benefit} when chasing
pointers\ch{~\cite{MutluKP05, DBLP:conf/isca/MutluKP05, DBLP:journals/micro/MutluKP06, DBLP:journals/tc/MutluKP06}}.
Second, as nodes can be inserted and removed dynamically, they can get 
allocated to different regions of memory. The irregular \chviii{memory} allocation causes
pointer chasing to exhibit irregular access patterns, which lead to
frequent cache and TLB
misses\change{~\cite{JosephG97, YangL00, KarlssonDS00, ebrahimi2009, MutluKP05}}.
Third, for data structures that diverge at each node, \kh{such as
B-trees}, link traversals often go down different paths during different 
iterations, as the inputs to the traversal function change. As a result,
lower-level nodes that were recently referenced during a link traversal are
unlikely to be reused in subsequent traversals, limiting the
effectiveness of \kh{many caching policies\kii{~\cite{kocberber2013meet,LipastiSKR95, LukM96}},} such
as LRU replacement.}

\begin{figure}[h]
  \centering
  \includegraphics[width=0.7\textwidth]{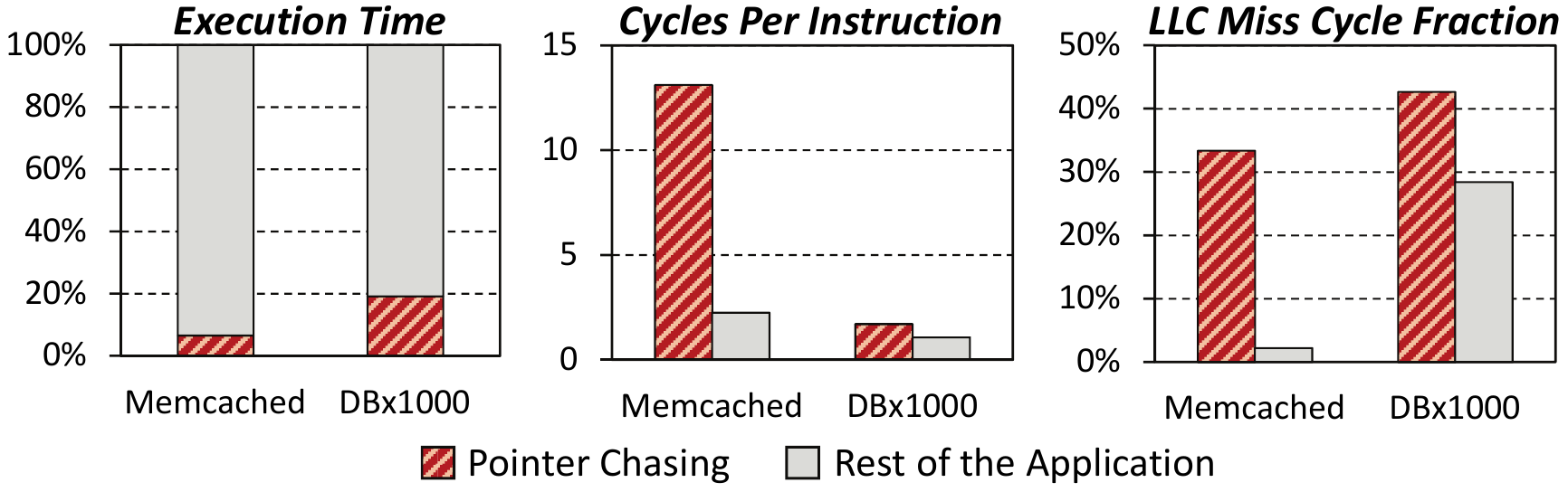}%
  \caption{Profiling results of pointer chasing portions of code vs.\ the rest of
    the application code in Memcached and DBx1000. \rachata{Figure \chiv{adapted} from \cite{impica}.}}
  \label{fig:profile}
\end{figure}

To quantify the performance impact of chasing pointers in real-world
workloads, we profile two popular applications that heavily depend on
linked data structures, using a state-of-art Intel Xeon
system:\footnote{\kh{We use the
    Intel\textregistered\ VTune\texttrademark\ profiling tool on a
    machine with a Xeon\textregistered\ W3550 processor (3GHz, 8-core,
    8~MB LLC)~\cite{xeonw3550}
    and 18~GB memory. We 
     profile \kii{each} application for 10 minutes after it reaches steady state.}} (1)~\emph{Memcached}~\cite{fitzpatrick2004distributed}, \rachata{an in-memory caching system,} \changes{using a
real Twitter dataset~\cite{ferdman.asplos12} as its input}; and 
(2)~\emph{DBx1000}~\cite{yustaring}, an in-memory database system, \kh{using the TPC-C benchmark~\cite{council2005transaction} as its input}.
\change{We profile the pointer chasing code within the application
  separately from other parts of the application code.
  Figure~\ref{fig:profile} shows \changesii{how pointer chasing compares to
the rest of the application in terms of} execution
  time, cycles per instruction (CPI), and the ratio of last-level
  cache (LLC) miss cycles to the total cycles.}

\changesii{
\kh{We make three major observations \rachata{from Figure~\ref{fig:profile}}.} First, both Memcached and DBx1000 spend a significant 
fraction of their total execution time (7\% and 19\%, respectively) on pointer 
chasing, as a result of 
dependent cache misses\ch{~\cite{MutluKP05, roth1998, hashemi.isca16,DBLP:journals/tc/MutluKP06}}. 
Though these percentages might sound small, real software often does \ch{\emph{not}} have a 
single type of operation that consumes this significant a fraction of the total
time.
Second, we find that pointer chasing is significantly more inefficient 
than the rest of the application, as it \kii{requires much higher} cycles per 
instruction \changes{(6$\times$ in Memcached, and 1.6$\times$ in DBx1000)}. 
Third, pointer chasing is largely memory-bound, as it exhibits much higher 
cache miss rates than the rest of the application and as a result
spends a \kh{much larger fraction of cycles waiting for LLC misses  (16$\times$ \kii{in Memcached,} and 1.5$\times$ \kii{in} DBx1000)}.}
From these observations, we conclude
that \change{(1)~pointer chasing consumes a significant fraction of
  execution time in two important \rachata{sophisticated} applications},
(2)~\kii{pointer chasing operations
% in important data-intensiveapplications 
are bound by memory}, and (3)~executing pointer chasing
code in a modern general-purpose processor is very inefficient and
thus can lead to a large performance overhead. \kh{Other works made
similar observations for different
workloads\ch{~\cite{hashemi.isca16,MutluKP05, roth1998, DBLP:journals/tc/MutluKP06,CookseyJG02}}.}

Prior works (e.g.,\ch{~\cite{LipastiSKR95, LukM96, JosephG97, roth1998, RothS99,
  YangL00, KarlssonDS00, ZillesS01, Luk01, CollinsWTHLLS01,
  SolihinTL02, Wu02, CollinsSCT02, CookseyJG02, DBLP:conf/hpca/HuMK03,
  HughesA05, MutluKP05, ebrahimi2009,
    DBLP:conf/micro/YuHSD15, DBLP:journals/tc/MutluKP06}}) proposed specialized prefetchers that
predict and prefetch the next node of a linked data structure to hide
memory \kh{latency}. \change{While prefetching can mitigate part of
  the \kh{memory latency problem}, it has three major
  shortcomings. First, the efficiency of \ch{a prefetcher} degrades
  significantly when the \kh{traversal of linked data structures
    diverges into multiple paths and the access order is
    irregular\ch{~\cite{KarlssonDS00, DBLP:journals/tc/MutluKP06, MutluKP05}}. Second, prefetchers can sometimes
    slow down the entire system due to contention caused by inaccurate \rachata{as well as accurate}
    prefetch requests\ch{~\cite{JosephG97,ebrahimi2009,SrinathMKP07,pa-micro08,ebrahimi-isca2011,lee-tc2011,cont-runahead,seshadri-taco2015, ebrahimi.micro09}}. Third, these \rachata{specialized} hardware prefetchers are usually
    designed for specific data structure implementations, and tend to
    be \kii{very} inefficient when dealing with other data structures. \rachata{For example, a Markov prefetcher~\cite{JosephG97} can \chviii{potentially} be very effective for static linked lists, but it becomes very inefficient for trees with dynamic access patterns.} It is
    \kii{difficult} to design a prefetcher that is efficient \kii{and effective
    for \emph{all} types of} linked data
    structures.} \textbf{Our goal} in this work is to improve the
  performance of pointer chasing applications \kh{\emph{without}
    relying on prefetchers, regardless of \kii{the types \ch{and access patterns} of linked data structures
    used} in an application}. \ignore{by minimizing the
    memory bottleneck by performing pointer chasing completely in
    memory.}}

\ignore{
Though this approach can mitigate part of the problem, it
has a major shortcoming.  It is very hard to accurately predict
irregular memory accesses for linked data structures that diverge into
\emph{multiple paths} at each node, e.g., hash tables and
B-trees~\cite{KarlssonDS00}.  
}

\subsubsection{Accelerating Pointer Chasing in 3D-Stacked Memory}

\kh{\changesii{
We propose to improve the performance of pointer chasing by leveraging 
processing-in-memory (PIM) to alleviate the memory bottleneck.}  Instead of \sg{sequentially fetching 
\emph{each node} from memory and sending it to the CPU when an application is looking
for a particular node, PIM-based pointer chasing} consists of
(1)~traversing the linked data structures \emph{in memory}, and 
(2)~returning only the final node found to the CPU.}

\begin{figure}[h!] 
\centering
\subfloat[Binary tree]{\includegraphics[height=85pt]{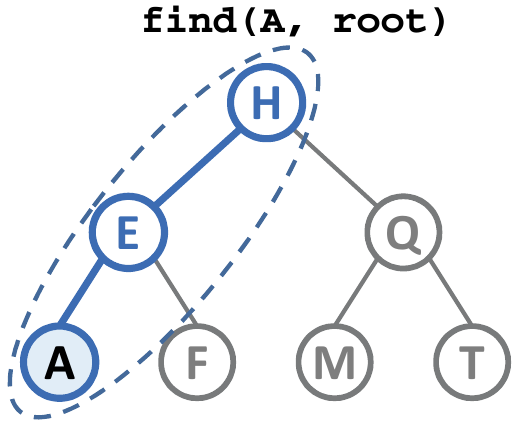}}\hfill%
\subfloat[Traditional architecture]{\includegraphics[height=85pt]{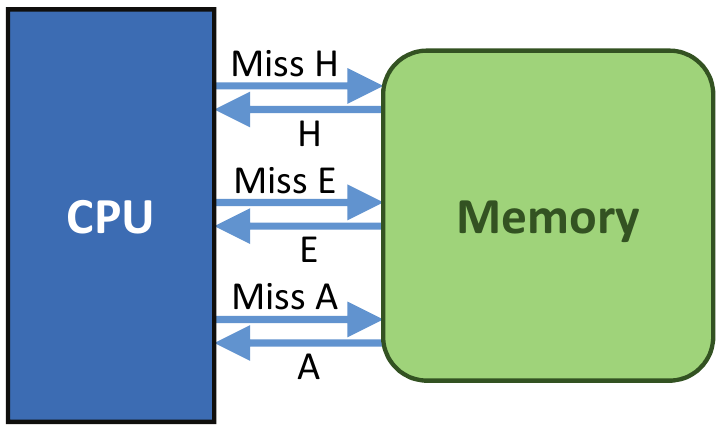}}\hfill%
\subfloat[\abbrv architecture]{\includegraphics[height=85pt]{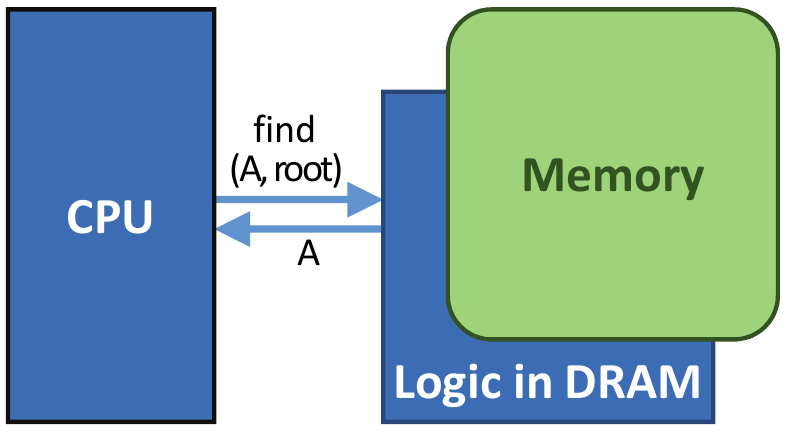}}%
\caption{\kh{Pointer chasing (a) in a traditional architecture (b) and in \abbrv with
  3D-stacked memory (c).} \rachata{Figure \chiv{adapted} from \cite{impica}.}}
\label{fig:high-level} \end{figure}

Unlike prior works that proposed general architectural models for
in-memory computation by embedding logic in main
memory\chiii{~\cite{zhang.hpdc14, farmahini-farahani.hpca15, ahn.pei.isca15,
    hsieh.isca16, pattnaik.pact16,
    DBLP:conf/hpca/GaoK16,
    asghari-moghaddam.micro16, boroumand2016pim,
    hashemi.isca16, cont-runahead, GS-DRAM, liu-spaa17,
    gao.pact15, guo2014wondp, sura.cf15,
    morad.taco15, hassan.memsys15, li.dac16, kang.icassp14, aga.hpca17,
    shafiee.isca16, seshadri2013rowclone, Seshadri:2015:ANDOR, 
    chang.hpca16, seshadri.arxiv16, seshadri.micro17, nai2017graphpim}}, we propose to design
a \emph{specialized \fullname} (\abbrv)
that exploits
the logic layer of 3D-stacked
memory\kiii{~\cite{jedec.hbm.spec, hmc.spec.1.1, hmc.spec.2.0, jeddeloh2012hybrid, lee.taco16}}. 3D die-stacked memory
achieves low latency \changes{(and high bandwidth)} by stacking memory dies on top
of a logic die, \kii{and} \changesii{interconnecting the layers using through-silicon vias} (TSVs). \ignore{ It provides a
unique opportunity to design a new and efficient processing in memory
architecture.  }Figure~\ref{fig:high-level} shows a binary tree traversal \sg{using \abbrv},
compared to a traditional architecture \change{where the CPU traverses
  the binary tree}. The traversal sequentially accesses the nodes from
the root to \kh{a particular node}
(e.g., \textbf{H}$\rightarrow$\textbf{E}$\rightarrow$\textbf{A} in Figure~\ref{fig:high-level}a).  In a
traditional architecture (Figure~\ref{fig:high-level}b), these \kii{serialized} accesses to the nodes miss in the
caches and \kh{three memory requests are sent to memory serially 
\sg{across a high-latency off-chip channel}. In contrast,}
\abbrv traverses the tree \sg{inside the \kii{logic layer of 3D-stacked memory}}, and as Figure~\ref{fig:high-level}c shows, only the \kh{final}
node (\textbf{A}) is sent from the memory to the host CPU \kh{in
  response to the traversal request}. \kh{Doing the
traversal in memory} minimizes
both traversal latency \change{(as queuing delays \kii{in} the on-chip
  interconnect and the CPU-to-memory bus are eliminated)} and off-chip
bandwidth consumption, \kh{as shown in Figure~\ref{fig:high-level}c}.

Our accelerator architecture has three major
advantages. \change{First, it improves performance and \kh{reduces \rachata{memory}
  bandwidth consumption} by
  eliminating the round trips required for memory accesses over the
  CPU-to-memory \rachata{interconnects}. Second, it frees the CPU to execute other
  \kh{work than linked data structure traversal}, \rachata{thereby} increasing system throughput. Third, it minimizes the cache
  contention caused by pointer chasing operations.}
% !TEX root=../../chapter.tex

\subsection{Design Challenges}
\label{sec:challenges}

\kh{We identify \kii{and describe} two new challenges
that are crucial to the performance and functionality of our new pointer chasing
accelerator in memory}: (1) the \textit{parallelism challenge},
and (2) the \textit{address translation challenge}. 
\sg{\kii{Section~\ref{sec:architecture} describes} our \abbrv architecture, which centers around two key ideas that
solve these \kii{two} challenges.}

\subsubsection{Challenge 1: Parallelism in the Accelerator}   
\label{subsec:parallelism}

\begin{figure}[h]
  \centering
  \subfloat[Pointer chasing on two CPU cores vs. one non-parallel accelerator]{\includegraphics[trim=5 225 5 0,clip,width=0.75\linewidth]{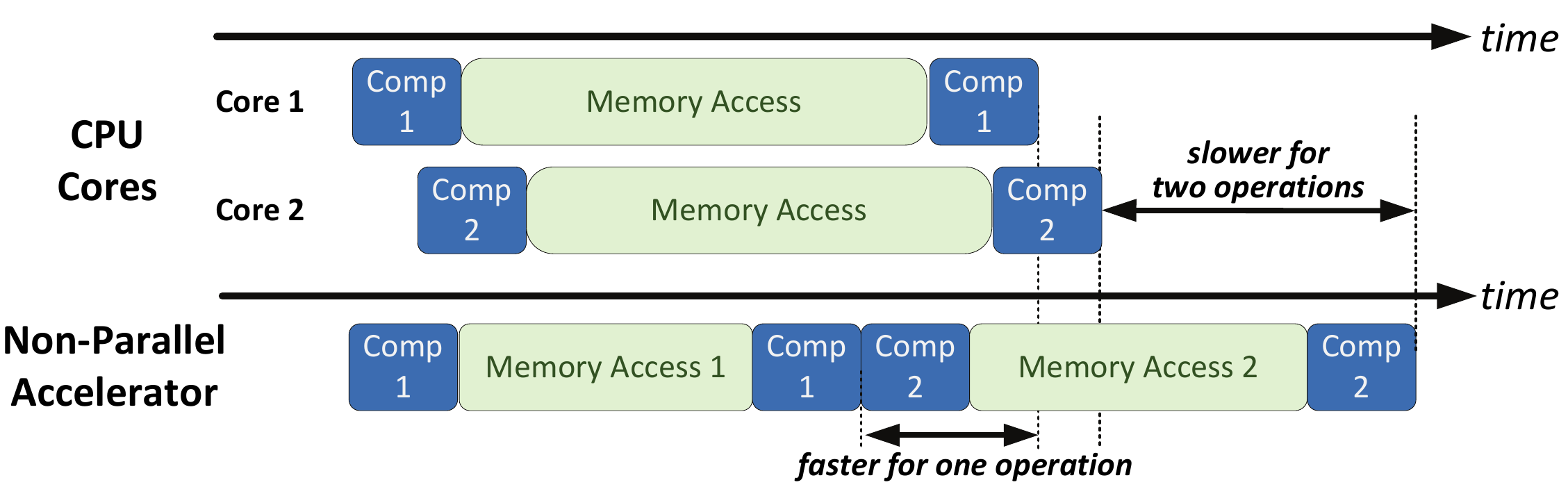}}\\%
  \subfloat[Pointer chasing using \rachata{our \abbrv proposal}]{\includegraphics[trim=5 25 5 265,clip,width=0.75\linewidth]{figures/IMPICA/parallel_timeline.pdf}}%
  \caption{Execution time of two independent pointer chasing operations\changes{, broken down into address computation time (\emph{Comp}) and memory access time}. \rachata{Figure \chiv{adapted} from~\cite{impica}.}}
  \label{fig:parallel_timeline}
\end{figure}

A pointer chasing accelerator supporting a multicore system needs to
handle \emph{multiple} link traversals \kh{(from different cores)} in parallel at low cost. A simple
accelerator that can handle only one request at a time \change{(which
  we call a \emph{non-parallel accelerator})} would 
serialize the requests and could potentially be slower than using multiple \kiii{CPU
cores to perform the multiple traversals}. As
depicted in Figure~\ref{fig:parallel_timeline}a, 
\sg{while a non-parallel accelerator speeds up each \emph{individual} pointer
chasing operation \kh{done by one of the \kiii{CPU} cores}
due to \kiii{its} shorter memory latency, the accelerator is slower
\emph{overall} \kiii{for two pointer chasing operations}, as
\emph{multiple} \emph{cores} can operate in \emph{parallel} on independent pointer chasing
operations}.

\sg{To overcome this deficiency, an in-memory accelerator needs to exploit
parallelism when it services requests.  However, the accelerator must do
this \emph{at low cost} \rachata{and \emph{low complexity}}, due to its placement within the logic layer of 
3D-stacked memory, where complex \kh{logic, such as out of order
  execution circuitry, is} \ch{currently not} feasible.}
The straightforward solution of adding
multiple accelerators to service independent pointer chasing operations
(e.g.,~\cite{kocberber2013meet}) does not scale well, and \kh{also}
\sg{can} lead to excessive energy dissipation \rachata{(and, thus, potentially thermal violations)} and die area usage in the logic layer.

A key observation we make is that pointer chasing operations are bottlenecked
by memory stalls, as shown in Figure~\ref{fig:profile}. \kh{In our
evaluation, the memory access time is \kiii{10--15$\times$ the
computation} time \rachata{(see Section~\ref{sec:impica-meth} for our methodology)}.}
As a result, the accelerator 
\sg{spends} \kh{a significant amount} of time waiting for memory, causing its
compute resources to sit idle. This makes typical in-order or out-of-order execution engines
\ch{\emph{inefficient}} for an in-memory pointer-chasing accelerator. 
If we utilize the hardware resources in a more efficient manner, we can enable 
parallelism by handling \ch{\emph{multiple}} pointer chasing operations \emph{within a single accelerator}.

\sg{Based on our observation, we \emph{decouple} address generation from
memory accesses in \abbrv~\kiii{using two engines (\ch{the address engine and the
  access engine}), allowing the accelerator} to generate addresses from one
pointer chasing operation while it \ch{\emph{concurrently}} performs memory accesses for a 
different pointer chasing operation \kiii{(as shown in Figure~\ref{fig:parallel_timeline}b)}.
We describe the details of our decoupled \kh{accelerator} design in 
Section~\ref{sec:architecture}.}

\subsubsection{Challenge 2: Virtual Address Translation}
\label{subsec:translation}

\sg{A second challenge arises when pointer chasing is moved out of the
  \kh{CPU \kii{cores, which are} equipped with facilities for address translation}.
Within the program data structures, \kh{each pointer is stored as a virtual
address, and requires \emph{translation} to a physical address before its
memory access can be performed}.}
This is a challenging task for
an in-memory accelerator, which has no easy access to the virtual address
translation engine that sits in the CPU \kii{core}. While \kii{sequential} array operations could potentially be constrained
to work within page boundaries or directly in physical memory, indirect memory
accesses that come with pointer-based data structures require
some support for virtual memory \ch{address} translation, as they \kh{might} touch many parts of the 
virtual address space.

There are two major \sg{issues} when designing a virtual address translation
mechanism for an in-memory accelerator. First, different processor architectures
have different page table implementations and formats. This lack of
compatibility makes it \kh{very expensive}
to simply replicate the CPU page table walker in the in-memory
accelerator \kh{as this approach requires replicating TLBs and page
  walkers for many architecture formats. Second,
a page table walk tends to be a high-latency operation involving multiple memory
accesses due to the heavily layered \ch{format} of \kii{a conventional} page table}. As a
result, TLB misses \kii{are} a major performance bottleneck in
 data-intensive applications\chviii{~\cite{basu2013efficient, power.hpca14, bhattacharjee.micro13, bhattacharjee.asplos10, pichai.asplos14, bhattacharjee.hpca10, srikantaiah.micro10, lustig.taco13, ausavarungnirun.micro17, ausavarungnirun.asplos18}}. \kh{If the
   accelerator requires many page table walks that are supported by
   the CPU's address translation mechanisms, \kii{which require
     high-latency off-chip accesses for the accelerator,} its performance can
   degrade greatly.} 

\sg{To address these issues, we \emph{completely decouple} the page table of
\abbrv from that of the \rachata{CPU, thereby} obviating the need for compatibility between the
two \rachata{page} tables.  This presents us with an opportunity to develop a new page table
design that is much more efficient for our in-memory accelerator.  We make two
key observations about the behavior of a pointer chasing accelerator.  First,
the accelerator}
operates
only on certain data structures that can be mapped to \kh{\emph{contiguous regions}} in the virtual address space, \sg{which we refer} to as
\emph{\abbrv regions}.  
As a result, it is possible to map contiguous \abbrv
regions with a \emph{smaller, region-based} page table without needing to duplicate
the page table mappings for the \emph{entire} address space. Second, we observe
that \sg{if we need to map \kii{\emph{only}} \abbrv regions, we can collapse the hierarchy
present in conventional page tables, \rachata{which allows us to limit the hardware and storage} overhead of the
\abbrv page table.  We describe the \abbrv page table in detail in 
Section~\ref{sec:virtual_memory}.}
% !TEX root=../../chapter.tex

\subsection{\abbrv Architecture}
\label{sec:architecture}

\sg{
We propose a new in-memory accelerator, \abbrv, that addresses the two design
challenges \rachata{that face in-memory} accelerators for pointer chasing.  The \abbrv architecture
consists of a single \kiii{specialized} core designed to decouple address 
generation from memory accesses. Our approach, which we call \emph{address-access 
decoupling}, allows us to \emph{efficiently} overcome
the parallelism challenge (Section~\ref{sec:core_architecture}). The \abbrv core
uses a novel \emph{region-based page table} design to perform efficient address 
translation locally \kh{in the accelerator}, \rachata{which allows} us to overcome the address translation challenge
(Section~\ref{sec:virtual_memory}).}

\subsubsection{\abbrv Core Architecture}
\label{sec:core_architecture}

\sg{
Our \abbrv core uses what we call address-access decoupling, where we separate
the core into \kii{two parts: (1)}~an \emph{address engine}, which generates the address specified
by the pointer; and \kii{(2)}~an \emph{access engine}, which performs memory access
operations using addresses generated by the address engine.  The key advantage
of this design is that the address engine supports fast context switching between 
multiple pointer chasing operations, allowing it to utilize the idle time 
during memory \changes{access(es)} to compute addresses from a different pointer
chasing operation. As Figure~\ref{fig:parallel_timeline}b \rachata{shows}, an \abbrv core can 
process multiple pointer chasing operations faster than multiple \kh{cores} because it has the ability to overlap 
address generation with memory accesses.}

\sg{
Our address-access decoupling has similarities to, and is in fact inspired by, 
the decoupled access-execute (DAE)
architecture\ch{~\cite{smith1982decoupled, smith.tc86, smith.computer89}}, \kii{with two key differences}. 
\kii{First}, the goal of DAE is to exploit instruction-level parallelism (ILP) 
within a \ch{\emph{single}} thread, whereas our goal is to exploit thread-level parallelism 
(TLP) \kh{across pointer chasing operations from \ch{\emph{multiple}} threads. Second}, unlike DAE, the decoupling in \abbrv does 
not require any programmer or compiler effort. Our approach is much simpler 
than both general-purpose DAE and out-of-order \kh{execution}\ch{~\cite{tomasulo.ibmjrd67, patt.hps.micro85, patt.critical.micro85}}, as it can switch \kii{between}
different \changes{independent} execution streams, without the need for dependency \kh{checking}.}

\begin{figure}[h]
  \centering
  \includegraphics[width=0.75\textwidth]{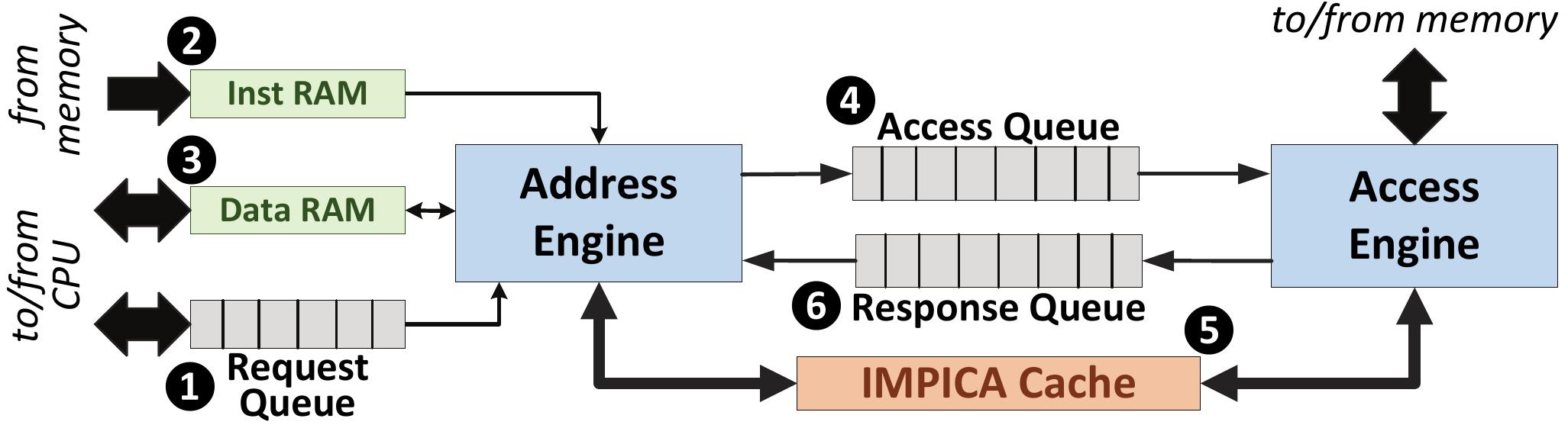}
  \caption{\abbrv core architecture. \rachata{Figure \chiv{adapted} from \cite{impica}}.}
  \label{fig:core_architecture}
\end{figure}

\sg{Figure~\ref{fig:core_architecture} shows the architecture of the
  \abbrv core.}
  \sg{The host CPU initializes \changes{a pointer chasing} operation by moving its code to main memory,
    and then enqueuing the request in the \emph{request queue}
    \kii{(\incircle{1} in Figure~\ref{fig:core_architecture})}. \kh{Section~\ref{sec:interface} describes the details of 
    the CPU interface}.}

  \sg{The \emph{address engine} services the enqueued request by loading the
    pointer chasing code into its \emph{instruction RAM} \kii{(\incircle{2})}.  This engine contains
    all of \abbrv's functional units, and executes the code in its instruction 
    RAM while using its \emph{data RAM} \kii{(\incircle{3})} as a stack. All instructions that do
    not involve memory accesses, such as ALU operations and control flow, are
    performed by the address engine.}
    The number of pointer chasing operations that can be processed in parallel is
    limited by the size of the stack \kii{in the data RAM}.

    \sg{When the address engine encounters a memory instruction, it
    enqueues the address (along with the data RAM stack pointer) into the 
    \emph{access queue} \kii{(\incircle{4})}, and then performs a
    \emph{context switch} \kii{to an independent stream}. For the switch, 
    the engine pushes the hardware context \kiii{(i.e., architectural
    registers and the program 
    counter)} into the data RAM stack. When this is done, the address 
    engine can work on a different pointer chasing operation.}

  \sg{The \emph{access engine} services requests waiting in the access
    queue. This engine translates the enqueued address from a virtual \kii{address} to
    a physical address, using the \abbrv page table (see 
    Section~\ref{sec:virtual_memory}). It then sends the physical address to the
    memory controller, which performs the memory access. Since the memory
    controller handles data retrieval, the access engine can
    issue multiple requests to the controller without waiting on the data, just
    as the CPU does today, \kii{\rachata{thereby} quickly servicing} \ch{the} queued requests.
    Note that the access engine does \ch{\emph{not}} contain any functional units.}

  \sg{When the access engine receives data back from the memory
    controller, it stores this data in the \emph{\abbrv cache} \kii{(\incircle{5})}, a small cache
    that contains data destined for the address engine. The access queue entry
    corresponding to the returned data is moved from the access queue to the
    \emph{response queue} \kii{(\incircle{6})}.}

  \sg{The address engine monitors the response queue. \kii{When a
      response queue entry is ready, the address engine reads it}, and
    uses the stack pointer to access and reload the registers and PC that were
    pushed onto the data RAM stack. It then resumes execution for the
    pointer chasing operation, continuing until it encounters the next memory instruction.

\subsubsection{\ch{\abbrv Cache}}

\rachata{\abbrv uses a cache to deliver data fetched by the 
access engine to the address engine. The cache employs three features
that cater to pointer-chasing applications. First, it uses
\emph{cache line locking} to guarantee that data is not 
displaced from the cache until  the address \chii{engine processes} the data. 
Cache line locking is achieved using a \emph{lock bit} in the tag that 
is set when the cache line is inserted, and is cleared only after the 
address engine processes the associated entry in the 
response queue. If all of the cache lines in a  set are locked,  the 
access engine stalls until one of the cache lines becomes unlocked. Second, 
when a traversal is completed, the \abbrv cache \emph{immediately} evicts cache lines fetched by that 
pointer-chasing operation. A \emph{request ID} associated with the tag is used 
to determine if a cache line belongs to a completed task. Third, the \abbrv 
cache prioritizes 
nodes that closer to the root of the data structure in the cache, by leveraging the observation that pointer-based 
structures traverse multiple paths and usually do \emph{not} re-reference the leaf 
nodes. To achieve this, the cache sets a \textbf{root bit} in the tag if a cache line is fetched by the 
first few memory accesses of a pointer-chasing operation. 
\chviii{Figure~\ref{fig:cache} shows the} structure of 
the \abbrv cache, including cache line metadata.

\begin{figure}[h]
  \centering
  \includegraphics[width=0.5\textwidth]{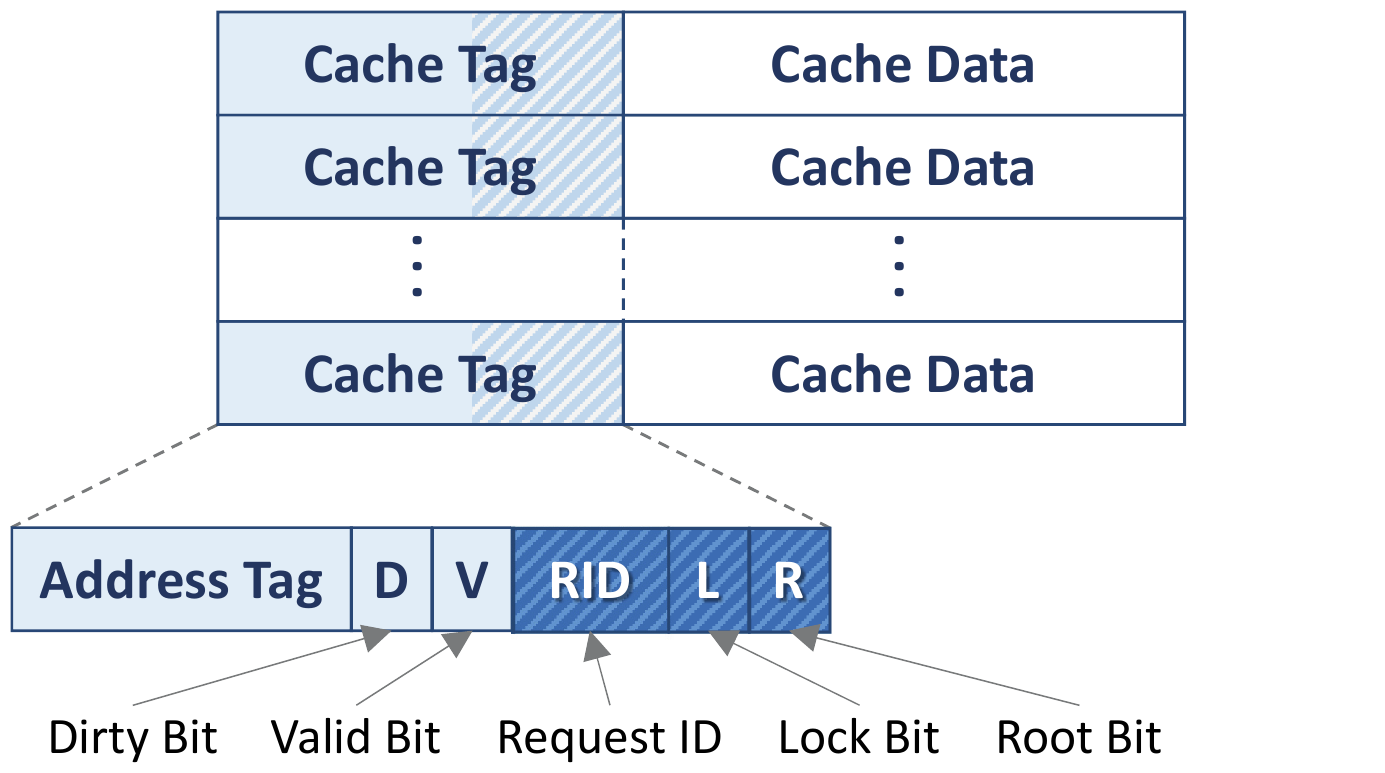}
  \caption{\chii{Structure of the \abbrv cache.}}
  \label{fig:cache}
\end{figure}

\subsubsection{\abbrv Page Table}
\label{sec:virtual_memory}

\sg{\abbrv uses a \emph{region-based} page table (RPT) design optimized for in-memory 
pointer chasing, leveraging the \kii{continuous} ranges of accesses (\emph{\abbrv regions})
discussed in Section~\ref{subsec:translation}. Figure~\ref{fig:virtual_memory}
shows the structure of the RPT in \abbrv. The RPT is split into three levels:
(1)~a first-level \emph{region table}, which needs to map only \kh{a
  small number of
the contiguously-allocated} \abbrv regions; (2)~a second-level \emph{flat page table}
for each region with a larger (e.g., 2MB) page size; and (3)~third-level \emph{small page 
tables} that use conventional \kii{small} (e.g., 4KB) pages. \kh{In the example in
Figure~\ref{fig:virtual_memory},} when a 48-bit virtual memory address arrives
for translation, bits 47--41 of the address are used to index the
region table \kii{(\incircle{1} in Figure~\ref{fig:virtual_memory})}
to find the corresponding flat page table. Bits 40--21 are used to index the 
flat page table {(\incircle{2}), providing the location of the small page table, which is 
indexed using bits 20--12 (\incircle{3}).} \kh{The entry in the small
page table provides the physical \kiii{page number} of the page, and bits 11--0
\kii{specify} the offset within the physical page (\incircle{4}).}

\begin{figure}[h!]
  \centering
  \includegraphics[width=0.8\textwidth]{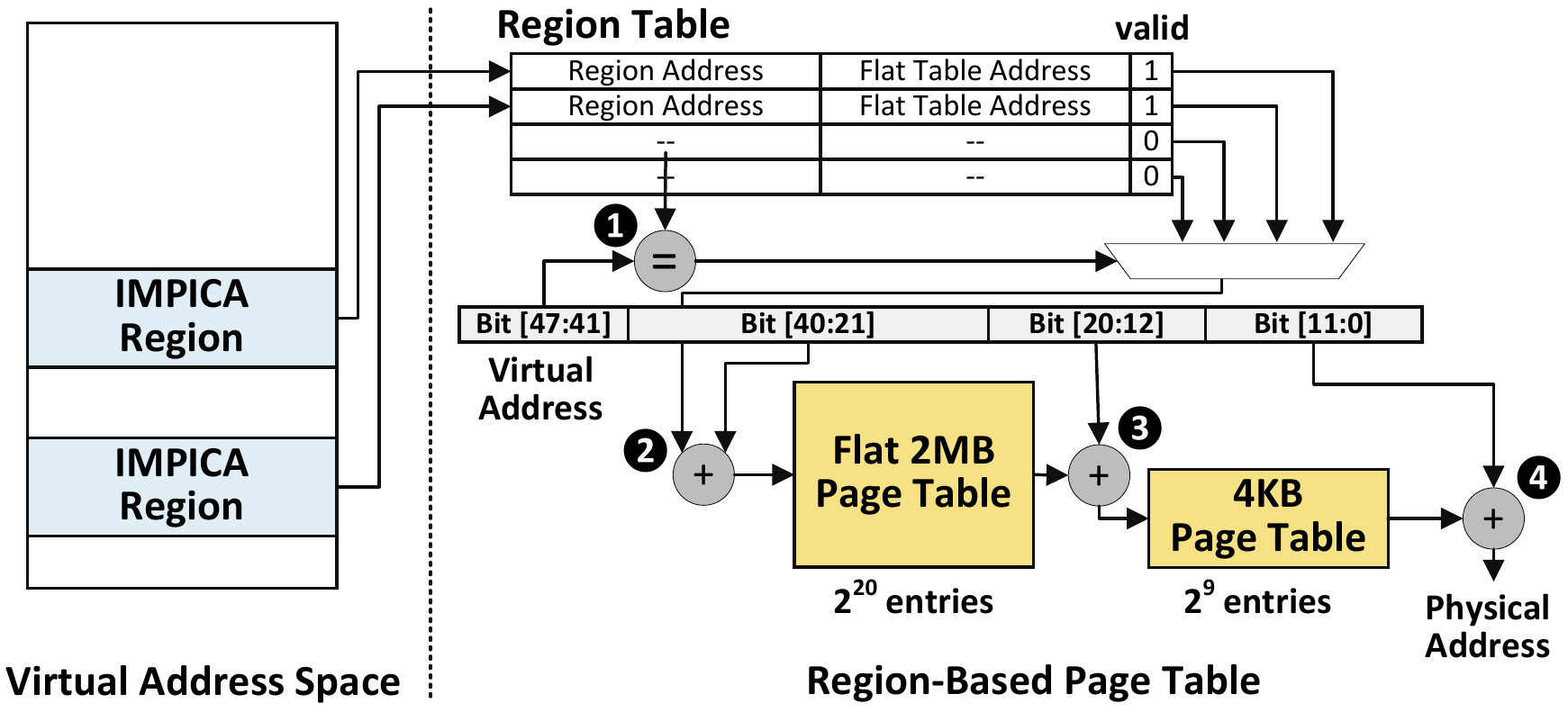}
  \caption{\abbrv virtual memory architecture. \rachata{Figure \chiv{adapted} from}~\cite{impica}.}
  \label{fig:virtual_memory}
\end{figure}

\sg{The RPT is optimized to take advantage of the properties of
  pointer chasing. The region table is almost always cached in the
  \abbrv cache, as the total number of \abbrv regions is small, 
  \kii{requiring} \changes{small} storage (e.g., a 4-entry region table \changes{needs only} 68B
  of cache space). We employ a flat table with large \kh{(e.g., 2MB)}
  pages at the second level in order to reduce the number of page
  misses, though this requires more memory capacity \kii{than the
  conventional 4-level page table structure}. As the number of
  regions touched by the accelerator is limited, this additional
  capacity overhead remains constrained. Our page table can optionally
  use traditional smaller page sizes to maximize memory management
  flexibility. The OS can freely choose \kii{large (2MB) pages} or
  \kii{small (4KB) pages} at the last level. Thanks to this design,
  a page walk in the RPT usually results in only two misses, one for
  the flat page table and another for the last-level \kh{small} page
  table. This represents a 2$\times$ improvement over \kii{a conventional
  four-level page table}, while our flattened page table still
  provides coverage for a 2TB memory range. \kh{\kii{The} size
    of the
    \abbrv region is configurable and can be increased to cover more
    virtual address space.
We believe that our RPT
  design is general enough for use in a variety of in-memory
  accelerators that operate on a specific range of memory regions.  }

\ignore{We maintain a page
table that uses traditional smaller page sizes at the last level, as multiple
accesses from different link traversals can touch a large number of pages,
which would require significant memory capacity if we only used large memory 
pages.}

We discuss how the OS manages the \abbrv RPT in 
Section~\ref{subsec:os_page_table}.
% !TEX root=../../chapter.tex

\subsection{\change{Interface and Design Considerations}}
\label{sec:interface_and_programming_model}

\sg{
In this section, we discuss how we expose \abbrv to the CPU and the \chiii{operating system (OS)}. 
Section~\ref{subsec:interface} describes the communication interface
between the CPU and \abbrv. Section~\ref{subsec:os_page_table} discusses how the
OS manages the page tables in \abbrv. In Section~\ref{subsec:coherence}, 
we discuss how cache coherence is maintained between the CPU and \abbrv caches.}

\subsubsection{CPU Interface and \chii{Communication} Model}
\label{sec:interface}
\label{subsec:interface}

We use a packet-based interface between the \kh{CPU and \abbrv}.
Instead of communicating individual operations or operands, the
packet-based interface buffers requests and sends them in a burst to
minimize the communication overhead.  Executing a function in \abbrv
consists of four steps on the interface.  (1)~\kh{The CPU sends \kii{to memory} a
  packet \kii{comprising}
the function call and parameters}.  (2)~This packet
is written to \kii{a specific location in memory}, which is memory-mapped to
the \textit{data RAM} in \abbrv and triggers \abbrv execution.  (3)~\abbrv
loads the specific function into the \chiii{\textit{instruction RAM}} with
appropriate parameters, by reading the values from predefined memory
locations.  (4)~Once \kh{\abbrv finishes the function execution, it
  writes the return value back to the memory-mapped locations in the \textit{data
  RAM}}. The CPU periodically polls these locations \kii{and receives
  the \abbrv output.} Note that the \abbrv interface is similar to the interface
proposed for the Hybrid Memory Cube (HMC)\chiii{~\cite{hmc.spec.1.1,hmc.spec.2.0, ahn.tesseract.isca15}}.  \ignore{We believe
that adopting a standard interface will help to solve any potential
portability issues.}

\subsubsection{\chii{\abbrv Programming Model}}
\label{sec:programming_model}

\rachata{The programming model for \abbrv is similar to the CPU programming model. 
An \abbrv program can be written in C with a new API that handles passing the parameters
and returning the results to the \abbrv accelerator.
Figure~\ref{fig:sample_code}a shows \chii{the pseudocode} for a B-tree traversal 
in the CPU, and Figure~\ref{fig:sample_code}b shows the equivalent pseudocode for \abbrv. 
We observe that the code fragments are very similar, differing in only two places.
First, the parameters passed in the function call of the CPU code are accessed 
with the \texttt{\_\_param} API call in \chviii{\abbrv (\incircle{1}} in Figure~\ref{fig:sample_code}).
The \texttt{\_\_param} API call ensures that the program explicitly 
reads the parameters from the predefined memory-mapped locations of the
data RAM. Second, instead of using the \texttt{return} statement, \abbrv
uses the same \texttt{\_\_param} API call to write the return value to a specific memory location \chii{(\incircle{2})}. 
This API call makes sure that the CPU can receive the output through the
\abbrv interface.}

\begin{figure}[h]
  \centering
  \subfloat[Conventional B-tree traversal pseudocode]{\includegraphics[height=140pt]{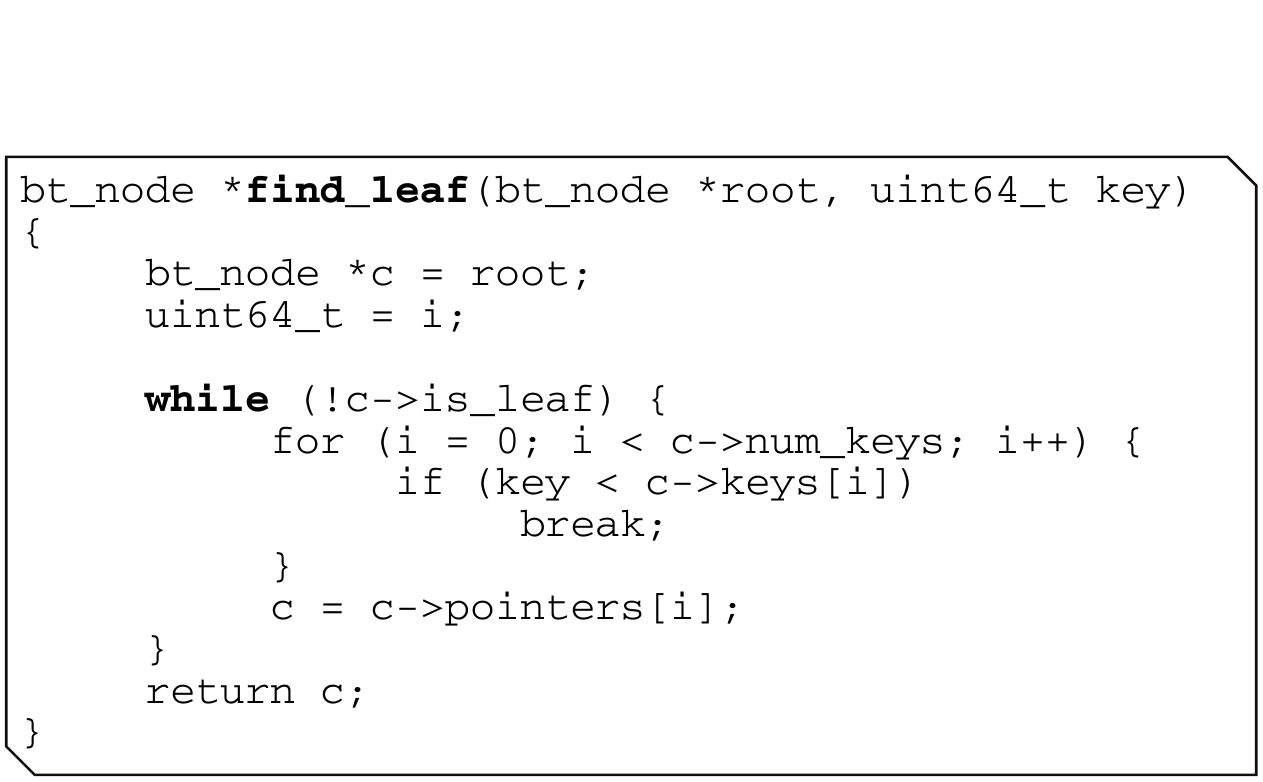}}\hfill%
  \subfloat[B-tree traversal pseudocode in \abbrv]{\includegraphics[height=140pt]{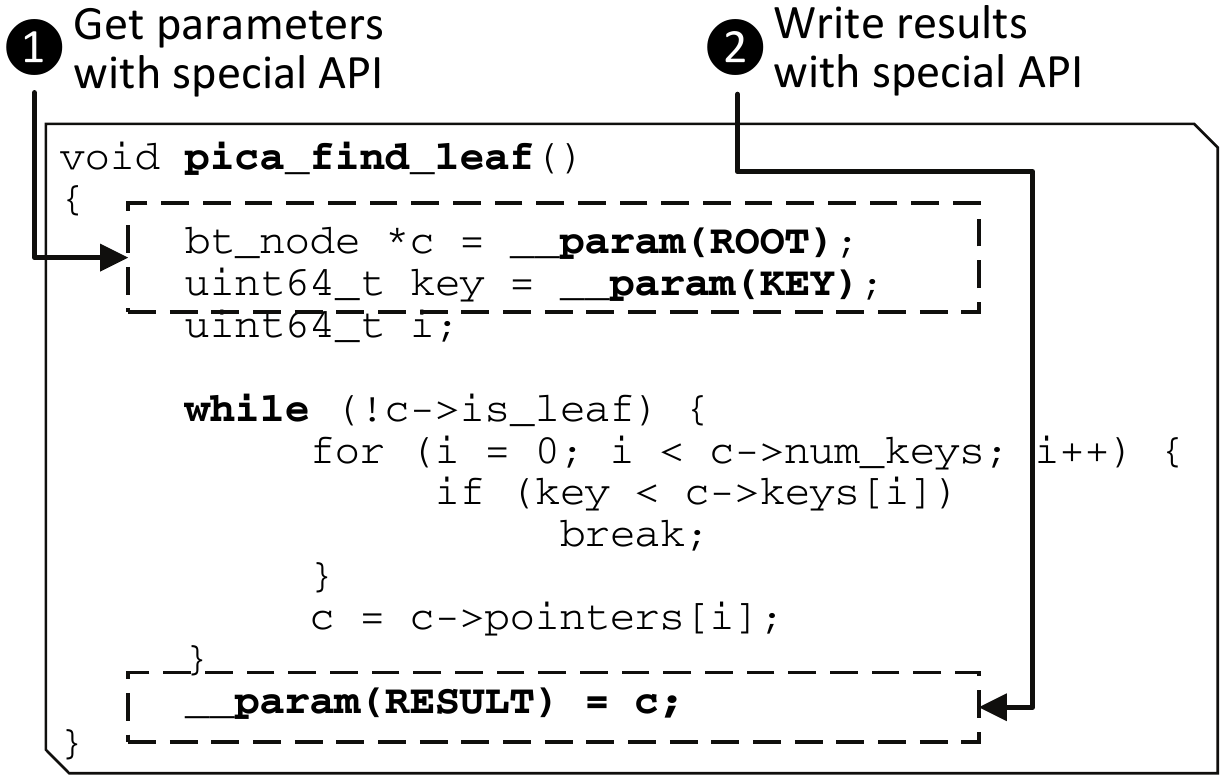}}%
  \caption{\rachata{B-tree traversal pseudocode \chii{demonstrating the differences 
    between the (a)~conventional and (b)~\abbrv programming models}.}}
  \label{fig:sample_code}
\end{figure}

\subsubsection{Page Table Management}
\label{subsec:os_page_table}

\sg{
In order for the RPT to identify \abbrv regions, the regions must be tagged by 
the application. \kh{For this, the application uses} a special API to allocate 
pointer-based data structures. This API allocates memory to a contiguous
virtual address space.  To ensure that all API allocations are contiguous, the 
OS reserves a \kii{portion} of the unused virtual address space for \abbrv, and always
allocates memory for \abbrv regions from this portion. \kii{The use
of such a special API} requires minimal changes to applications,
\kii{and} it allows the system to
provide more efficient virtual address translation. This also allows us to
ensure that when multiple memory stacks are present within the system, the OS
can allocate all \abbrv regions belonging to a single application (along with
the associated \abbrv page table) into one memory stack, \kii{thereby} avoiding the need
for \kii{the} accelerator to communicate with a remote \kh{memory} stack.}

\sg{
\kh{The OS maintains coherence between the \abbrv RPT and the \kii{CPU} page table}. When memory is allocated in the \abbrv region, the OS allocates the \abbrv
page table.  The OS also shoots down TLB entries in \abbrv if the CPU performs any
updates to \abbrv regions.  While this makes the OS page fault handler more
complex, the additional complexity does not cause a noticeable performance
impact, \kh{\kii{as} page faults occur rarely and take a long time to service in the CPU}.}

\subsubsection{Cache Coherence}
\label{subsec:coherence}

\sg{ Coherence must be maintained between the CPU and \abbrv caches,
  and with memory, to avoid using stale data and \kii{thus ensure
    correct execution}. We maintain coherence by executing
  \emph{every} function that operates on the \abbrv regions in the
  accelerator. This solution guarantees that no data is shared between
  the CPU and \abbrv, and that \abbrv always works on up-to-date data.
  \kii{Other PIM coherence solutions (e.g., LazyPIM in Section~\ref{sec:lazypim}, 
  \chii{or those proposed by} prior works~\cite{ahn.pei.isca15, gao.pact15}) can also be
  used} to allow the CPU \ignore{functions }to update the linked data structures,
  but we choose not to employ these solutions in our evaluation, as
  our workloads do \chviii{\emph{not}} perform any such updates.}

 \subsubsection{\rachata{Handling Multiple Memory Stacks}}
 \label{subsec:multiple_stack}
 
\rachata{
 Many systems need to employ multiple memory stacks to have enough memory
 capacity, as the \chii{current die-stacking} technology can integrate \chviii{only} a limited number of DRAM dies into
 a single memory stack~\cite{DBLP:conf/hpca/MeswaniBRSIL15}. In systems that use multiple memory stacks,
 the efficiency of an in-memory accelerator such as \abbrv could be
 significantly degraded whenever the data that the accelerator accesses is placed on different
 memory stacks. Without any modifications, \abbrv would have to go through the off-chip memory channels to access the
 data, which would effectively eliminate the benefits of in-memory computation.}

\rachata{ 
 Fortunately, this challenge can be tackled with our proposed modifications
 to the \chii{operating system (OS)} in Section~\ref{subsec:os_page_table}. As we can
 identify the memory regions that \abbrv needs to access, the OS can
 easily map \emph{all \abbrv regions} of an application into the same memory stack. In
 addition, the OS can allocate all \abbrv page tables into the same memory
 stack. This ensures that an \abbrv accelerator can access all of that data that it needs from within 
 the memory stack that it resides in
 without incurring any additional hardware cost or latency overhead.}
% !TEX root=../../chapter.tex

\subsection{\rachata{Evaluation Methodology for \abbrv}}
\label{sec:impica-meth}

\change{We use the gem5~\cite{GEM5} full-system simulator with 
DRAMSim2~\cite{DRAMSim2} to evaluate our proposed design. 
We choose
\kii{the 64-bit ARMv8 architecture,} the accuracy of which
has been validated against real
hardware~\cite{gutierrez2014sources}. We \chv{conservatively} model the internal memory
bandwidth of the memory stack to be 4$\times$ that of the external bandwidth,
similar to the configuration used in prior
works~\cite{zhang.hpdc14, farmahini-farahani.hpca15}. Our simulation
parameters are summarized in Table~\ref{table:sim_param}.}
\ch{Our source code is available openly at our research group's GitHub site~\cite{safari.github, impica.github}.
This distribution includes the source code of our microbenchmarks as well.}

\begin{table}[h]
\vspace{0pt}
\centering 
\small
\centering \caption{\kiii{Major simulation parameters \chiii{used} \rachata{for \abbrv evaluations}.}}
\vspace{-3pt}
\label{table:sim_param}	
\begin{tabular}{|c|l|} \hline
\multicolumn{2}{|c|}{\textbf{\rachata{Baseline Main Processor (CPU)}}}\\ \hhline{|=|=|}
\textbf{ISA} & ARMv8 (64-bits) \\ \hline 
\textbf{Core Configuration} & 4 OoO cores, 2 GHz, \changes{\chiii{8-wide issue}, 128-entry ROB} \\ \hline
\textbf{Operating System} & 64-bit Linux from Linaro~\cite{linaroGem5} \\ \hline 
	
\textbf{L1 I/D Cache} & 32KB/2-way each, 2-cycle \\ \hline 
\textbf{L2 Cache} & 1MB/8-way, shared, 20-cycle \\ \hline \hline

\multicolumn{2}{|c|}{\textbf{\rachata{Baseline Main Memory} Parameters}} \\ \hhline{|=|=|}
\textbf{Memory Configuration} & DDR3-1600, 8 banks/device, FR-FCFS scheduler\chii{~\cite{rixner.isca00, zuravleff.patent97}} \\ \hline 
\textbf{DRAM Bus Bandwidth} & 12.8 GB/s for CPU, 51.2 GB/s for \abbrv \\ \hline \hline

\multicolumn{2}{|c|}{\textbf{\abbrv Accelerator}} \\ \hhline{|=|=|}
\textbf{Accelerator Core} & \kh{500 MHz, 16 entries for each queue} \\ \hline
\textbf{Cache\footnotemark} & 32KB / 2-way  \\ \hline 
\textbf{Address Translator} & 32 TLB entries with region-based page table \\  \hline 
\textbf{RAM} & 16KB \chiii{data} RAM and 16KB \chiii{instruction} RAM \\ \hline 

\end{tabular}

\end{table}

\subsubsection{Workloads}

\sg{We use three data-intensive microbenchmarks, \kii{which} are essential building blocks
in a wide range of workloads, to evaluate the native performance of 
\rachata{pointer-chasing} operations: linked lists, hash tables, and B-trees.  We also evaluate
the performance improvement in a \kh{real data-intensive workload}, measuring the transaction 
latency and \rachata{transaction throughput} of DBx1000~\cite{yustaring}, an in-memory OLTP
database. We modify all four workloads} to offload each 
pointer chasing request to \abbrv. To minimize communication
overhead, we map the \abbrv registers to user mode address space,
\kh{thereby avoiding} the
need for \kii{costly} kernel code intervention.

\paratitle{Linked List} We use the linked list traversal
microbenchmark~\cite{zilles2001benchmark} derived from the \textit{health}
workload in the Olden benchmark suite~\cite{rogers1995supporting}. The parameters
are configured to approximate the performance of the \textit{health} workload.
We measure the performance of the linked list traversal after 30,000 iterations.

\paratitle{Hash Table} We create a microbenchmark from the hash table
implementation of \textit{Memcached}~\cite{fitzpatrick2004distributed}. The hash
table in Memcached resolves hash collisions using chaining \kh{via
  linked lists. When}
there are more than 1.5\textit{n} items in a table of \textit{n} buckets, it
doubles the number of buckets. We follow this rule by inserting
$1.5\times2^{20}$ random keys into a hash table with $2^{20}$
buckets. We run evaluations for 100,000 random key look-ups.

\paratitle{B-Tree}
We use the B-tree implementation of DBx1000 for our B-tree
microbenchmark. \ignore{The B-tree is used as the index of the database. }It is a 16-way
B-tree that uses a 64-bit integer as the key of each node. We randomly generate
3,000,000 keys and insert them into the B-tree. After the insertions, we measure
the performance of the B-tree traversal with 100,000 random keys. This is the
most time consuming operation in the database index lookup.

\paratitle{DBx1000}
We run DBx1000~\cite{yustaring} with the TPC-C
benchmark~\cite{council2005transaction}.
We set up the TPC-C tables
with 2,000 customers and 100,000 items. For each run, we spawn 4
threads and bind them to 4 different CPUs to achieve maximum
throughput. We run each thread for a warm-up period for the duration
of 2,000 transactions, and then record the software and hardware
statistics for the next 5,000 transactions per thread,\footnote{Based on our
  experiments on a real \kii{Intel} Xeon machine, we find that this is large enough to satisfactorily represent the behavior
  of 1,000,000 transactions.}
which takes 300--500 million CPU cycles.

\footnotetext{We sweep the size of the \abbrv cache from 32KB to
  128KB, and find that it has negligible effect on \kii{our} results.}

\subsubsection{Die Area and Energy Estimation}

We estimate the die area of the \abbrv processing logic at the 40nm process node based on
recently-published work~\cite{lotfi2012scale}. We include the most important
components: processor cores, L1/L2 caches, and the memory controller. 
We use the area of ARM Cortex-A57\kii{~\cite{CortexA57, filippo2012technology}}, a small embedded processor, for the
\rachata{baseline} main CPU.
We \emph{conservatively} estimate
the die area of \abbrv using the area of the Cortex-R4~\cite{CortexR4}, an 8-stage dual issue RISC
processor with 32 KB I/D
\changes{caches}. \rachata{We believe the actual area of an optimized \abbrv design can be much smaller.} \kh{Table~\ref{table:area_estimate} lists the
area estimate of each component.}

\begin{table}[h] \vspace{0pt} %\scriptsize 
\small
\caption{Die area estimates
    using a 40nm process \rachata{for \abbrv evaluations}.}
\vspace{-3pt}
\label{table:area_estimate}		\centering \begin{tabular}{|c||c|}
\hline \textbf{\kh{Baseline CPU \rachata{Core} (Cortex-A57)}} & 5.85 mm$^2$ per core \\ \hline
\textbf{L2 Cache} & 5 mm$^2$ per MB \\ \hline \textbf{Memory Controller} & 10
mm$^2$ \\ \hline 
\textbf{\ch{Complete Baseline Chip}} & \ch{38.4 mm$^2$} \\ \hline \hline
\textbf{\kh{\abbrv Core (including 32 KB I/D caches)}} & 0.45 mm$^2$ \rachata{(1.2\% of the baseline chip area)} \\
\hline \end{tabular}%
\end{table}

\kh{\abbrv comprises only 7.6\% the area of a single baseline \rachata{main} CPU core, or only 1.2\%
  the total area of the baseline chip (which includes four CPU cores, \kii{1MB} L2
  cache, and \kii{one} memory controller).}
Note that we conservatively model \abbrv as a RISC core. A much more specialized 
engine can be designed for \abbrv to solely execute pointer chasing code.
Doing so would reduce the area and energy overheads of \abbrv greatly, but can 
reduce the generality of the pointer chasing access patterns that \abbrv can 
accelerate. \changes{We} leave \rachata{such optimizations, evaluations, and analyses} for future work.

We use
McPAT~\cite{LiASBTJ13} to estimate the energy consumption of the CPU, caches, memory controllers,
and \abbrv. We conservatively use the configuration of the Cortex-R4 to estimate the
energy consumed by \abbrv. We use DRAMSim2~\cite{DRAMSim2} to analyze DRAM energy.

% !TEX root=../../chapter.tex

\subsection{Evaluation \rachata{of \abbrv}}
\label{sec:results}

\kh{We first evaluate the effect of \abbrv on system performance,}
using both our microbenchmarks (Section~\ref{sec:microbenchmark}) and the
DBx1000 database (Section~\ref{sec:database}). We investigate the impact of
different \abbrv page table designs in Section~\ref{sec:sensitivity_page_table},
and examine system energy consumption in Section~\ref{sec:energy}.
We compare a system containing \abbrv
to an accelerator-free baseline that includes an additional
128KB of L2 cache (which is equivalent to the area of \abbrv) \kh{to
  ensure area-equivalence across evaluated systems.}

\subsubsection{Microbenchmark Performance}
\label{sec:microbenchmark}

Figure~\ref{fig:microbenchmark_performance} shows the speedup of
\abbrv~\kh{and the baseline with extra 128KB of L2 cache}
over the baseline for each microbenchmark. 
\abbrv achieves significant speedups across all three
data structures --- 1.92$\times$ for the linked list, 1.29$\times$ for the hash table,
and 1.18$\times$ for the B-tree. \ignore{As the efficiency of \abbrv does not rely on any
prediction accuracy, it can improve the performance of a large hash table and a
16-way B-tree.} \kh{In contrast, \kii{the extra 128KB} of
L2 cache provides very small speedup (1.03$\times$, 1.01$\times$,
and 1.02$\times$, respectively). We conclude \kii{that} \abbrv is much more
\kii{effective} than the area-equivalent \kii{additional} L2 cache for pointer chasing operations. }

\begin{figure}[h]
  \centering
  \includegraphics[width=0.65\textwidth]{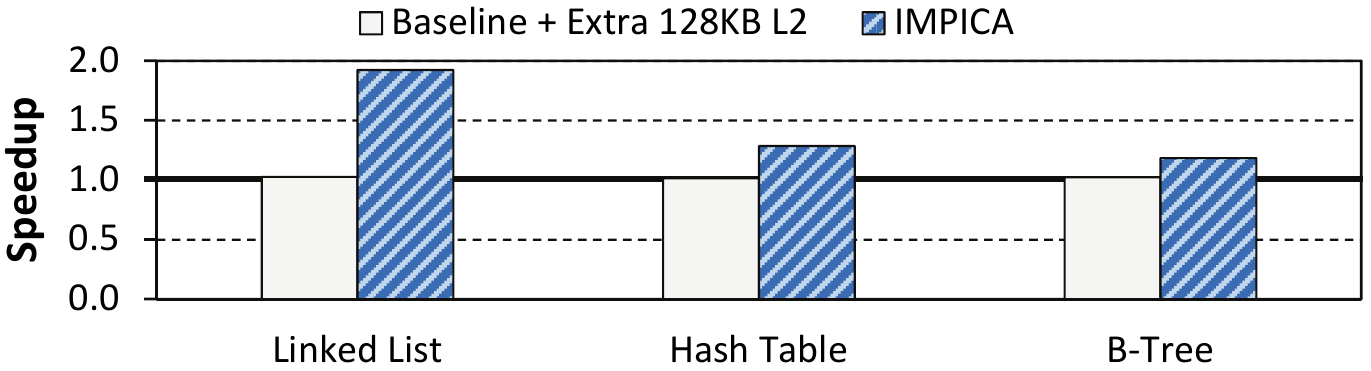}
  \caption{Microbenchmark performance with \abbrv. \rachata{Figure \chiv{adapted} from}~\cite{impica}.}
  \label{fig:microbenchmark_performance}
\end{figure}

To provide insight into \kh{why \abbrv improves performance}, we
present \kiii{total (\rachata{i.e., combined} CPU and \abbrv)} TLB misses
per kilo instructions (MPKI),
cache miss latency,
and total memory bandwidth usage
for these microbenchmarks in Figure~\ref{fig:microbenchmark}. We make three
observations.

First, a major factor contributing to the performance improvement is
the reduction in TLB misses. The TLB MPKI in
Figure~\ref{fig:microbenchmark}a depicts the \kiii{total (i.e.,
  combined CPU and
  \abbrv)} TLB misses in both the baseline system
and \abbrv.
The pointer chasing operations have low locality and pollute the CPU TLB. This
leads to a higher overall TLB miss rate in the application. With \abbrv, the
pointer chasing operations are offloaded to the accelerator. This reduces the
pollution and contention at the CPU TLB, reducing the overall number of
TLB misses. The linked list has \kii{a significantly higher
TLB MPKI than the other data structures} \ignore{ linked list has much more TLB misses per kilo
instructions than the other data structures. The MPKI of the linked list is 150, or
1.5 TLB misses per 10 instructions. This is }because linked list traversal
requires far fewer instructions \kii{in an iteration}. It simply accesses
the next pointer, while \kiii{a hash table or a B-tree} \kii{traversal} \kiii{needs} to compare the keys
in the node to determine the next step.

\begin{figure}[h]
  \centering
  \subfloat[Total TLB misses per kilo-instruction (MPKI)]{\includegraphics[height=100pt, trim = 0 0 495 0, clip]{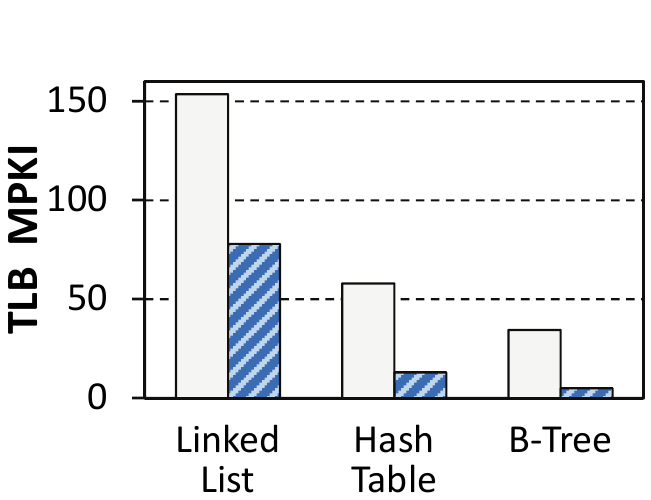}}\hfill%
  \subfloat[Average last-level cache miss latency]{\includegraphics[height=100pt, trim = 219 0 248 0, clip]{figures/IMPICA/microbenchmark.pdf}}\hfill%
  \subfloat[Total DRAM memory bandwidth utilization]{\includegraphics[height=100pt, trim = 481 0 0 0, clip]{figures/IMPICA/microbenchmark.pdf}}%
  \caption{Key architectural statistics for the \ch{evaluated} microbenchmarks. \rachata{Figure \chiv{adapted} from~\cite{impica}}.}
  \label{fig:microbenchmark}
\end{figure}

Second, we observe a significant reduction in \rachata{last-level} cache miss latency \kh{with}
\abbrv. Figure~\ref{fig:microbenchmark}b compares the \kh{average} cache miss
latency between the baseline last-level cache and the \abbrv
cache. \kh{On average, the cache miss latency of \abbrv is only
  60--70\% of the baseline cache miss latency}. This is because \abbrv leverages the
faster and wider TSVs in 3D-stacked memory as opposed to the \rachata{narrow, high-latency 
DRAM} interface used by the CPU.

Third, as Figure~\ref{fig:microbenchmark}c shows, \abbrv effectively
utilizes the internal memory bandwidth \ch{of} 3D-stacked memory, which is
cheap and abundant. \kh{There are two reasons for high bandwidth
  utilization: (1)~\abbrv runs much faster than the baseline so it
  generates more traffic within the same amount time; and (2)~\abbrv
  always accesses memory at a larger granularity, \kii{retrieving each
    full node in a linked data structure with a single memory
    request}, while a CPU issues multiple requests for each node as it
  can fetch only one cache line at a time. \kii{The CPU can avoid
    using some of its limited memory bandwidth by skipping some fields
    in the data structure that are not needed for the current loop
    iteration. For example, some keys and pointers in a B-tree node
    can be skipped whenever a match is found. In contrast, \abbrv
    utilizes the \ignore{cheap, }wide internal bandwidth of 3D-stacked memory
    \kiii{to retrieve a full node on each access}}.}

We conclude that \abbrv is effective at significantly improving the performance
of \kh{important} linked data structures.

\subsubsection{Real Database Throughput and Latency}
\label{sec:database}

Figure~\ref{fig:database} presents two key performance metrics \kh{for
our evaluation of DBx1000}:
\emph{database throughput} and \emph{database latency}. \emph{Database
throughput} represents how many transactions \kii{are completed} within a certain
period, while \emph{database latency} is the average time to complete a
transaction. We normalize the results of three configurations to the baseline.
As mentioned earlier, the die area increase of \abbrv is similar to a
128KB cache. To understand the effect of additional LLC space better, we also
show the results of adding 1MB of cache\kh{, which takes about 8$\times$ the area of 
\abbrv, to the baseline}. We make two observations from our analysis of DBx1000.

\begin{figure}[h]
  \centering
  \subfloat[Database transaction throughput]{\includegraphics[height=90pt, trim = 0 0 325 0, clip]{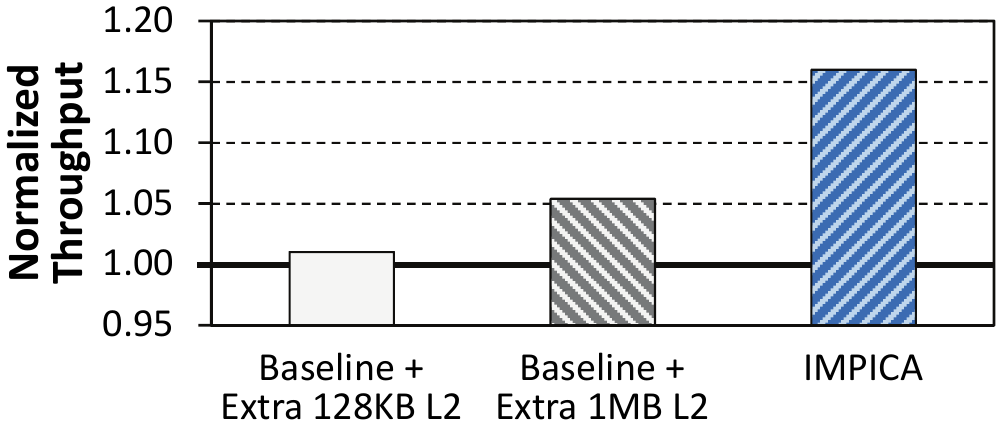}}\hfill%
  \subfloat[Database transaction latency]{\includegraphics[height=90pt, trim = 325 0 0 0, clip]{figures/IMPICA/database.pdf}}%
  \caption{Performance results for DBx1000, normalized to the baseline. \rachata{Figure \chiv{adapted} from~\cite{impica}.}}
  \label{fig:database}
\end{figure}

\kh{First, \abbrv improves the overall database
throughput by 16\% and reduces the average database transaction latency by
13\%. The performance improvement is due to three reasons}:
(1) database indexing becomes faster with \abbrv, (2) offloading database
indexing to \abbrv reduces the TLB and cache contention due to pointer
chasing \kh{in the CPU}, and (3) the CPU can do other useful tasks in parallel while
waiting for \abbrv. Note that our profiling results in
Figure~\ref{fig:profile} show that DBx1000 spends 19\% of its time on
pointer chasing. Therefore, a 16\% overall improvement is very close to the upper
bound that \emph{any} pointer chasing accelerator can achieve for this database.

Second,
\abbrv yields much higher \kh{database throughput than simply providing} additional cache
capacity. \kh{\abbrv improves the database throughput by 16\%, while
  an extra 128KB of cache (with a similar area \kiii{overhead as} \abbrv) \kii{does so} by only 2\%, and
  an extra 1MB of cache (8$\times$ \kii{the} area of \abbrv) by only
  5\%.}

We conclude that by accelerating the fundamental pointer chasing \kh{operation},
\chviii{\abbrv efficiently improves} the performance of \kii{a sophisticated
  real \rachata{data-intensive} workload}.

\subsubsection{Sensitivity to the \abbrv TLB Size \ch{and} Page Table Design}
\label{sec:sensitivity_page_table}

\kh{To understand the effect of different TLB sizes and page table
  designs in \abbrv, we evaluate the \kii{speedup in the amount of time spent on address translation} for
  \abbrv when different \abbrv TLB sizes (32 and 64 entries) and
  accelerator page table structures (the baseline 4-level page table;
  and the region-based page table, or RPT) are
  used \kiii{inside the accelerator}. Figure~\ref{fig:sensitivity_page_table} shows the speedup
  \kii{in address translation time}
  relative to \abbrv with a 32-entry TLB and the \kii{conventional}
  4-level page table. Two observations are in order.}

\begin{figure}[h]
  \centering
  \includegraphics[width=0.8\textwidth]{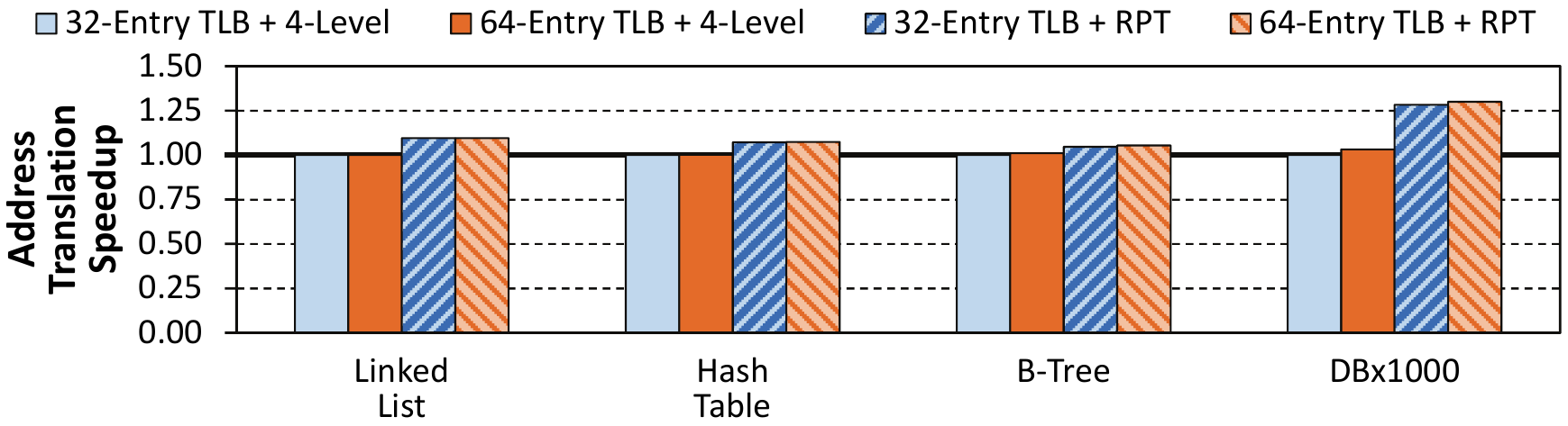}
  \caption{Speedup of address translation with different \rachata{TLB sizes and page table designs. Figure \chiv{adapted} from~\cite{impica}.}}  
  \label{fig:sensitivity_page_table}
\end{figure}

First, the performance of \abbrv is largely
unaffected \rachata{from small changes in} the \abbrv TLB size.  Doubling the \abbrv TLB
entries from 32 to 64 barely improves the address translation time.
This observation reflects the irregular nature of pointer
chasing. Second, the benefit of the RPT is much more significant in a
sophisticated workload (DBx1000) than in microbenchmarks. This is because the
working set size of the microbenchmarks is much smaller than that of
the database system. When the working set is small, the operating
system needs only a small number of page table entries in the first and second
levels of a traditional page table. These
entries are used frequently, so they stay in the \abbrv cache much
longer, reducing the address translation
overhead. This caching benefit goes away with a larger working set,
which would require a significantly larger TLB and \abbrv cache to reap
locality benefits. \kii{The benefit of RPT is more significant in
such \kiii{a} case because RPT does not \kiii{rely} on this caching effect. Its
region table is \emph{always} small irrespective of the workload \rachata{working set} size
and it has \kiii{fewer} page table levels.} \rachata{Thus, we conclude that RPT is a much more efficient and high-performance
page table design for our \abbrv accelerator than conventional page table design.}

\subsubsection{Energy Efficiency} \label{sec:energy}

Figure~\ref{fig:energy} \kh{shows} the \chiii{system power and} 
system energy consumption for the microbenchmarks and DBx1000. 
We observe that the overall \chiii{system \emph{power}
increases by 5.6\% on average}, due \kh{to the addition of \kii{\abbrv}}
and higher utilization of internal memory bandwidth. However, as \abbrv
significantly reduces the execution time of the evaluated workloads,
the \kii{overall} system \emph{energy}
consumption \kh{reduces by 41\%, 24\%, and 10\% for the 
microbenchmarks}, and by 6\% for DBx1000. We conclude that \abbrv is an
energy-efficient accelerator for \kh{pointer chasing.}

\begin{figure}[h]
  \centering
  \includegraphics[width=0.8\textwidth, trim = 0 110 0 0, clip]{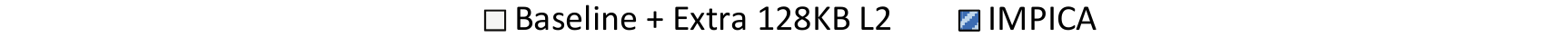}\\%
  \vspace{-12pt}\qquad\quad%
  \subfloat[System power]{\includegraphics[height=84pt, trim = 0 0 262 12, clip]{figures/IMPICA/energy.pdf}}\hfill%
  \subfloat[System energy consumption]{\includegraphics[height=84pt, trim = 262 0 0 12, clip]{figures/IMPICA/energy.pdf}\qquad\quad}%
  \caption{Effect of \abbrv on \chiii{system power (a) and system energy consumption (b)}. \rachata{\chiii{Figure~\ref{fig:energy}b} \chiv{adapted} from~\cite{impica}.}}
  \label{fig:energy}
\end{figure}
% !TEX root=../../chapter.tex

\subsection{Summary \rachata{of \abbrv}}

We introduce the design and evaluation of an \textit{in-memory
  accelerator}\kii{, called \abbrv,} for performing pointer chasing operations in 3D-stacked
memory. We identify two major challenges in the design
of such an in-memory accelerator: (1) the \textit{parallelism
  challenge} and (2) the \textit{address translation challenge}. We
provide new solutions to these two challenges: (1)~{\em
  address-access decoupling} solves the parallelism challenge
by decoupling the address generation from memory accesses in pointer
chasing operations and exploiting the idle time during memory accesses
to execute multiple pointer chasing operations in parallel, and (2)~the
{\em region-based page table} in 3D-stacked memory solves the
address translation challenge by tracking only those limited set of
virtual memory regions that are accessed by pointer chasing
operations. Our evaluations show that for both commonly-used linked
data structures and a \kh{real database application}, \abbrv
significantly improves both performance and energy efficiency. We
conclude that \abbrv is an efficient and effective accelerator design
for pointer chasing. We also believe that the two challenges we identify
(parallelism and address translation) exist in various forms in other
in-memory accelerators (e.g., for graph processing), \kh{and, therefore,} our
solutions to these challenges can be adapted for use by a broad
class of \changes{(in-memory)} accelerators. \rachata{We believe ample
future work potential exists on examining other solutions for these two
challenges as well as our solutions for them within the context of other in-memory
accelerators, such as those described in \chviii{\cite{ahn.tesseract.isca15, hsieh.isca16, seshadri2013rowclone, Seshadri:2015:ANDOR, seshadri.arxiv16, seshadri.micro17, kim.bmc18, kim.arxiv17, boroumand.asplos18}}.}
\chii{We also believe that examining solutions like \abbrv for other, non-in-memory accelerators is a promising direction to examine.}

% LazyPIM
% !TEX root=../../chapter.tex

\section{LazyPIM: An Efficient Cache Coherence Mechanism for Processing-in-Memory}
\label{sec:lazypim}

As discussed in Section~\ref{sec:keyissues:coherence}, cache coherence is a
major challenge for PIM architectures, as traditional coherence cannot be 
performed along the off-chip memory channel without potentially undoing the 
benefits of high-bandwidth \rachata{and low-energy} PIM execution.  
To work around the limitations presented by cache coherence,
most prior works assume a limited amount of sharing between 
the PIM kernels and the processor threads of an application.
Thus, they sidestep coherence by employing solutions \changes{that
restrict} PIM to execute on non-cacheable data
(e.g., \chii{\cite{ahn.tesseract.isca15, farmahini-farahani.hpca15, zhang.hpdc14, 
morad.taco15, DBLP:conf/hpca/GaoK16}}) or \changes{force} processor cores to
flush or not access any data that could \emph{potentially} be used by
PIM (e.g., \chiii{\cite{farmahini-farahani.hpca15, gao.pact15, Seshadri:2015:ANDOR, 
seshadri2013rowclone, guo2014wondp, ahn.pei.isca15, impica, nai2017graphpim,
hsieh.isca16, seshadri.arxiv16, seshadri.micro17, chang.hpca16, pattnaik.pact16,
DBLP:conf/isca/AkinFH15}}). \rachata{In fact, \chii{the} \abbrv accelerator design, described in Section~\ref{sec:impica}, falls into the latter
category.}

To understand the trade-offs that can occur by sidestepping coherence, we 
analyze several data-intensive applications.
We make two \emph{key observations} based on our analysis: 
(1)~some portions of the applications are
better suited for execution in processor threads, and these portions often concurrently access the same region
of data as the PIM kernels, leading to \emph{significant data sharing}; and 
(2)~poor handling of coherence eliminates a significant portion of the performance benefits of PIM.
As a result, we find that a good coherence mechanism is \emph{required} to ensure
the correct execution of the program while maintaining the benefits of PIM
(see Section~\ref{sec:motivation}).
\textbf{Our goal} in this \rachata{section is to describe} a cache coherence mechanism for PIM architectures that
\emph{logically behaves} like traditional coherence, but retains all of the benefits of PIM.

To this end, we propose \emph{LazyPIM}, a new cache coherence mechanism that
efficiently batches coherence messages sent by the \chiii{PIM processing logic}.  During PIM kernel execution, a PIM core 
\emph{speculatively} assumes that it has acquired coherence \mbox{permissions} without 
sending a coherence message, and maintains all data updates speculatively in its 
cache.  Only when the kernel finishes execution, the processor receives compressed information from
the PIM core, and checks if any coherence conflicts occurred. 
If a conflict exists
(see Section~\ref{sec:mech:conflicts}), the dirty cache lines in the processor 
are flushed, and the PIM core rolls back and re-executes the kernel. Our execution model \emph{for \chiii{PIM processing logic}}
is similar to \emph{chunk-based execution}~\cite{bulksc} (i.e., each 
\emph{batch} of consecutive instructions executes atomically), which prior 
work has harnessed for various purposes\ch{~\cite{bulksc,tcc,dmp, muzahid.isca09, sanchez.micro07, pokam.micro09}}.
Unlike past works, however, the processor in LazyPIM executes conventionally and \emph{never rolls back},
which can make it easier to enable PIM.

We make the following key contributions in this \rachata{section}:
\begin{itemize}
  \itemsep 0pt
  \item We propose a new hardware coherence mechanism for PIM.  Our approach
    (1)~reduces the off-chip traffic between the PIM 
    cores and the processor, (2)~avoids the costly overheads of prior 
    approaches to provide coherence for PIM, and (3)~retains the same 
    logical coherence behavior as architectures without PIM to keep programming 
    simple.
  \item LazyPIM improves average performance by 49.1\% (coming within 5.5\% of an ideal PIM mechanism), and
reduces off-chip traffic by 58.8\%, over the best prior coherence approach.
\end{itemize}
% !TEX root=../paper.tex

\subsection{Baseline PIM Architecture}
\label{sec:background}

In our evaluation, we assume that the compute units \chiii{inside memory} consist of simple 
\emph{in-order} cores. These \chiii{PIM cores}, which are ISA-compatible with the 
out-of-order processor cores, are much weaker in terms of performance, as
they lack large caches and sophisticated ILP techniques, but are more 
practical to implement within the DRAM logic layer, \rachata{as we discussed earlier in Section~\ref{sec:pimarch:logic}}.
Each PIM core has private L1 I/D caches, which are kept coherent using
a MESI directory\ch{~\cite{papamarcos.isca84, censier.tc78}}
within the DRAM logic layer.
A second directory in the 
processor acts as the main coherence point for the system, interfacing 
with both the processor cache and the PIM coherence directory.
Like prior PIM works~\chii{\cite{ahn.tesseract.isca15, farmahini-farahani.hpca15,ahn.pei.isca15,hsieh.isca16,pattnaik.pact16, DBLP:conf/isca/AkinFH15,
DBLP:conf/hpca/GaoK16}},
we assume that direct segments~\cite{basu2013efficient}
are used for PIM data, and that PIM kernels operate only on physical addresses.

% !TEX root=../../chapter.tex

\subsection{Motivation \rachata{for Coherence Support in PIM}}
\label{sec:motivation}

Applications benefit the most from PIM execution when their memory-intensive parts, 
which often exhibit poor locality and contribute to a large portion of 
execution time, are dispatched to \chiii{PIM processing logic}. On the other hand, compute-intensive
parts or those parts that exhibit high locality \emph{must remain 
on the processor cores} to maximize performance~\cite{ahn.pei.isca15, hsieh.isca16}.

Prior work mostly assumes that there is only a limited amount of sharing between the PIM kernels and the processor. 
However, \emph{this is not the case} for many important applications, such as graph and database workloads. For example, in multithreaded graph  
frameworks, each thread performs a graph algorithm (e.g., connected 
components, PageRank) on a shared graph\ch{~\cite{Seraph,ligra, ahn.tesseract.isca15}}. We study a number of these algorithms~\cite{ligra}, and find that (1)~only 
certain portions of each algorithm are well suited for PIM, and (2)~the PIM kernels and processor threads access the shared graph and intermediate 
data structures concurrently. 
Another example is modern in-memory databases that support Hybrid Transactional/Analytical
 Processing (HTAP) workloads\ch{~\cite{SAP,stonebraker2013voltdb,MemSQL,GS-DRAM}}. 
The analytical portions of these databases are well suited for PIM execution~\cite{kocberber2013meet,Hash:NME,JAFAR}. 
In contrast, even though transactional queries access the \emph{same} data, they perform better if they are executed
 on the \rachata{main processor (i.e., the CPU)}, 
as they are short-lived and latency sensitive, accessing only a few rows each. 
Thus, concurrent accesses from both PIM kernels (analytics) and processor threads (transactions) are inevitable.

The shared data needs to remain
 coherent between the processor and PIM cores.
Traditional, or \emph{fine-grained}, coherence protocols (e.g., MESI\ch{~\cite{censier.tc78, papamarcos.isca84}})
have several qualities well suited for 
 pointer-intensive data structures, such as those in graph workloads and databases.
Fine-grained coherence allows the processor or PIM to acquire permissions for only
the pieces of data that are \ch{\emph{actually accessed}}.
In addition, fine-grained coherence 
can ease programmer effort when developing PIM applications, as multithreaded programs already use this programming model.
Unfortunately, if a PIM core participates in traditional coherence, it would have to send
 a message for \emph{every cache miss} to the processor over a narrow \chvi{shared interconnect} (we call this \rachata{type of \chvi{interconnect} traffic as} \emph{PIM
 coherence traffic}). 

\rachata{We study four mechanisms to evaluate how coherence protocols impact 
PIM: (1)~\emph{CPU-only}, a baseline where PIM is disabled; (2)~\emph{FG}, fine-grained coherence, which is
the MESI protocol, variants of which are employed in many state-of-the-art systems; (3)~\emph{CG}, coarse-grained
lock based coherence, where PIM cores gain \emph{exclusive} access to all PIM data during PIM kernel execution; and (4)~\emph{NC}, non-cacheable,
where the PIM data is not cacheable in the CPU. We describe CG and NC in more detail below.}
Figure~\ref{fig:motiv} shows the speedup of PIM with \ch{these four mechanisms} for certain graph workloads,
normalized to \ch{CPU-only}.\footnoteref{foot:meth}
To illustrate the impact of inefficient \ch{coherence} mechanisms, we also show the performance of 
an \emph{ideal} mechanism where there is no performance penalty for coherence (\emph{Ideal-PIM}).
 As shown in Figure~\ref{fig:motiv}, \rachata{employing PIM with a state-of-the-art fine-grained coherence (\emph{FG}) mechanism \emph{always}} performs worse than CPU-only execution.

\begin{figure}[h]
    \centering
        \centering
        \includegraphics[width=0.75\linewidth]{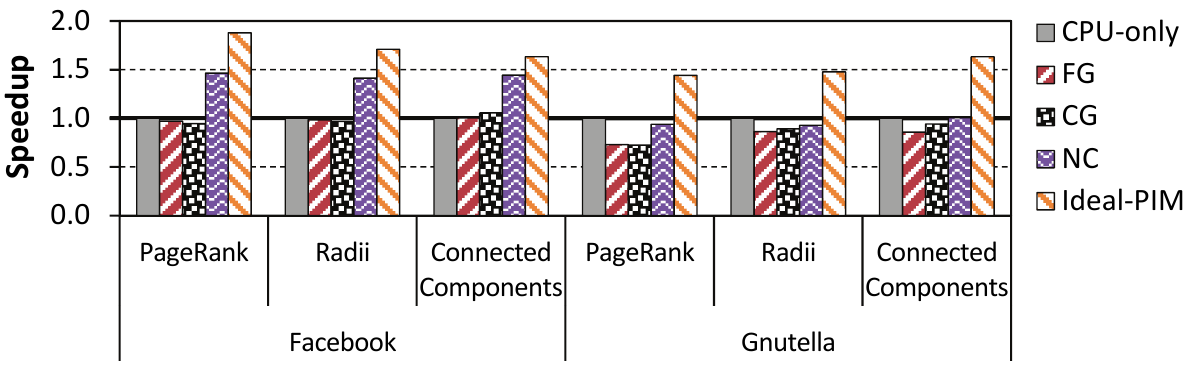}%
    \caption{PIM speedup with 16 threads, normalized to CPU-only\ch{, with three different and ideal coherence mechanisms}.{\protect\footnoteref{foot:meth}} \rachata{Figure \chiv{adapted} from~\cite{boroumand2016pim}}.}
    \label{fig:motiv}
\end{figure}

To reduce the impact of PIM coherence traffic, there are three general alternatives to fine-grained
coherence for PIM execution: (1) coarse-grained coherence, (2) coarse-grained locks,
 and (3) making PIM data non-cacheable in the processor. \rachata{We briefly examine these alternatives.}
 
\paratitle{Coarse-Grained Coherence}
One approach to reduce PIM coherence traffic is to 
maintain a single
coherence entry for \emph{all} of the PIM data.  Unfortunately, this can 
still incur high overheads, as the processor
must \changes{flush} \emph{all} of the dirty cache lines within the PIM data 
region \emph{every time} the PIM core acquires permissions, 
\emph{even if the PIM kernel may not access most of the data}.
For example, with just four processor threads, the number of cache lines flushed for PageRank is 
227x the number of lines \emph{actually required by the PIM kernel}.\footnote{\label{foot:meth}See
Section~\ref{sec:methodology} for our \rachata{experimental evaluation} methodology.}
 Coherence at a smaller granularity, such as page-granularity~\cite{gao.pact15}, 
does not cause flushes for pages not
accessed by the PIM kernel.  However, many data-intensive applications perform
\emph{pointer chasing}, where a large number of pages are
accessed non-sequentially, but only a \emph{few lines} in each page are used, forcing
the processor to flush \emph{every} dirty page.

\paratitle{Coarse-Grained Locks}
Another drawback of coarse-grained coherence is that data can ping-pong
between the processor and the PIM cores whenever the PIM data region is concurrently
accessed by both.  \emph{Coarse-grained
locks} avoid ping-ponging by having the PIM cores acquire \emph{exclusive} access to a region for
the duration of the PIM kernel.  However, 
coarse-grained locks greatly restrict performance. 
Our application study shows that PIM kernels and processor threads often
work in parallel on the same data region, and coarse-grained locks frequently
cause thread serialization. % in the processor. 
 PIM with coarse-grained locks (\emph{CG} in Figure~\ref{fig:motiv}) 
performs 8.4\% \emph{worse}, on average, than CPU-only execution.
We conclude that using coarse-grained locks is not suitable for many
important applications for PIM execution.

\paratitle{Non-Cacheable PIM Data}
Another approach sidesteps coherence by marking the PIM data region 
as \emph{non-cacheable} in the processor\chii{~\cite{ahn.tesseract.isca15, farmahini-farahani.hpca15, zhang.hpdc14, 
morad.taco15, DBLP:conf/hpca/GaoK16}}, 
so that DRAM always contains up-to-date data.  For applications where PIM data
is almost exclusively accessed by the \chiii{PIM processing logic}, this incurs little
penalty, but for many applications, the processor also accesses PIM data
often.  For our graph applications with a representative input (arXiV),\footnoteref{foot:meth}
the processor cores generate 42.6\% of the total number of accesses to PIM data.
With so many processor accesses,
making PIM data non-cacheable results in \ch{a} high performance and 
bandwidth overhead. As shown in Figure~\ref{fig:motiv}, though 
marking PIM data as non-cacheable (\emph{NC}) sometimes performs better than CPU-only,
it still loses up to 62.7\% (\changes{on average, 39.9\%}) of the improvement of Ideal-PIM.
 Therefore, while this approach avoids the overhead of
coarse-grained mechanisms, it is a poor fit for applications
that rely on processor involvement, and thus restricts \rachata{the applications where} PIM is effective.

We conclude that prior approaches to PIM coherence eliminate a significant
portion of the benefits of PIM when data sharing occurs, due to their high 
coherence overheads. In fact, they sometimes cause PIM execution to \ch{\emph{consistently 
degrade performance}}. Thus, an \emph{efficient} alternative to fine-grained 
coherence is necessary to retain PIM benefits across a wide range
of applications. 

% !TEX root=../../chapter.tex

\subsection{LazyPIM Mechanism \rachata{for Efficient PIM Coherence}}
\label{sec:mech}

Our goal is to design a coherence mechanism that maintains the logical behavior
of traditional coherence while retaining the large performance benefits of PIM.
%but without the overheads of prior PIM coherence approaches. 
To this end, we propose 
\emph{LazyPIM}, a new coherence mechanism that lets PIM kernels
\emph{speculatively} assume that they have the required permissions from the
coherence protocol, \emph{without} actually sending off-chip messages to the main
(processor) coherence directory during execution.
Instead, coherence states are updated only
\emph{after} the PIM kernel completes, at which point the PIM core transmits a
single batched coherence message (i.e., a compressed \emph{signature} containing \emph{all} addresses that the PIM
kernel read from or wrote to) back to the processor coherence
directory. The directory checks to see whether any \emph{conflicts} occurred. 
If a conflict exists, the PIM kernel
\emph{rolls back} its changes, conflicting cache lines are written back by the processor to 
DRAM, and the kernel re-executes.  If no conflicts exist, speculative data within 
the PIM core is \emph{committed}, and the processor coherence directory is
updated to reflect the data held by the PIM core.  Note that in LazyPIM, the
processor \emph{always} executes \emph{non-speculatively}, which ensures
minimal changes to the processor design, thereby \rachata{likely} enabling easier adoption of 
PIM.

LazyPIM avoids the pitfalls of the \rachata{coherence} mechanisms discussed in
Section~\ref{sec:motivation} (\rachata{FG, CG, NC}).  With its compressed signatures, LazyPIM causes much
less PIM coherence traffic than traditional \rachata{fine-grained} coherence.  Unlike coarse-grained
coherence and coarse-grained locks, LazyPIM checks coherence \emph{only after} 
it completes PIM execution, avoiding the need to
unnecessarily flush a large amount of data.  
\ch{Unlike non-cacheable, LazyPIM allows processor threads to cache the data 
used by PIM kernels within the processor cores as well, avoiding the need for
the processor to perform a large number of off-chip accesses that can hurt 
performance greatly.}
LazyPIM also allows for efficient concurrent
execution of processor threads and PIM kernels: by executing speculatively,
the PIM cores do \emph{not} invoke coherence requests during concurrent execution,
\ch{avoiding data ping-ponging between the PIM cores and the processor.}

\paratitle{Conflicts} %Due to Speculation}
\label{sec:mech:conflicts}
In LazyPIM, a PIM kernel \emph{speculatively} assumes during execution that it
has coherence permissions on a cache line, without checking the processor
coherence directory.  In the meantime, the processor continues to execute
\emph{non-speculatively}.  To resolve \changes{PIM kernel} speculation, % without violating the memory consistency model, 
LazyPIM
provides \emph{coarse-grained atomicity}, where all PIM memory updates are
treated as if they \emph{all} occur \emph{at the moment that a PIM kernel
finishes execution}.  (We explain how LazyPIM 
enables this in Section~\ref{sec:mech:arch}.) 
\ch{If, before the PIM kernel finishes, the processor updates a cache line that the PIM 
kernel read during its execution, a \emph{conflict} occurs.  LazyPIM detects and handles
all potential conflicts once the PIM kernel finishes executing.}

\ch{Figure~\ref{fig:lazypimtimeline}} shows an example timeline where a PIM kernel is launched on PIM
core PIM0 while execution continues on processor cores CPU0 and CPU1.
Due to the use of coarse-grained atomicity, PIM kernel execution behaves as if
\emph{the entire kernel's memory accesses} take place at the moment coherence is checked (i.e., at the end of 
kernel execution), \emph{regardless of the actual time at which the kernel's accesses are
performed}.  Therefore, for \emph{every} cache line read by PIM0,
if CPU0 or CPU1 modify the line before the coherence
check occurs, PIM0 unknowingly uses stale data, leading to incorrect execution.
\ch{Figure~\ref{fig:lazypimtimeline}} shows
two examples of this: (1)~CPU0's write to line~C \emph{during} kernel execution; and
(2)~CPU0's write to line~A \emph{before} kernel execution, which was not 
written back to DRAM.
To detect such conflicts, we record the addresses of processor writes and 
PIM kernel reads into two signatures, and then check to see if any 
addresses \changes{in them} match \rachata{(i.e., conflict) \emph{after} the PIM} kernel finishes (see 
Section~\ref{sec:mech:arch:spec}).

\begin{figure}[h]
\includegraphics[width=0.5\linewidth, trim=0 0 300 0, clip]{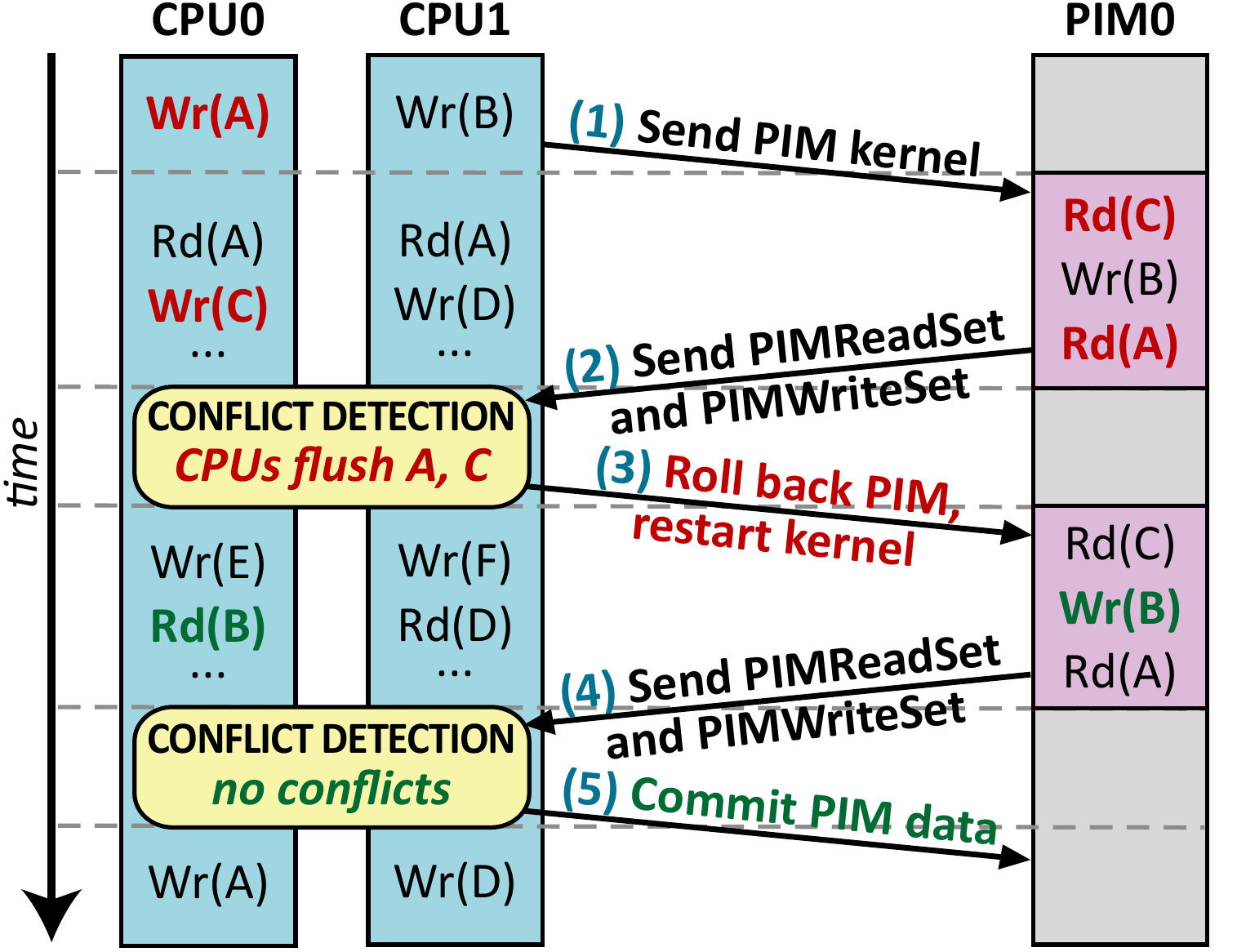}%
\caption{\ch{Example timeline of LazyPIM coherence sequence.  Figure reproduced from \cite{boroumand2016pim}.}}
\label{fig:lazypimtimeline}
\end{figure}

If the PIM kernel writes to a cache line that is subsequently 
read by the processor before the kernel finishes (e.g., the second write by PIM0 to line~B in Figure~\ref{fig:lazypimtimeline}), this is \emph{not} 
a conflict. With coarse-grained atomicity, any read by the processor during 
PIM execution is ordered \emph{before} the PIM kernel's write. LazyPIM ensures
that the processor cannot read the \rachata{PIM} kernel's writes, by marking \rachata{the PIM kernel writes} as speculative \rachata{inside the \chiii{PIM processing logic}}
until the kernel finishes (see Section~\ref{sec:mech:arch:spec}).
This is also the case when the processor and a PIM kernel write 
to the \ch{\emph{same}} cache line. Note that this ordering does not violate consistency models, such as sequential consistency.\ch{\footnote{\ch{A thorough treatment of memory
consistency\chviii{~\cite{lamport.tc79}} is outside the scope of this work. Our goal is to deal with the coherence problem in PIM, not handle consistency issues.}}}

\ch{If the PIM kernel writes to a cache line that is subsequently 
\emph{written to} by the processor before the kernel finishes, this is \emph{not} 
a conflict. With coarse-grained atomicity, any
write by the processor during PIM kernel execution is ordered before the PIM
core's write since the PIM core write effectively
takes place \emph{after the PIM kernel finishes}. When the two writes modify
different words in the same cache line, LazyPIM uses a per-word dirty bit mask
in the PIM L1 cache to merge the writes, similar to prior work~\cite{lee.hpca01}.
Note that the dirty bit mask is only in the PIM L1 cache; processor caches
remain unchanged.}

\chiii{More details on the operation of the LazyPIM coherence mechanism are provided in our arXiv paper~\cite{boroumand.arxiv17}.}

\subsection{\ch{Architectural Support for LazyPIM}}
\label{sec:mech:arch}

\subsubsection{\chii{LazyPIM Programming Model}}
\label{sec:mech:arch:program}

We provide a simple %two-step 
interface
to port applications to 
LazyPIM.  
\ch{We show the implementation of a simple LazyPIM kernel within a program in
\chii{Code Example}~\ref{code:program_lazypim}.
The programmer} selects the portion(s) of the code to
execute on PIM cores, 
using two macros (\ch{\texttt{\#PIM\_begin} and \texttt{\#PIM\_end}}).
The compiler converts the macros into instructions
that we add to the ISA, which \emph{trigger} and \emph{end} PIM kernel execution. 
\ch{LazyPIM also needs to know which parts of the 
allocated data \emph{might} be accessed by the PIM cores,
which we refer to as the \emph{PIM data region}.}\chii{\footnote{\chii{The programmer should be conservative in identifying PIM data regions, and should not miss \emph{any possible data} that may be touched by a PIM core.  If any data \emph{not marked} as PIM data is accessed by the PIM core, the program can produce incorrect results.}}}
\ch{We assume that either the
programmer or the compiler can annotate all of the \ch{PIM
data region} using compiler directives or a PIM memory allocation API (\texttt{@PIM}).} This information is saved in the
page table using per-page flag bits, \rachata{via communication to the system software using the system call interface.}

\begin{lstlisting}[frame=tb, language=C, captionpos=b, caption=Example PIM program implementation. \chiii{Modifications for PIM execution are shown in bold.},
label=code:program_lazypim, basicstyle=\small\ttfamily, float=h, numbers=left, framexleftmargin=2em, xleftmargin=2em,
keywords={@PIM, PIM_begin, PIM_end}, breaklines=true, escapechar = ^, commentstyle=\color{gray}, keywordstyle=\color{Blue}\bfseries]
PageRankCompute(@PIM Graph GA) { // GA is accessed by PIM cores in edgeMap()
  const int n = GA.n;            // not accessed by PIM cores
  const double damping = 0.85, epsilon = 0.0000001; // not accessed by PIM cores
  @PIM double* p_curr, p_next;   // accessed by PIM cores in edgeMap()
  @PIM bool* frontier;           // accessed by PIM cores in edgeMap()
  @PIM vertexSubset Frontier(n, n, frontier); // accessed by PIM in edgeMap()
  double L1_norm;                // not accessed by PIM cores
  long iter = 0;                 // not accessed by PIM cores
  ...
  while(iter++ < maxIters) {
    #PIM_begin
      // only the edgeMap() function is offloaded to the PIM cores
      vertexSubset output = ^\break^ edgeMap(GA, Frontier, @PIM PR_F<vertex>(p_curr, p_next, GA.V), 0);
      // PR_F<vertex> object allocated during edgeMap() call, needs annotation
    #PIM_end
    vertexMap(Frontier, PR_Vertex_F(p_curr, p_next, damping, n));  // run on CPU

    // compute L1-norm between p_curr and p_next
    L1_norm = fabs(p_curr - p_next);  // run on CPU
    if(L1_norm < epsilon) break;      // run on CPU
    ...
  }
  Frontier.del();
}
\end{lstlisting} 

\chiii{Code Example~\ref{code:program_lazypim} shows a portion of the compute
function used by PageRank, as modified for execution with LazyPIM.  
All of our modifications are shown in bold.  In this example, 
we want to execute only the \texttt{edgeMap()} function (Line~13) on
the PIM cores.  To ensure that LazyPIM tracks all data accessed during the \texttt{edgeMap()}
call, we mark all of this data using \texttt{@PIM}, 
including any objects passed by value (e.g., \texttt{GA} on Line~1), 
any objects allocated in the function (e.g., those on Lines 4--6),
and any objects allocated during functions that are executed on the PIM cores
(e.g., the \texttt{PR\_F{\textless}vertex\textgreater} object on Line~13).  To tell the compiler
that we want to execute only \texttt{edgeMap()} on the PIM cores, we surround
it with the \texttt{\#PIM\_begin} and \texttt{\#PIM\_end} compiler directives 
on Lines~11 and 15, respectively.  No other modifications are needed to execute our example
code with LazyPIM.}

\subsubsection{Speculative Execution}
\label{sec:mech:arch:spec}

When an application reaches a \emph{PIM kernel trigger} instruction, the processor
dispatches the kernel's starting PC to a free PIM core.
The PIM core \emph{checkpoints} the starting PC \ch{and registers},
and starts executing the kernel.  The kernel 
\emph{speculatively} assumes that it has coherence permissions for \emph{every} 
line it accesses, without \emph{actually} 
checking the processor directory.  We add 
a one-bit flag to each line in the PIM core cache,
to mark all data updates as speculative. 
If a speculative line is selected for eviction, the core rolls back 
to the starting PC and discards the updates.

LazyPIM tracks three sets of addresses during PIM kernel execution.  These are
recorded into three \emph{signatures}, as shown in \ch{Figure~\ref{fig:lazypimhw}}: 
(1)~the \emph{CPUWriteSet} (all \emph{CPU writes} to the PIM data region), 
(2)~the \emph{PIMReadSet} (all \emph{PIM reads}), and
(3)~the \emph{PIMWriteSet} (all \emph{PIM writes}).
When the kernel starts, the dirty lines in the processor cache containing PIM 
data are recorded in the CPUWriteSet, by scanning the tag store (potentially
using a Dirty-Block Index~\cite{seshadri2014dirty}).  The processor uses the page
table flag bits from Section~\ref{sec:mech:arch:program} to identify which writes 
need to be added to the CPUWriteSet during kernel execution.  The PIMReadSet
and PIMWriteSet are updated for \emph{every} read and write performed by the
PIM kernel.  When the kernel finishes execution, the three signatures are used
to resolve speculation (see Section~\ref{sec:mech:arch:detect})

\begin{figure}[h]
\includegraphics[width=0.85\linewidth]{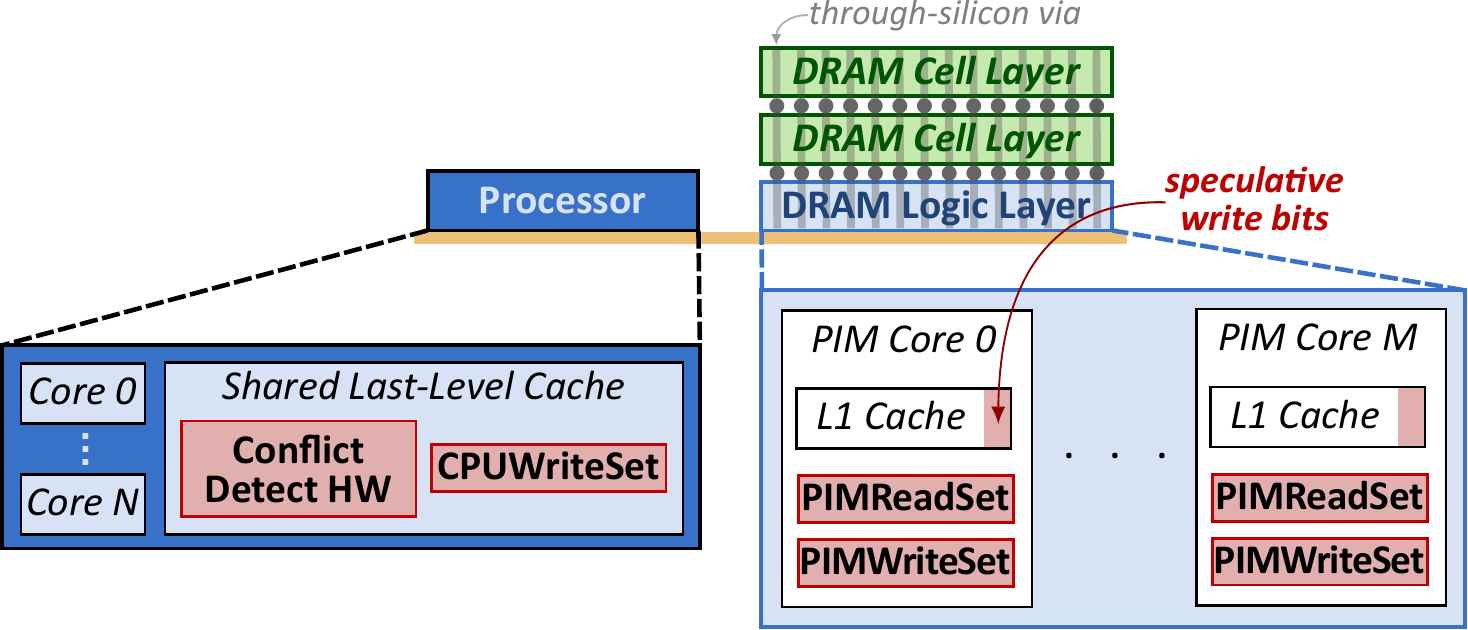}%
\caption{\ch{High-level additions (in bold) to PIM architecture to support LazyPIM.  Figure \chiv{adapted} from \cite{boroumand2016pim}.}}
\label{fig:lazypimhw}
\end{figure}

\chv{The} signatures use parallel
Bloom filters~\cite{Bloom:1970:STH:362686.362692},
which employ simple
Boolean logic to hash multiple addresses into a single (256B) fixed-length
register.  
\chv{If the speculative coherence requests were sent back to the processor
without any sort of compression at the end of PIM kernel execution, the
coherence messages would still consume a large amount of off-chip traffic, 
nullifying most of the benefits of the speculation.
Bloom filters allow LazyPIM to compress these coherence messages into a
much smaller size, while guaranteeing that there are no false 
negatives~\cite{Bloom:1970:STH:362686.362692}
(i.e., no coherence messages are lost during compression).
The addresses of \emph{all} data accessed speculatively by the PIM cores} can be 
extracted and compared from the \chv{Bloom filter}~\cite{Bloom:1970:STH:362686.362692, 
bulksc}. The hashing 
introduces a limited number of false positives,
\chv{with the false positive rate increasing as we store more addresses
in a single fixed-length Bloom filter.
In our evaluated system, each 
signature is 256B long, and can store up to 607 addresses
without exceeding a 20.0\% false positive rate (with \emph{no} false 
negatives).
%Each 256B~filter can store up to 607~addresses.
To store more addresses, we use multiple filters to guarantee an upper bound
on the false positive rate.}

\subsubsection{Handling Conflicts}
\label{sec:mech:arch:detect}

As \ch{Figure~\ref{fig:lazypimtimeline}} shows,
we need to detect conflicts that occur during PIM kernel execution.
In LazyPIM, when the kernel finishes executing, both the PIMReadSet 
and PIMWriteSet are sent back to the processor.

If no matches are detected between the PIMReadSet and the CPUWriteSet (i.e., no 
conflicts have occurred), PIM kernel \emph{commit} starts.  
Any addresses (including false positives) in the PIMWriteSet are invalidated 
from the processor cache. A message is then sent to the PIM core, allowing it 
to write its speculative cache lines back to DRAM.  During the commit \rachata{process}, 
all coherence directory entries for the PIM data region are locked to ensure
atomicity \rachata{of commit}. Finally, all signatures are erased.

If an overlap is found between the PIMReadSet and the CPUWriteSet,
a conflict may have occurred.
 %All
\rachata{At this point,} only the dirty lines in the processor that match 
in the PIMReadSet are flushed back to DRAM.
During this flush, all
PIM data directory entries are locked to ensure atomicity.  Once 
the flush completes, a message is sent to the PIM core, telling it to
invalidate all speculative cache lines, and to \emph{roll back} the PC to the checkpointed
value.  Now that all possibly conflicting cache lines are written back to DRAM, all signatures
are erased, and the PIM core \ch{\emph{restarts}} the kernel.  After
re-execution \rachata{of the PIM kernel} finishes, conflict detection is performed again.

\ch{Note that during the commit process, processor cores do not stall
unless they access the same data accessed by \chiii{PIM processing logic}.} LazyPIM guarantees
forward progress by acquiring a lock for each line in the PIMReadSet after a
number of rollbacks \ch{(we empirically set this number to three rollbacks).
This simple mechanism ensures there is no livelock even if the sharing of
speculative data among PIM cores might create a cyclic dependency. Note that
rollbacks are caused by CPU accesses to conflicting addresses, and not by the
sharing of speculative data between PIM cores. As a result, once we lock
conflicting addresses following three rollbacks, the PIM cores will not roll back
again as there will be no conflicts, guaranteeing forward progress.}

\subsubsection{Hardware Overhead}
\label{sec:mech:arch:overhead}

LazyPIM's overhead consists mainly of 
(1)~1~bit per page (0.003\% of DRAM capacity) and 1~bit per TLB entry for the page table
flag bits (Section~\ref{sec:mech:arch:program}); 
(2)~a 0.2\% increase in PIM core L1 size to mark speculative data (Section~\ref{sec:mech:arch:spec}); 
\ch{(3)~a 1.6\% increase in PIM core L1 size for the dirty bit mask (Section~\ref{sec:mech:conflicts});} and 
(4)~in the worst case, 12KB for the signatures per PIM core (Section~\ref{sec:mech:arch:spec}).
This overhead
can be greatly optimized (as part of future work):
for PIM kernels that
need multiple signatures, we could instead divide the kernel into smaller chunks where each chunk's addresses fit in a single signature, lowering signature overhead to \ch{512B}.
\rachata{We leave a detailed evaluation of LazyPIM hardware overhead optimization to future work. Some ideas
related to this and a detailed hardware overhead analysis are presented in our arXiv paper~\cite{boroumand.arxiv17}.}
% !TEX root=../../chapter.tex

\subsection{Methodology \rachata{for LazyPIM Evaluation}}
\label{sec:methodology}

We study two types of data-intensive applications: graph workloads and databases.  
We use three Ligra~\cite{ligra} graph applications (PageRank, Radii, Connected Components), with input
graphs constructed from real-world network datasets~\cite{snap}: Facebook, arXiV High
Energy Physics Theory, and
Gnutella25 (peer-to-peer). We also use an in-house prototype of a modern in-memory database
 (IMDB)\ch{~\cite{SAP,stonebraker2013voltdb,MemSQL, GS-DRAM}} that supports HTAP workloads. 
 Our transactional workload consists of 200K transactions, each randomly performing
 reads or writes on a few randomly-chosen tuples. Our analytical workload 
 consists of 256 analytical queries that use the select and join operations on 
 randomly-chosen tables and columns.

PIM kernels are selected from these applications with the help of OProfile~\cite{OProfile}.
We conservatively select candidate PIM kernels, choosing \ch{portions of functions}
where the application (1)~spends the majority of its
cycles, and (2)~generates the majority of its last-level cache misses.
From these candidates, we pick kernels that we believe minimize 
the coherence overhead for each evaluated mechanism, by minimizing
data sharing between the processor and \chiii{PIM processing logic}.
We modify each application to ship the \rachata{selected PIM kernels} to
the PIM cores.  We manually annotate the PIM data set.

For our evaluations, we modify the gem5 simulator~\cite{GEM5}.
We use the x86-64 architecture in full-system mode, and use DRAMSim2~\cite{DRAMSim2}
to perform detailed timing simulation of DRAM.  Table~\ref{tbl:config}
shows our system configuration.

\begin{table}[h]
    \small
    \centering
    \caption{Evaluated system configuration \rachata{for LazyPIM evaluation}.}
    \vspace{-3pt}
    \label{tbl:config}%
\begin{tabular}{|c|l|} \hline
\multicolumn{2}{|c|}{\textbf{\rachata{Main Processor (CPU)}}}\\ \hhline{|=|=|}
\textbf{ISA} & x86-64 \\ \hline 
\textbf{Core Configuration} & 4--16 cores, 2 GHz, 8-wide issue \\ \hline
\textbf{Operating System} & 64-bit Linux from Linaro~\cite{linaroGem5} \\ \hline 
\textbf{L1 I/D Cache} & 64KB per core, private, 4-way associative, 64B blocks, 2-cycle lookup \\ \hline 
\textbf{L2 Cache} & 2MB, shared, 8-way associative, 64B blocks, 20-cycle lookup \\ \hline
\textbf{Cache Coherence} & MESI directory~\cite{papamarcos.isca84, censier.tc78} \\ \hline \hline

\multicolumn{2}{|c|}{\textbf{\rachata{PIM Cores}}}\\ \hhline{|=|=|}
\textbf{ISA} & x86-64 \\ \hline 
\textbf{Core Configuration} & 4--16 cores, 2 GHz, 1-wide issue \\ \hline
\textbf{L1 I/D Cache} & 64KB per core, private, 4-way associative, 64B blocks, 2-cycle lookup \\ \hline 
\textbf{Cache Coherence} & MESI directory~\cite{papamarcos.isca84, censier.tc78} \\ \hline \hline

\multicolumn{2}{|c|}{\textbf{\rachata{Main Memory} Parameters}} \\ \hhline{|=|=|}
\multirow{2}{*}{\textbf{Memory Configuration}} & HMC 2.0~\cite{hmc.spec.2.0}, one 4GB cube, 16 vaults per cube, 16 banks per vault,\\
& FR-FCFS scheduler\chii{~\cite{rixner.isca00, zuravleff.patent97}} \\ \hline 
    \end{tabular}%
\end{table}
% !TEX root=../../chapter.tex

\subsection{Evaluation \ch{of LazyPIM}}
\label{sec:eval}

We first analyze the off-chip traffic reduction of LazyPIM. \rachata{This off-chip reduction}
leads to bandwidth and energy savings. We then
analyze LazyPIM's \rachata{effect on system}
performance.
We show \rachata{system performance} results normalized to a processor-only baseline (\emph{CPU-only}, as defined in Sec.~\ref{sec:motivation}),
and compare \rachata{LazyPIM's performance} with using fine-grained coherence (\emph{FG}), coarse-grained locks 
(\emph{CG}), or non-cacheable data (\emph{NC}) for PIM data.

\subsubsection{Off-Chip \rachata{Memory} Traffic}
\label{sec:eval:offchip}

\chiii{Figure~\ref{fig:traffic}a} shows the normalized off-chip \rachata{memory} traffic of the PIM coherence mechanisms for a
16-core architecture (with 16 processor cores and 16~PIM cores)
\chiii{Figure~\ref{fig:traffic}b} shows the normalized off-chip \rachata{memory} traffic as the 
number of threads increases, for PageRank using the Facebook graph.  
LazyPIM significantly reduces \ch{the} \emph{overall} off-chip traffic 
(up to 81.2\% over CPU-only, 70.1\% over FG, 70.2\% over CG, and 97.3\% over NC), and 
scales better with thread count.
LazyPIM reduces \rachata{off-chip memory} traffic by 58.8\%, on average, over CG, the best prior approach in terms of \rachata{off-chip} traffic.

\begin{figure}[h]
    \centering
    \subfloat[\chiii{16-thread off-chip memory traffic}]{\includegraphics[width=0.63\textwidth]{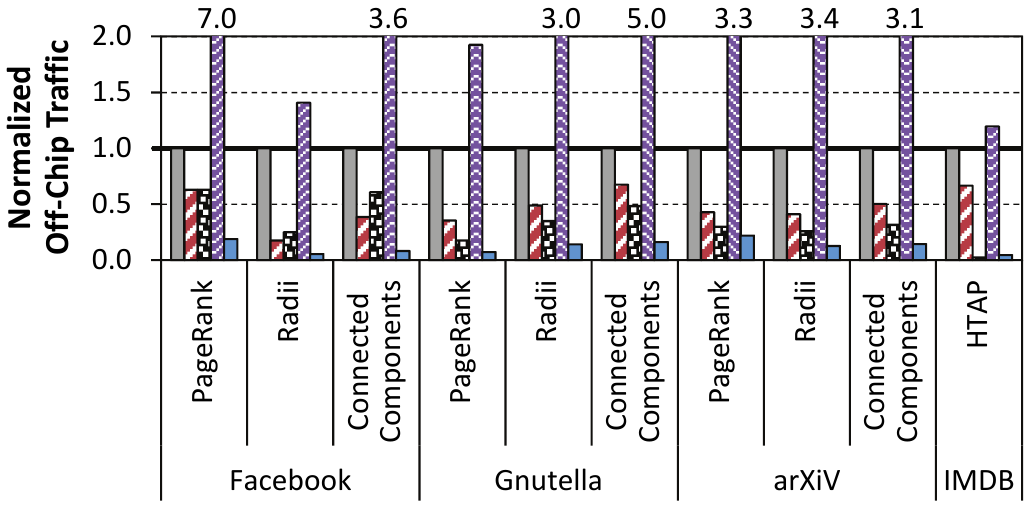}}%
    \hfill%
    \subfloat[\chiii{Off-chip memory traffic sensitivity to thread count for PageRank}]{\includegraphics[width=0.31\textwidth]{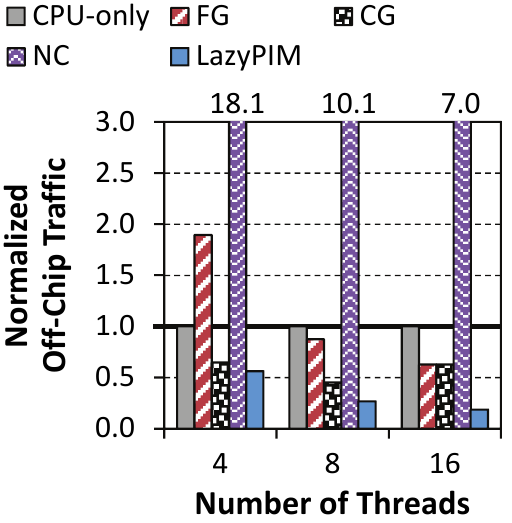}}%
    \caption{\chiii{Effect of LazyPIM on off-chip memory traffic, normalized to CPU-only.} \rachata{Figure \chiv{adapted} from~\cite{boroumand2016pim}.}} 
    \label{fig:traffic}
\end{figure}

CG has greater traffic than LazyPIM, the majority of which is 
due to having to flush dirty cache lines before each PIM kernel invocation.  
Due to false sharing, the number of flushes scales \emph{superlinearly} with 
thread count (not shown),
increasing 13.1x from 4 to 16~threads.  LazyPIM avoids this cost
with speculation, as \emph{only} the \emph{necessary} flushes are performed \emph{after} the
PIM kernel finishes execution. As a result, it reduces the flush count (e.g., by 94.0\% for
16-thread PageRank \changes{using} Facebook), and thus lowers overall 
off-chip \rachata{memory} traffic (by 50.3\% for our example).

NC suffers from the fact that \emph{all} processor accesses
to PIM data must go to DRAM, increasing average off-chip \rachata{memory} traffic by 3.3x over CPU-only.
NC off-chip \rachata{memory} traffic also scales poorly with thread count, as more processor threads generate a greater number of accesses.  In
contrast, LazyPIM allows processor cores to cache PIM data, by enabling
coherence efficiently.

\subsubsection{Performance}
\label{sec:eval:perf}

\chiii{Figure~\ref{fig:performance}a} shows the performance improvement for 16~threads.
Without any coherence overhead, Ideal-PIM 
significantly outperforms CPU-only across \ch{\emph{all}} applications, showing PIM's potential on these workloads. 
Poor handling of coherence by FG, CG, and NC leads to drastic performance losses compared to Ideal-PIM,
indicating that an efficient coherence mechanism is essential for PIM performance.
For example, in some cases, NC and CG actually perform \emph{worse} than CPU-only, and for
PageRank running on the Gnutella graph, all prior mechanisms degrade performance.  
In contrast, LazyPIM consistently retains most of Ideal-PIM's benefits for all applications, coming within
5.5\% on average. LazyPIM outperforms all of the other approaches, improving over the best-performing prior approach (NC) by 
49.1\%, on average.

\begin{figure}[h]
    \centering
    \subfloat[\chiii{Speedup for all applications with 16~threads}]{\includegraphics[width=0.63\textwidth]{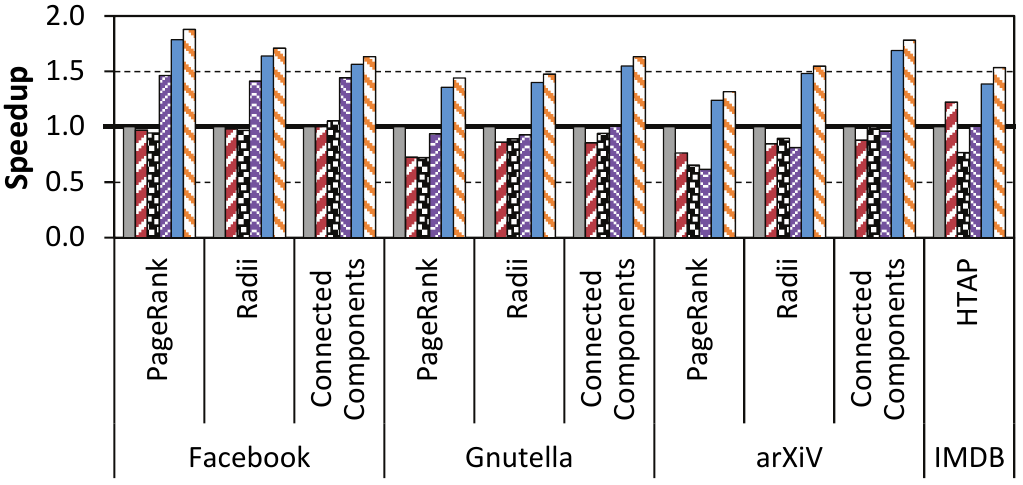}}%
    \hfill%
    \subfloat[\chiii{Speedup sensitivity to thread count for Gnutella}]{\includegraphics[width=0.31\textwidth]{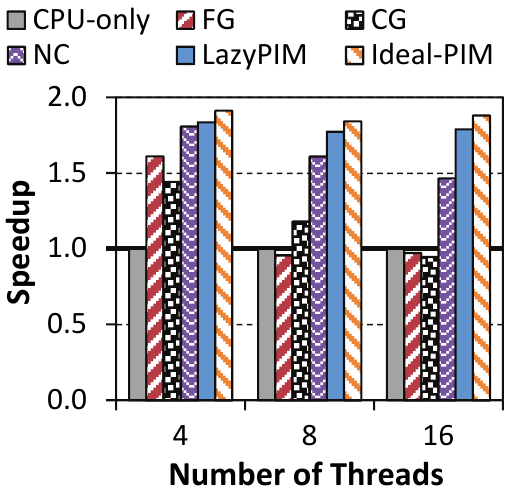}}   % \caption{Speedup sensitivity to thread count for PageRank with Facebook graph (left), and speedup for all applications with 16~threads (right) for coherence mechanisms, normalized to CPU-only.}
    \caption{\chiii{Speedup of cache coherence mechanisms,} normalized to CPU-only. \rachata{Figure \chiv{adapted} from~\cite{boroumand2016pim}.}}
    \label{fig:performance}
\end{figure}

\chiii{Figure~\ref{fig:performance}b} shows the performance 
 of PageRank using Gnutella as we increase the thread count.  
 LazyPIM comes within 5.5\% of Ideal-PIM, \rachata{which has no coherence overhead} (as defined in Sec.~\ref{sec:motivation}), and improves performance by 73.2\% over FG, 47.0\%
 over CG, 29.4\% over NC, and 39.4\% over CPU-only, on average. With NC, the processor threads
 incur a large penalty \changes{for} \rachata{accessing} DRAM frequently.
 CG suffers greatly due to (1) flushing dirty cache lines, and (2) blocking all processor
 threads that access PIM data during execution. In fact, processor threads 
 are blocked for up to 73.1\% of the total execution time with CG. With more threads, the \rachata{negative} effects of 
 blocking worsen CG's performance. FG also loses a significant portion of Ideal-PIM's 
 improvements, as it sends a large amount of off-chip messages.
Note that NC, despite its high off-chip traffic, performs better 
than CG and FG, as it neither blocks processor cores nor slows down PIM execution.

\chv{One reason for the difference in performance between LazyPIM and
Ideal-PIM is the number of conflicts that are detected at the end of PIM kernel
execution.  As we discuss in Section~\ref{sec:mech:arch:detect}, any detected
conflict causes a rollback, where the PIM kernel must be re-executed.
We study the number of commits that contain conflicts for
two representative 16-thread workloads: Components using the Enron graph, 
and HTAP-128 (results not shown).  If we
study an idealized version of full kernel commit, where no
false positives exist, we find that a relatively high percentage of
commits contain conflicts (47.1\% for Components and 21.3\%
for HTAP). Using realistic signatures for full kernel commit,
which includes the impact of false positives, the conflict rate
increases to 67.8\% for Components and 37.8\% for HTAP.
Despite the high number of commits that induce rollback, the overall performance
impact of rollback is low, as LazyPIM comes within 5.5\% of the performance
of Ideal-PIM.  We find that for all of our applications, a kernel never rolls
back more than once, limiting the performance impact of conflicts.
We can further improve the performance of LazyPIM by optimizing the
commit process to reduce the rollback overhead, which we explore in our
arXiv paper~\cite{boroumand.arxiv17}.}
% !TEX root=../paper.tex

\subsection{Summary \ch{of LazyPIM}}
\label{sec:lazypim:conclusion}

We propose LazyPIM, a new cache coherence mechanism  
for PIM architectures.  Prior approaches to PIM
coherence generate very high off-chip traffic for important
data-intensive applications.  LazyPIM avoids this by avoiding coherence lookups
\emph{during} PIM kernel execution.
\ch{The key idea is to use compressed coherence \emph{signatures} to batch the lookups and verify correctness \emph{after} 
the kernel completes. 
As a result of the more efficient approach to coherence employed by LazyPIM,
applications that performed poorly under prior approaches to PIM coherence
can now take advantage of the benefits of PIM execution.}
LazyPIM improves average performance by 49.1\% (coming within 5.5\% of an ideal PIM mechanism), and
reduces off-chip traffic by 58.8\%, over the best prior approach to PIM coherence 
while retaining the conventional multithreaded  programming model.

\vspace{-2pt}

% !TEX root=../chapter.tex

\section{Related Work}
\label{sec:related}

\ch{We briefly survey related work in processing-in-memory, accelerator design, mechanisms for handling pointer chasing, and techniques for pointer chasing.}

\paratitle{\chii{Early Processing-in-Memory (PIM) Proposals}}
\chii{Early proposals for PIM architectures had limited to no adoption, as the
proposed logic integration was too costly and did not solve many of the 
obstacles facing \chiii{the adoption of} PIM.}
\chii{The earliest such proposals date} from the 1970s, where small processing elements
\chiii{are} combined with small amounts of RAM to provide a distributed array of
memories that perform computation~\cite{shaw1981non, stone1970logic}.
\chii{Some of the \chiii{other} early works, such as EXECUBE~\cite{kogge1994execube}, Terasys~\cite{gokhale1995processing},
IRAM~\cite{patterson1997case}, and Computational RAM~\cite{elliott1992computational, elliott.dt99},}
 \chiii{add} logic within DRAM
 to perform vector operations. \chii{\chiii{Yet other} early works, such as FlexRAM~\cite{kang2012flexram},
DIVA~\cite{Draper:2002:ADP:514191.514197}, Smart Memories~\cite{Mai:2000:SMM:339647.339673},
and Active Pages~\cite{oskin1998active}, \chiii{propose} more versatile substrates 
that tightly \chiii{integrate} logic and reconfigurability within DRAM itself to increase} flexibility and \chiii{the available compute power}.

\paratitle{\chii{Processing in 3D-Stacked Memory}}
With the advent of 3D-stacked memories, we have seen a resurgence of PIM
proposals\chiii{~\cite{loh2013processing, seshadri.bookchapter17}}.
\chii{Recent PIM proposals 
add} compute units within the logic layer to exploit the high bandwidth available.
These works \chiii{primarily focus} on the design of the underlying logic that is
placed within memory, and in many cases propose special-purpose PIM architectures
that cater only to a limited set of applications.
\chii{These works include accelerators for
MapReduce~\cite{pugsley2014ndc}, matrix
multiplication~\cite{zhu2013accelerating}, data
  reorganization~\cite{DBLP:conf/isca/AkinFH15}, graph
  processing~\cite{ahn.tesseract.isca15, nai2017graphpim},
  databases~\cite{DBLP:conf/sigmod/BabarinsaI15}, 
  in-memory analytics~\cite{gao.pact15}, 
\chv{genome sequencing~\cite{kim.bmc18, kim.arxiv17},}
data-intensive processing~\cite{gu.isca16}, 
\chviii{consumer device workloads~\cite{boroumand.asplos18},}
and machine learning
  workloads~\cite{chi.isca16, kim.isca16,
    DBLP:conf/IEEEpact/LeeSK15}.}
Some works propose more generic
architectures by adding PIM-enabled
instructions~\cite{ahn.pei.isca15}, GPGPUs\kh{~\cite{zhang.hpdc14,
    hsieh.isca16, pattnaik.pact16}, \chii{single-instruction multiple-data (SIMD)
processing units~\cite{morad.taco15}}, or reconfigurable
  hardware\chii{~\cite{farmahini-farahani.hpca15, DBLP:conf/hpca/GaoK16, guo2014wondp}}} to
memory.

\paratitle{\chii{Processing Using Memory}}
\chii{A number of recent works have examined how to perform memory operations
directly within the memory array itself, which we refer to as \emph{processing
using memory}~\cite{seshadri.bookchapter17}.  These works take advantage of inherent
architectural properties of memory devices to perform operations in bulk.  While
such works can significantly improve computational efficiency within memory,
they still suffer from many of the same programmability \chiii{and adoption} challenges that PIM
architectures face, such as the address translation and cache coherence challenges
that we focus on in this chapter.  Mechanisms for processing using memory 
can perform a variety of functions, such as bulk copy and data initialization for 
\chvii{DRAM~\cite{seshadri2013rowclone, chang.hpca16,kevinchang-thesis,seshadri.thesis16}}, bulk \chviii{bitwise operations} for 
DRAM\chv{~\cite{Seshadri:2015:ANDOR, seshadri.arxiv16, seshadri.micro17, li.micro17}} and phase-change memory (PCM)~\cite{li.dac16}, and
simple arithmetic operations for SRAM\chiii{~\cite{kang.icassp14, aga.hpca17}} and memristors\chviii{~\cite{shafiee.isca16, levy.microelec14, kvatinsky.tcasii14, kvatinsky.iccd11, kvatinsky.tvlsi14}}.}

\paratitle{\chii{Processing in the DRAM Module or Memory Controller}}
\chii{Several works have examined how to embed processing functionality near
memory, but not within the DRAM chip itself.  Such an approach can reduce the
cost of PIM manufacturing, as the DRAM chip does not need to be modified or
specialized for any particular functionality.  However, these works (1)~are often
unable to take advantage of the high internal bandwidth of 3D-stacked DRAM,
which reduces the efficiency of PIM execution, and (2)~\chiii{may} still suffer from many of the
same challenges faced by architectures that embed logic within the DRAM chip.
Examples of this work include
Chameleon~\cite{asghari-moghaddam.micro16}, which proposes a method of
integrating logic within the DRAM module but outside of the chip to reduce
manufacturing costs, Gather-Scatter DRAM~\cite{GS-DRAM}, which
embeds logic within the memory controller to remap a single memory request
across multiple rows and columns within DRAM, and work by 
Hashemi et al.~\cite{hashemi.isca16, cont-runahead} to embed logic in the
memory controller that accelerates dependent cache misses and performs
runahead execution\chiii{~\cite{DBLP:conf/hpca/MutluSWP03, DBLP:conf/isca/MutluKP05, 
DBLP:journals/micro/MutluSWP03, DBLP:journals/micro/MutluKP06}}.}

\paratitle{\chii{Addressing Challenges to PIM Adoption}}
\chii{Recent work has examined design challenges for systems with PIM support
that can affect PIM adoption.  A number of these works improve PIM programmability,
such as LazyPIM~\cite{boroumand2016pim, boroumand.arxiv17}, which 
\chiii{provides efficient cache coherence support for PIM (as we described in
detail in Section~\ref{sec:lazypim})}
the study by Sura et al.~\cite{sura.cf15},
which optimizes how programs access PIM data, and work by Liu et al.~\cite{liu-spaa17},
which designs \chiii{PIM-specific concurrent data structures} to improve PIM performance.
\chiii{Other works} tackle hardware-level design challenges, including IMPICA~\cite{impica},
which introduces in-memory support for address translation and pointer chasing
\chiii{(as we described in detail in Section~\ref{sec:impica})},
work by Hassan et al.~\cite{hassan.memsys15} to
optimize the 3D-stacked DRAM architecture for PIM,}
\chv{and work by Kim et al.~\cite{kim.sc17} to enable the distribution of PIM data
across multiple memory stacks.}

\paratitle{Coherence for PIM Architectures}
In order to avoid the overheads of fine-grained coherence, \chiii{many} prior works on
PIM architectures design their systems in such a way that they do not need to
utilize traditional coherence protocols.  
Instead, these works use one of two alternatives.
Some works restrict \chiii{PIM processing logic} to execute on only non-cacheable
data (e.g., \chii{\cite{ahn.tesseract.isca15, farmahini-farahani.hpca15, zhang.hpdc14, 
morad.taco15, DBLP:conf/hpca/GaoK16}}),
which forces cores within the CPU to read PIM data directly from DRAM.
Other works use coarse-grained coherence or coarse-grained locks, 
which force processor cores to not access any data that could 
\emph{potentially} be used by the \chiii{PIM processing logic}, 
or to flush this data back to
DRAM before the PIM kernel begins executing (e.g.,
\chiii{\cite{farmahini-farahani.hpca15, gao.pact15, Seshadri:2015:ANDOR, 
seshadri2013rowclone, guo2014wondp, ahn.pei.isca15, impica, nai2017graphpim,
hsieh.isca16, seshadri.arxiv16, seshadri.micro17, chang.hpca16, pattnaik.pact16,
DBLP:conf/isca/AkinFH15, impica}}).
Both of these approaches generates significant coherence overhead, as discussed
in Section~\ref{sec:motivation}.
Unlike these approaches, LazyPIM \chiii{(Section~\ref{sec:lazypim})} places no restriction on the way in which 
processor cores and \chiii{PIM processing logic} can access \chiii{data.}
Instead, LazyPIM uses PIM-side coherence speculation and
efficient coherence message compression to \chiii{provide} cache coherence, which
avoids the communication overheads associated with traditional coherence
protocols.

\paratitle{Accelerators in CPUs}
There have been various CPU-side accelerator
proposals for database systems (e.g.,~\cite{kocberber2013meet, ChungDL13, WuBKR13,
WuLPKR14}) and key-value stores~\cite{LimMSRW13}. 
\chiii{Widx~\cite{kocberber2013meet} is a database indexing accelerator that uses}
a set of custom RISC cores in the CPU to
accelerate hash index lookups. While a hash table is one of \kii{our} data
structures of interest, \chvi{IMPICA (Section~\ref{sec:impica})} differs from Widx in three \chviii{major} ways. First,
\kii{it is} an \emph{in-memory} (as opposed to CPU-side) accelerator, which poses very different design
challenges. Second, we solve the address translation challenge for
in-memory accelerators, while Widx uses the CPU \kii{address
  translation structures}. Third, we enable parallelism within a single accelerator core, while Widx
achieves parallelism by replicating several RISC cores.

\paratitle{Prefetching for Linked Data Structures}
Many works 
propose mechanisms to prefetch data in linked data structures to hide
memory latency. These proposals \kh{are}
hardware-based (e.g.,\kii{~\cite{JosephG97, CollinsSCT02, CookseyJG02,
  RothS99, DBLP:conf/hpca/HuMK03, MutluKP05, HughesA05, DBLP:conf/micro/YuHSD15, DBLP:journals/tc/MutluKP06}}), software-based
(e.g.,~\cite{LipastiSKR95, LukM96, roth1998, YangL00, Wu02}),
pre-execution-based (e.g.,~\cite{ZillesS01, Luk01, CollinsWTHLLS01,
  SolihinTL02}), or software/hardware-\kii{cooperative}
(e.g.,\kiii{~\cite{RothS99, ebrahimi2009, KarlssonDS00}}) mechanisms. 
These approaches have two
major drawbacks. First, they \chiii{usually} rely on predictable traversal sequences
to prefetch accurately. \chiii{As a result, many of these} mechanisms \kii{can become very} inefficient
\ignore{and inaccurate }if the linked data structure is \kh{complex} or
when access patterns are less regular.  Second, the pointer chasing \chiii{or prefetching} is
performed at the \kii{CPU cores or at the} memory controller, which likely leads to
pollution of the CPU caches and TLBs by these irregular memory
accesses.

% !TEX root=../chapter.tex

\section{Other System-Level Challenges for PIM Adoption}
\label{sec:future}

\ch{IMPICA (Section~\ref{sec:impica}) and LazyPIM (Section~\ref{sec:lazypim})
demonstrate the need for and gains that can be \chiii{achieved} by designing system-level
solutions that \chiii{are applicable} across a wide variety of PIM architectures.
In order for PIM to achieve widespread adoption, we believe there are a number
of other system-level challenges that must be addressed. In this section,
we discuss \chix{six} research directions that aim towards solving these
challenges:
(1)~the PIM programming model,
(2)~data mapping,
(3)~runtime scheduling support for PIM,
(4)~the granularity of PIM scheduling,
(5)~evaluation infrastructures and benchmark suites for PIM, \chix{and
(6)~applying PIM to emerging memory technologies}.}

\paratitle{\ch{PIM Programming Model}}
\ch{
Programmers need a well-defined interface to incorporate PIM functionality
into their applications.
Determining the programming model for how a programmer should invoke and 
interact with \chiii{PIM processing logic} is an open research direction. Using a set of special
instructions allows for very fine-grained control of when \chiii{PIM processing logic is} invoked, 
which can potentially result in a significant performance improvement. 
However, this approach can potentially \chii{introduce overheads while taking} advantage 
of PIM, due to the need to frequently exchange information between \chiii{PIM processing logic} and the CPU. Hence,
there is a need for researchers to investigate how to integrate PIM instructions with other
compiler-based methods or library calls that can support PIM integration, and
how these approaches can ease the burden on the programmer.
\chii{\chiii{For example, one} of our recent works~\cite{hsieh.isca16} examines compiler-based
mechanisms to decide what portions of code should be offloaded to \chiii{PIM processing logic}
in a GPU-based system.}
\chiii{Another recent work~\cite{pattnaik.pact16} examines system-level techniques
that decide which GPU application kernels are suitable for PIM execution.}
}

\paratitle{\ch{Data Mapping}}
\ch{
\chvii{Determining the ideal memory mapping for data used by \chiii{PIM processing logic} 
is another important research direction. To maximize the benefits of PIM, data that needs 
to be read from or written to by a single PIM kernel instance should be mapped to the same
memory stack. Hence, it is
important to examine both static and adaptive data mapping mechanisms to
intelligently map (or remap) data. }
Even with such data mapping mechanisms, it is beneficial to
provide low-cost and low-overhead data migration mechanisms to facilitate
easier PIM execution, in case the data mapping needs to be adapted to execution and
access patterns at runtime. One of our recent works provides a mechanism that
provides programmer-transparent data mapping support for
PIM~\cite{hsieh.isca16}.  Future work can focus on developing \chvii{new}
data mapping mechanisms, as well as designing systems that can
take advantage of these new data mapping mechanisms.}

\paratitle{\ch{PIM Runtime Scheduling Support}}
\ch{
At least four key runtime issues in PIM are to decide (1)~when to enable PIM
execution, (2)~what to execute near data, (3)~how to map data to multiple
(hybrid) memory modules such that PIM execution is viable and effective, and
(4)~how to effectively share/partition PIM mechanisms/accelerators at runtime
across multiple threads/cores to maximize performance and energy efficiency.
It is possible to build on our recent works that employ locality
prediction~\cite{ahn.pei.isca15} and combined compiler and dynamic code
identification and scheduling in GPU-based systems
~\cite{hsieh.isca16,pattnaik.pact16}. 
Several key research questions that should be investigated include:
\begin{itemize}
\item What are simple mechanisms to enable and disable PIM execution? How can PIM execution be throttled for highest
performance gains? How should data locations and access patterns affect where/whether PIM execution should occur?
\item Which parts of the application code should be executed on PIM? What are simple mechanisms to identify such code?
\item What are scheduling mechanisms to share PIM accelerators between multiple requesting cores to maximize \chvii{PIM's benefits}? 
\end{itemize}
}

\paratitle{\ch{Granularity of PIM Scheduling}}
\chii{To enable the widespread adoption of PIM, we must understand the ideal
granularity at which PIM operations can be scheduled without sacrificing 
\chvii{PIM execution's efficiency and limiting changes} to the shared
memory programming model. Two key issues for scheduling code for PIM 
execution are (1)~how large each part of the code should be (i.e., the 
granularity of PIM execution), and (2)~the frequency at which code executing on a PIM 
engine should synchronize with code executing on the CPU cores (i.e., the 
granularity of PIM synchronization).}

\chii{The optimal granularity of PIM execution remains an open question.
For example, is it best to offload only a single instruction to the \chiii{PIM processing logic}?
Should PIM kernels consist of a set of instructions, 
and if so, how large is each set?  Do we limit PIM execution to work only on
entire functions, entire threads, or even entire applications?
If we offload too short a piece of code, the benefits
of executing the code near memory may be unable to overcome the overhead of 
invoking PIM execution (e.g., communicating registers or data, taking checkpoints).}

\chii{Once code begins to execute on \chiii{PIM processing logic, there} may be times where the code needs to
synchronize with code executing on the CPU.  For example, many shared memory
applications employ locks, barriers, or memory fences to coordinate access to
data and ensure correct execution.  PIM system architects must determine
(1)~whether code executing on PIM should allow the support of such
synchronization operations; and
(2)~if they do allow such operations, how to perform them efficiently.
Without an efficient mechanism for synchronization, \chiii{PIM processing logic} may need to
communicate frequently with the CPU when synchronization takes place, 
which can introduce overheads and undermine the benefits of PIM execution.
Research on PIM synchronization can build upon our prior work, where we \chiii{limit}
PIM execution to atomic instructions to avoid the need for synchronization~\cite{ahn.pei.isca15},
or where \chiii{provide} support within LazyPIM to perform synchronization
during PIM kernel execution~\cite{boroumand.arxiv17}.}

\paratitle{\ch{PIM Evaluation Infrastructures and Benchmark Suites}}
\ch{
\chii{To ease adoption, it is critical that \chiii{we accurately} assess the benefits
of PIM.  Accurate assessment for PIM requires
(1)~a set of real-world memory-intensive applications that have the potential to benefit significantly 
when executed near memory, and
(2)~a \chiii{simulation/evaluation} infrastructure that allows architects and system designers to
precisely analyze the benefits and overhead of adding \chiii{PIM processing logic} to memory and
executing code on \chiii{this processing logic}.}

\ch{\chii{In order to identify what \chiii{processing logic} should be introduced near memory, and 
to know what properties are ideal for PIM kernels, we must begin by developing a
real-world benchmark suite of applications that can potentially benefit from PIM.
\chvii{While many data-intensive applications, such as pointer chasing and bulk memory copy,
can potentially benefit from PIM,} it is crucial to examine important candidate applications for PIM
execution, and for researchers to agree on a common set of these candidate applications
to focus the efforts of the community.  
We believe that these applications should come from a number of popular and emerging
domains.  Examples of potential domains include} data-parallel applications, neural networks, 
\chiii{machine} learning, graph processing, \chiii{data analytics,} search/filtering, \chviii{mobile workloads,}
bioinformatics, Hadoop/Spark programs, and in-memory data stores.  
\chiii{Many of these} applications have large data sets and can
benefit from high memory bandwidth and low memory latency benefits provided by
PIM mechanisms.}
\chii{As an example, in
our prior work, we have started analyzing mechanisms for accelerating graph
processing~\cite{ahn.pei.isca15,ahn.tesseract.isca15};
pointer chasing~\cite{cont-runahead,impica};
\chiii{\chviii{databases}~\cite{boroumand2016pim, boroumand.arxiv17, impica, GS-DRAM};} 
\chviii{consumer workloads~\cite{boroumand.asplos18}, including web browsing,
video encoding/decoding, and machine learning;}
and GPGPU workloads~\cite{hsieh.isca16,pattnaik.pact16}.}

\chii{Once we have established a set of applications to explore,
it is essential for researchers to develop an extensive and flexible application profiling and simulation
infrastructure and mechanisms that can (1)~identify parts of these applications
for which PIM execution can be beneficial; and
(2)~simulate in-memory acceleration.
A systematic process for identifying potential PIM kernels within an application can not only ease
the burden for performing PIM research, but could also inspire tools that 
programmers and compilers can use to automate the process of offloading portions
of existing applications to \chiii{PIM processing logic}.
Once we have identified potential PIM kernels, we need a simulator to accurately model 
the energy and performance consumed by PIM hardware structures, 
available memory bandwidth, and communication overhead when
we execute the kernels near memory.}
\chii{Highly-flexible memory} simulators (e.g., Ramulator\chiii{~\cite{ramulator, ramulator.github}}, 
SoftMC\chiii{~\cite{hassan2017softmc, softmc.github}})
\chii{can be combined} with full-system simulation infrastructures (e.g., gem5~\cite{GEM5})
\chii{to} provide a robust environment \chii{that can} evaluate how various PIM architectures
affect the \chii{\emph{entire compute stack}, and can allow designers to
identify memory characteristics (e.g., internal bandwidth, trade-off between number
of PIM engines and memory capacity) that affect the efficiency of PIM execution}.

\paratitle{\chviii{Applicability to Emerging Memory Technologies}}
\chviii{As DRAM scalability issues are becoming more difficult to work 
around\chix{~\cite{dean.cacm13, kanev.isca15, mckee.cf04, mutlu.imw13,
mutlu.superfri15, wilkes.can01,kim-isca2014,salp,kang.memoryforum14,yoongu-thesis,raidr,mutlu2017rowhammer,ahn.tesseract.isca15,ahn.pei.isca15,hsieh.isca16,donghyuk-ddma,lee-isca2009,rbla,yoon-taco2014,lim-isca09, wulf1995hitting, chang.sigmetrics16, lee.hpca13, lee.hpca15, chang.sigmetrics17, lee.sigmetrics17,
luo.dsn14, luo.arxiv17,hassan2017softmc,chargecache}},
there has been a growing amount of work on emerging non-volatile memory
technologies to replace DRAM.  Examples of these emerging memory technologies
include \emph{phase-change memory} (PCM)~\cite{lee-isca2009, qureshi.isca09,
wong.procieee10, lee.ieeemicro10, zhou.isca09, lee.cacm10, yoon-taco2014},
\emph{spin-transfer torque magnetic RAM} (STT-MRAM)~\cite{naeimi.itj13, kultursay.ispass13},
\emph{metal-oxide resistive RAM} (RRAM)~\cite{wong.procieee12},
and \emph{memristors}~\cite{chua.tct71, strukov.nature08}.
These memories have the potential to offer
much greater memory capacity and high internal memory bandwidth.
Processing-in-memory techniques can take advantage of this potential, by
exploiting the high available internal memory bandwidth, and by making use of
the underlying memory device behavior, to perform computation.}

\chviii{PIM can be especially useful in single-level store settings\chix{~\cite{bensoussan.cacm72,
kilburn.iretec62, shapiro.sosp99, shapiro.usenixatc02, meza.weed13, zhao.micro14, ren.micro15}},
where multiple memory and storage technologies (including emerging
memory technologies) are presented to the system as a single monolithic
memory, which can be accessed quickly
and at high volume by applications.  By performing some of the computation
in memory, PIM can take advantage of the high bandwidth and capacity available
within a single-level store without being bottlenecked by the limited off-chip
bandwidth between the various memory \chix{and system software} components of the single-level store
and the CPU.}

\chviii{Given the worsening DRAM scaling issues, and the limited bandwidth
available between memory and the CPU, we believe that there is a growing
need to investigate PIM processing logic that is designed for emerging
memory technologies.  We believe that many PIM techniques can be applicable
in these technologies.
Already, several prior works propose to exploit memory device behavior to
perform processing using memory, where the memory consists of
PCM~\cite{li.dac16} or memristors~\cite{shafiee.isca16, levy.microelec14, kvatinsky.tcasii14, kvatinsky.iccd11, kvatinsky.tvlsi14}.
Future research should explore how PIM can take advantage of emerging
memory technologies in other ways, and how PIM can work effectively in
single-level stores.}

% !TEX root=../chapter.tex

\section{Conclusion}
\label{sec:conclusion}

\ch{\chii{Circuit and device technology scaling for main memory, built
predominantly with DRAM, is already showing signs of coming to an end,
with three major issues emerging.
\chiii{First, the reliability and \chiv{data} retention capability of DRAM have been 
decreasing, as shown by various error characterization and analysis 
studies\chix{~\cite{raidr, liu.isca13, mukundan.isca13, kang.memoryforum14, 
chang.hpca14, khan.sigmetrics14, qureshi.dsn15, patel.isca17,
khan.dsn16, mandelman.ibmjrd02, schroeder.sigmetrics09, sridharan.asplos15, 
meza.dsn15, mutlu2017rowhammer, hassan2017softmc, khan.micro17, khan.cal16,
lee.thesis16, kim.hpca18}},
and new failure mechanisms have been slipping into devices in the field (e.g.,
Rowhammer\chiv{~\cite{mutlu2017rowhammer, kim-isca2014, kim.arxiv16, yoongu-thesis}}).}
Second, main} memory
performance improvements have not grown as rapidly as logic performance
improvements have for several years now, 
\chii{resulting in significant performance bottlenecks\chiv{~\cite{mutlu.imw13, mutlu.superfri15, 
mckee.cf04, wulf1995hitting, chang.sigmetrics16,kevinchang-thesis, lim-isca09,
lee.hpca15, lee.hpca13, donghyuk-ddma, lee.sigmetrics17, seshadri.thesis16, lee.thesis16, 
ahn.tesseract.isca15}}.
Third, the} increasing application demand
for memory places even greater pressure on the main memory system \chii{in terms of
both performance and energy efficiency\chviii{~\cite{dean.cacm13, kanev.isca15, mckee.cf04, 
mutlu.imw13, mutlu.superfri15, wilkes.can01,kim-isca2014,salp,kang.memoryforum14,
yoongu-thesis,raidr,mutlu2017rowhammer,ahn.tesseract.isca15,ahn.pei.isca15,
hsieh.isca16,donghyuk-ddma,lee-isca2009,rbla,yoon-taco2014,lim-isca09,
wulf1995hitting, chang.sigmetrics16, lee.hpca13, lee.hpca15, chang.sigmetrics17, lee.sigmetrics17,
luo.dsn14, luo.arxiv17, boroumand.asplos18}}}.}
\chii{To solve these issues, there is an increasing
need for architectural and system-level approaches\chiii{~\cite{mutlu.imw13, mutlu.superfri15, mutlu2017rowhammer}}.}

\chii{A major hindrance to memory performance and energy efficiency is the high 
cost of moving data between the CPU and \ch{memory}.  Currently, this cost \chiii{must be} paid
\emph{every time} an application needs to perform an operation on data that is stored
within \ch{memory}.}
The recent advent of 3D-stacked \ch{memory} architectures, which contain
a layer dedicated for logic within the same stack as \ch{memory} layers, \chiii{open}
new possibilities to reduce unnecessary data movement by allowing architects to
shift some computation into \ch{memory}.  Processing-in-memory (PIM), or near-data 
processing, allows architects to introduce simple \chiii{PIM processing logic} \ch{(which can be 
\chvi{specialized acceleration logic, general-purpose cores, or reconfigurable logic})} into the logic 
layer of the \ch{memory}, where the \chiii{PIM processing logic has} access to the high internal bandwidth
\chii{and low memory access latency that exist} within 3D-stacked \ch{memory}.  
As a result, PIM architectures can reduce \chii{costly data movement over the 
memory \chiii{channel, lower memory access latency, and thereby also} reduce energy consumption.}

\chii{A number of challenges exist in enabling PIM at the system level, such 
that PIM can be adopted easily \chiii{in many system designs}.}
In this work, we examine two \chii{such key design issues}\ch{, which
we believe require efficient and elegant solutions to enable widespread adoption 
of PIM in real systems.  First, because applications store memory references as
virtual addresses, \chiii{PIM processing logic} needs to perform \emph{address translation} to 
determine the physical addresses of these references during execution.  However,
\chiii{PIM processing logic does not have} an efficient way of accessing to the translation lookaside buffer 
or the page table walkers that reside in the CPU.
Second, because \chiii{PIM processing logic} can often access the same data structures that are
being accessed and modified by the CPU, 
\chii{a system that incorporates PIM cores needs to support cache coherence between the CPU and PIM cores}
to ensure that \chii{all of the cores are} using the correct version of the data.}
Naive solutions to overcome \chii{the address translation and cache coherence challenges} either place significant 
restrictions on the types of computation that can be performed by \chiii{PIM processing logic},
which can break the existing multithreaded programming model and prevent the 
widespread adoption of PIM, or force \chiii{PIM processing logic} to communicate with the
CPU frequently, which can undo the benefits of moving computation to \ch{memory}.
Using key observations about the behavior of address translation and cache
coherence for several memory-intensive applications, we propose two solutions
that \ch{(1)~provide general purpose support for translation and
coherence in PIM architectures, (2)~maintain the conventional multithreaded
programming model, and (3)~do not incur high communication overheads.}
The first solution, IMPICA, provides \chiii{an efficient in-memory
accelerator} for pointer chasing that can perform efficient address translation
from within \ch{memory}.  The second solution, LazyPIM, provides an efficient cache
coherence protocol that does not \ch{restrict how \chiii{PIM processing logic} and the CPU
share data,} by using speculation and coherence
message compression to minimize the overhead of PIM coherence requests.

\chii{We hope that our solutions to the address translation and cache
coherence challenges can ease the adoption of PIM-based
architectures, by easing both the design and programmability of such
systems. We also hope that the challenges \chiii{and ideas} discussed
in this chapter can inspire other researchers to develop \chiii{other} novel
solutions that can ease the adoption of PIM architectures.}

\section*{Acknowledgments}
\ch{We thank all of the \chiii{members of} the SAFARI Research Group, and our 
collaborators at Carnegie Mellon, \chiii{ETH Z{\"u}rich}, and other universities, who have
contributed to the \chiii{various} works we describe in this chapter.
Thanks also goes to our research group's industrial sponsors over the past
\chiii{nine} years, \chiii{especially Google, Huawei, Intel, Microsoft, NVIDIA, 
Samsung, Seagate, and VMware}.
This work was also partially supported by}
the Intel Science and Technology Center for Cloud Computing, 
the Semiconductor Research Corporation, 
the Data Storage Systems Center at Carnegie Mellon University,
and NSF grants 1212962, 1320531, and 1409723.

% Bibliography
\newpage
\bibliographystyle{IEEEtranS}
\bibliography{refs}

\end{document}